\documentclass[11pt]{article}
\usepackage[letterpaper, margin=2.54cm]{geometry}
\usepackage{graphicx} 
\usepackage{rotating}
\usepackage{tabularx}
\usepackage{array}
\usepackage{arydshln}
\usepackage{amsmath,bm}
\usepackage{changepage}

\usepackage[natbib=true,style=authoryear,backend=biber,useprefix=true,maxbibnames=99,mincitenames=1,maxcitenames=2]{biblatex}
\usepackage{threeparttable}
\usepackage{tablefootnote}
\usepackage{color}
\usepackage{hyperref}
\usepackage{caption}
\usepackage{subcaption}
\usepackage{graphicx}
\usepackage{array}
\usepackage{float}

\usepackage{pifont}
\usepackage{setspace}

\usepackage{comment}

\makeatletter
\let\TPT@hookin\@gobble
\let\TPT@hookarg\@gobble
\makeatother

\title{
  \makebox[\textwidth][c]{%
    \parbox{1\textwidth}{\centering\huge
    \vspace{-2.7cm} Generative AI and Sales Productivity:\\
    Field Experiments in Online Retail\thanks{The authors thank Kinshuk Jerath, Carl Mela, and Andrey Simonov for valuable feedback, as well as seminar and conference participants at the Marketing Science Conference, BSE Quantitative Marketing Workshop, Theory and Practice in Marketing, Duke University, Imperial Business School, BIG AI Conference, MSI AI Forum, TSE Digital Economics Conference, AIML Conference, Operational Innovation Network Summit, University of Rochester, UC Davis, TUM Workshop on Generative AI in Marketing, UCL School of Management, Columbia Business School, Business \& Generative AI Conference, and Zhejiang University School of Management for helpful comments and discussions. The authors thank the partner e-commerce company for its support with this project and are grateful to its employees for patiently answering many of our questions. The paper only reflects the opinion of the authors, and not the opinion of any organizations. All remaining errors are the responsibility of the authors. Lu Fang acknowledges financial support from the National Natural Science Foundation of China [Grants 72501258 and 72192803], the National Social Science Fund of China [Grant 22\&ZD081]. Zhe Yuan acknowledges financial support from the National Key Research and Development Project [Grant 2024YFB3312900], the National Natural Science Foundation of China [Grants 72141305 and 72203202], Major Project of the National Social Science Fund of China [Grant 25\&ZD149], the Fundamental Research Funds for the Central Universities. Kaifu Zhang, Dante Donati, and Miklos Sarvary have no funding to report. Lu Fang and Zhe Yuan served as consultants for the partner company during the duration of the project. Kaifu Zhang was an employee of the partner company during the duration of the project. Dante Donati and Miklos Sarvary have no conflicts of interest to report.
    }}
  }\vspace{-0.9cm}
}
\author{{\Large Lu Fang}\thanks{School of Management, Zhejiang University of Finance \& Economics; Laboratory for Human-AI Collaboration and Digital Intelligence Innovation, Zhejiang University; fl.fanglu.fl@gmail.com}\hspace{25pt} 
{\Large Zhe Yuan}\thanks{Corresponding author. Center for Research of Private Economy, School of Economics, Future Regional Development Laboratory, and Academy of Financial Research, Zhejiang University; yyyuanzhe@gmail.com}\hspace{25pt}
{\Large Kaifu Zhang}\thanks{Partner Company, zhangkaifu314@gmail.com}\vspace{10pt} 
\\{\Large Dante Donati}\thanks{Columbia Business School; dd3137@gsb.columbia.edu}\hspace{25pt} 
{\Large Miklos Sarvary}\thanks{Columbia Business School; miklos.sarvary@columbia.edu}\vspace{5pt}}

\date{\vspace{-0.1cm}June 30, 2026}

\begin{document}

\maketitle

\vspace{-1cm} \begin{abstract}
We quantify the short-term impact of Generative Artificial Intelligence (GenAI) on sales performance through a series of large-scale randomized field experiments involving millions of users and products at a leading cross-border online retail platform. Over 2023-2024, the platform integrated GenAI into seven business workflows spanning customer service, consumer-product matching, advertising, and seller services. We find that GenAI adoption increases sales in most workflows, with effects ranging from no detectable impact to 16.3\%, depending on GenAI’s marginal contribution relative to baseline firm practices. Across the four GenAI applications with positive sales effects, the implied annual incremental value is roughly \$5 per consumer—an economically meaningful impact given the retailer’s scale and the early stage of GenAI adoption. The gains operate primarily through higher conversion rates rather than larger cart values, consistent with GenAI improving the shopping experience by reducing search, information, communication, and personalization frictions. Importantly, these effects are not associated with worse post-purchase outcomes, as product return rates and customer ratings do not deteriorate. Finally, we document substantial demand-side heterogeneity, with larger gains for less experienced consumers. Our findings provide novel, large-scale causal evidence on how GenAI shapes sales productivity in online retail, highlighting both its immediate value and broader potential.
\end{abstract}

\hspace{+0.2cm}\makebox[\textwidth][l]{\small \textbf{Keywords}: AI, Consumer Experience, Field Experiments,  Productivity, Retail Platforms, Sales }

\hspace{+0.2cm}\makebox[\textwidth][l]{\small  \textbf{JEL codes}: C93, D24, L81, O33}

\newpage

\onehalfspacing
\noindent {\small ``\emph{The next industrial revolution has begun},'' Nvidia Chief Executive Officer Jensen Huang said. \emph{``AI will bring significant productivity gains to nearly every industry and help companies be more cost- and energy-efficient, while expanding revenue opportunities.''} \textit{Nvidia Stock Surges as Sales Forecast Delivers on AI Hopes}, Bloomberg, May 22, 2024.}\\

\vspace{-0.2cm}
\section{Introduction}
 The rapid diffusion of Generative Artificial Intelligence (GenAI) has sparked growing interest in its potential to improve productivity across sectors of the economy \citep{acemoglu2025simple}. At the same time, investors and industry practitioners have increasingly questioned whether large AI investments will translate into sustained business returns.\footnote{See, for example, recent articles by \textit{Sequoia Capital}: \href{https://www.sequoiacap.com/article/ais-600b-question/}{www.sequoiacap.com/article/ais-600b-question}; and \textit{The Economist}: \href{https://www.economist.com/leaders/2025/09/11/what-if-the-3trn-ai-investment-boom-goes-wrong}{www.economist.com/leaders/2025/09/11/what-if-the-3trn-ai-investment-boom-goes-wrong}.}  A growing literature shows that GenAI can enhance performance in a range of worker-level tasks, including software development, customer support, education, law, and other professional services \citep[e.g.,][]{Bjorn2025, Noy2023, Cui2026, DellAcqua2024, Choi2023, Kestin2025}. Yet, despite rapid adoption, there is still limited causal evidence on whether GenAI improves sales performance in real-world, consumer-facing business workflows. This question is especially important in online retail platforms, which, over the past thirty years, have reduced consumers' search, information, and communication costs  throughout the customer journey \citep{belleflamme2021economics}.\\

This paper provides large-scale experimental evidence on the short-run effect of GenAI on sales performance in online retail. We study seven randomized field experiments conducted on one of the world’s largest cross-border e-commerce platforms between September 2023 and June 2024. The experiments evaluate GenAI deployments in seven business workflows across three broad functional areas: consumer and seller services, consumer-product matching, and advertising and promotion. This setting allows us to compare the effects of GenAI across multiple business functions within the same firm, under a common implementation environment but against different baseline conditions. Each application was evaluated through randomized field experiments, with sample sizes ranging from tens of thousands to tens of millions of users or products. Leveraging granular consumer- and product-level data, we examine the impact of GenAI on sales, measured by consumer expenditure, as well as on conversion, cart value, intermediate engagement outcomes, and post-purchase outcomes. This design allows us to assess not only whether GenAI generates measurable business value, but also where and for whom these gains are most likely to arise.\\

We document three main findings. First, most GenAI deployments improve sales, but the effects vary substantially across workflows, ranging from no detectable impact to gains of up to 16.3\%, with the largest improvements observed in customer service and search-related applications. Aggregating across the four GenAI applications with positive sales effects, we estimate an annual incremental value of approximately \$5 per consumer.
Second, the evidence suggests that GenAI improves sales primarily by increasing purchase incidence rather than spending intensity: across several workflows, engagement and conversion rates rise, while average cart values remain largely unchanged, consistent with an enhanced consumer experience. We also find no evidence that these gains come at the expense of ex post consumer satisfaction, as return rates and customer ratings do not deteriorate following GenAI adoption. Third, the effects are heterogeneous across market participants, with less experienced consumers benefiting disproportionately more from GenAI-powered workflows. Taken together, these results show that GenAI can generate economically meaningful improvements in online retail performance, partly by further reducing frictions, while also revealing substantial heterogeneity in where these gains arise and who benefits from them.\\

A key distinction of our study is that it examines GenAI in real-world, consumer-facing retail workflows, focusing on front-office sales and related business outcomes rather than back-office worker-level task efficiency alone \citep[e.g.,][]{Bjorn2025,Noy2023,Cui2026}. Using seven large-scale field experiments within a single online retail platform, we identify the short-run impact of GenAI under a common organizational environment, holding other observed inputs and prices fixed. Because all experiments take place within the same firm, this setting allows us to compare GenAI across workflows that differ in function and baseline conditions without conflating those differences with variation in implementation environments across studies. The results point to heterogeneous gains across applications and users, and are consistent with GenAI improving consumer experience by reducing search, information, and personalization frictions, though our design does not separately identify those channels from other possible mechanisms.\\

We partnered with a leading cross-border e-commerce platform to identify and quantify the sales productivity gains from GenAI adoption in online retail. The platform enables consumers worldwide to purchase directly from manufacturers at competitive prices.\footnote{The platform connects hundreds of thousands of predominantly small-business sellers with hundreds of millions of active buyers across more than 100 countries and regions. The platform supports around 20 languages, providing localized services that facilitate global communication and accessibility.} Between September 2023 and June 2024, the firm deployed GenAI solutions across seven business workflows: (1) Pre-sale Service Chatbot, (2) Search Query Refinement, (3) Product Description Generation, (4) Marketing Push Message Creation, (5) Google Advertising Title Optimization, (6) Chargeback Defense, and (7) Live Chat Translation. These workflows span three broader functional areas: (i) consumer and seller services (1, 6, 7); (ii) consumer–product matching (2, 3); and (iii) advertising and promotion (4, 5). Each workflow corresponds to a distinct stage of the customer journey, allowing us to assess GenAI’s sales productivity impact across a wide range of retail operations.\\

Our setting involves multiple experiments applying GenAI to distinct business workflows. Several features enhance comparability across these experiments: all applications were developed and deployed by the same technical team and operated within the same firm under similar organizational and competitive conditions. Despite this common implementation environment, the effect of GenAI on workflow performance is likely to depend on its marginal contribution relative to baseline conditions, which differ across workflows. In each case, the control group reflects the firm’s standard practices prior to GenAI adoption. For example, in the Pre-sale Service Chatbot workflow, the control group received no live or interactive pre-sale customer support; in Search Query Refinement, the control condition relied on standard machine-learning-based search algorithms with only lexical translation; and in most other workflows, the benchmark was human input. \\

Within each workflow, the GenAI deployment was evaluated through a large-scale randomized field experiment that compared the GenAI-enhanced workflow in the treatment group to a baseline version used in the control group. Notably, baseline workflows often included automation or human input but did not incorporate GenAI technologies. The treatment condition differed solely through the integration of GenAI, while prices, other inputs and the broader workflow environment remained constant across conditions. Randomization occurred at the level of consumers—and, in one case, products—with minimal overlap (less than 1\%) across experiments, enabling causal identification of the effects of GenAI on consumer behavior. For five of the seven experiments, we obtained detailed consumer- or product-level engagement and transaction data, including views, clicks, number of orders, expenditure, conversions, product returns, and customer ratings. For analysis, we aggregate these data to the consumer level (and to the product level in one experiment) and leverage consumer, seller, and product characteristics to study treatment-effect heterogeneity. Our primary outcome is sales value, measured by total consumer expenditure. We also examine conversion rates, a standard measure of consumer experience in online retail; cart value, which captures changes in purchase intensity among consumers who make a purchase; and returns and ratings, reflecting customer decision quality and ex post satisfaction.\\

Our results reveal that most GenAI deployments yield economically significant short-term gains in sales, though the magnitudes differ across workflows. Across the five processes with detailed data, sales effects range from no detectable impact in the advertising workflows to gains of up to 16.3\% in the Pre-sale Service Chatbot. In the Search Query Refinement and Product Description workflows, the effects are smaller, typically in the range of 2--3\%, yet still substantial for a platform of this scale and maturity. In percentage terms, these effects are smaller than estimates from GenAI applications in worker-level tasks, ranging from about 15\% in customer support to 55.8\% in coding \citep{Bjorn2025,Peng2023}. But the comparison should be interpreted cautiously: we measure downstream sales outcomes in a mature retail platform, rather than worker-level efficiency such as task completion time, so even a 2--3\% increase can imply economically meaningful gains when applied at scale. Back-of-the-envelope calculations based on the four deployments with detailed transaction data and positive effects suggest that these GenAI applications generate an annual incremental value of approximately \$4.6--\$5.2 per consumer. These effects correspond to roughly 5.5--6.2\% of the per-user revenue growth observed in global e-commerce between 2023 and 2024. We also document notable improvements in workflows without granular data: a 15\% increase in success rate for Chargeback Defense, and a 5.2\% increase in self-reported consumer satisfaction from Live Chat Translation.\\

Taken together, these results show that GenAI generates sizable gains in targeted workflows and  meaningful effects for a large, mature retailer, with further potential as adoption broadens and  targets revenue-critical processes. For example, while in 2023 the platform applied GenAI to only a handful of workflows, by 2024 it had expanded to more than 40 applications and by 2025 to over 60. Meanwhile, API calls to proprietary GenAI tools increased twentyfold between 2024 and 2025, reflecting the rapid scaling of GenAI adoption across the platform. The long-run impact will ultimately depend on equilibrium forces, specifically whether complementarities across workflows amplify these gains or industry-wide adoption offsets them through intensified competition. \\

We next study the potential mechanisms underlying the sales effects. Across workflows, sales gains are accompanied by higher conversion rates—and, where available, by improvements in intermediate engagement metrics such as click-through rates—while cart values and purchase intensity conditional on purchase remain largely unchanged. Conversion rates increase by 1--22\%, with little evidence of changes in spending intensity among buyers. Importantly, we also find no deterioration in the post-purchase outcomes available to us: product return rates and customer ratings do not decline following GenAI adoption, and in some cases returns fall or ratings improve. Taken together, these results suggest that GenAI primarily affects purchase incidence rather than spending conditional on purchase, while providing no evidence in our measured post-purchase outcomes of lower ex post consumer satisfaction.\\

The overall pattern is consistent with GenAI affecting sales through both informational and persuasive channels, with their relative importance differing across workflows. In some applications, especially Search Query Refinement and Live Chat Translation, the most natural interpretation is friction reduction: GenAI improves semantic understanding, matching, or communication in ways that help consumers and service agents interact more effectively. In others, such as Pre-sale Service Chatbot, Product Description, Marketing Push Message, and Google Advertising Title, GenAI may operate through both channels: providing richer and more relevant information on the one hand, while also making content more salient, polished, or engaging on the other hand. These channels are especially plausible in our multilingual cross-border setting, where search, information, and communication frictions are naturally high. Although our data do not allow us to separately identify the contribution of each channel, we quantify their combined effect on workflow-level sales performance.\\

Finally, we examine heterogeneity in treatment effects across buyers, sellers, and products. A natural hypothesis is that GenAI may generate larger gains for participants who either face greater frictions to begin with or respond more strongly to changes in consumer-facing content, such as less experienced buyers, smaller and newer sellers, and products that are harder to find or evaluate. We find statistically robust heterogeneity on the demand side: newer, less active, and lower-spending consumers benefit disproportionately more from GenAI-powered workflows. On the supply side, point estimates often suggest larger gains for smaller and newer sellers, although these differences are imprecisely estimated. Heterogeneity across product groups is more mixed and context-dependent.\\

This paper contributes to the emerging literature on the economic effects of GenAI. Existing work shows that GenAI can improve performance in a range of individual tasks, including writing, coding, marketing, and legal analysis, but most of that evidence comes from worker-level settings and emphasizes supply-side outcomes such as time savings or task completion \citep{Noy2023,Peng2023,DellAcqua2024,Cui2026}. By contrast, we study large-scale GenAI deployment within a single firm and evaluate its effects using demand-side business outcomes. In doing so, we provide evidence on whether GenAI creates measurable value through higher sales and conversion, rather than only through labor-saving or efficiency effects. Our setting also allows us to compare multiple workflows within the same organizational environment and to study heterogeneity across users with different levels of experience, complementing prior work on the uneven effects of GenAI across tasks and users \citep{Calvino2025,Noy2023,Bjorn2025,Hui2024,Otis2024}.\\

The paper also contributes to the literature on how digital platforms shape consumer decisions by reducing frictions and, in some settings, by influencing the content consumers see. A large body of work shows that platforms can improve outcomes by reducing information asymmetries, search costs, and targeting frictions through reputation systems, ranking algorithms, filtering tools, personalization, and targeted recommendations \citep{JinKato2006,tadelis2016reputation,CabralHortacsu2010,Dinerstein2018,donati2026end,Ursu2018,Chen2017,fradkin2017search,Yoganarasimhan2020,Blake2015,sun2024value}. More broadly, some consumer-facing technologies may affect outcomes not only by improving information and matching, but also by making platform content more salient, polished, or engaging, consistent with classic informative and persuasive views of advertising \citep{bagwell2007economic, GoldfarbTucker2011}. We extend this literature to GenAI by showing how it affects several consumer-facing retail workflows along the customer journey, increasing conversion in ways consistent with improved matching and a smoother consumer experience, while also leaving room for persuasive channels that may matter more in some applications than in others.\\

The remainder of the paper is structured as follows. Section 2 describes the study setting, experimental design, and data. Section 3 presents the main results. Section 4 examines heterogeneity in treatment effects. Section 5 concludes. The Appendix provides additional details and results.

\section{Study Setting, Experimental Design, and Data}{\label{setting}}

\subsection{Study Setting}
\label{subsec:study-setting}
The seven field experiments analyzed in this paper were conducted over roughly nine months, from September 2023 to June 2024. The company’s GenAI initiatives, however, began earlier in 2023, with initial efforts devoted to model training, implementation planning, and experimental preparation. The selection of workflows for GenAI re-engineering was not systematic but instead reflected managerial judgment, with platform managers prioritizing workflows viewed as most promising in terms of technical feasibility, organizational cost, and potential business value. The selected workflows cover several core modules of e-commerce operations, including customer service, consumer-product matching, advertising, and seller service. Table \ref{table:business processes} provides an overview of these workflows, highlighting the business need in each case and the specific way GenAI was integrated.\\

Table \ref{table:business processes} also helps conceptualize how GenAI may affect retail sales performance across workflows. Broadly speaking, GenAI may operate through both informational and persuasive channels, consistent with classic work on information and advertising in markets \citep{stigler1961economics,nelson1974advertising}. \citet{bagwell2007economic} provides a more recent comprehensive review. On the informational side, GenAI may improve semantic understanding, translation, responsiveness, and the richness and relevance of product information. In online retail, these improvements may lower search costs, reduce information asymmetries, improve matching, and facilitate communication between consumers, sellers, and service agents. This mechanism is especially relevant in multilingual and cross-border settings, where search, information, and communication frictions are naturally high. On the persuasive side, GenAI may affect outcomes through consumer-facing text and communication that is more salient, polished, or engaging, that increases the perceived professionalism or credibility of the platform’s communication, or that generates short-run novelty effects that raise attention and purchase propensity.\\

\begin{table}[t]
\centering
\scriptsize
\caption{Business Workflows Re-engineered with Generative AI}
\label{table:business processes}
\hspace*{-0.5cm}\begin{tabular}{p{0.1em} >{\raggedright}p{1.5cm} >{\raggedright}p{1.5cm} >{\raggedright}p{4.8cm} >{\raggedright}p{2cm} >{\raggedright}p{4.9cm} p{0.01em}}
\hline\hline
& Functional Area & Business \newline Workflow & Business \newline Needs/Objectives & GenAI \newline Capability & Description of \newline GenAI Application &  \\
\hline
1 & Customer Service & Pre-sale \newline Service \newline Chatbot  & Addressing each individual service request, providing unique, accurate, and content-rich answers. & AI agent & Deploying a GenAI-powered, 24/7 customer service chatbot that can respond to idiosyncratic consumer inquiries in all languages. & \\
 &  &  &  &  &  &  \\
2 & Consumer-product Matching & Search \newline Query \newline Refinement & Accurately decoding and translating the latent demands behind multilingual consumer search queries to improve consumer-product match. & Translation, content comprehension and generation & Using GenAI to improve consumers' demand expression by understanding, refining and translating their search queries, thus enhancing the matching accuracy of the search algorithm. & \\
 &  &  &  &  &  &  \\
3 & Consumer-product Matching & Product \newline Description & Creating comprehensive, structured product descriptions tailored to diverse linguistic preferences and cultural norms (e.g., currently, nearly half of the self-sold products have no or limited description). & Content recognition, comprehension and generation & Using GenAI to produce comprehensive and structured textual descriptions for the product detail page’s description module, adapted to each market. &  \\
 &  &  &  &  &  &    \\
4 & Advertising & Marketing \newline Push \newline Message  & Individual targeting of hundreds of millions of users with customized messages. & Content comprehension and generation & GenAI allows the generation of millions of messages, thereby enhancing the personalization of messages for precision marketing. & \\
 &  &  &  &  &  &   \\
5 & Advertising & Google \newline Advertising \newline Title & Creating product advertisement titles that closely match user interest and demands. & Content optimization and generation & Using GenAI to optimize product titles for Google ads for better user interest and engagement. & \\
 &  &  &  &  &  &  \\
6 & Seller Service & Chargeback \newline Defense & Streamlining the complicated process in a cross-border context with language barriers and diverse regulations and customs (e.g., over half of chargeback disputes go unaddressed by sellers). & AI agent & Developing a GenAI-driven agent that offers a one-stop, automated solution for sellers to streamline the intricacies of chargeback defense. & \\
 &  &  &  &  &  &  \\
7 & Customer Service & Live \newline Chat \newline Translation & Delivering native-language customer services to a diverse, multilingual consumer base & Real-time translation & Integrating GenAI into the platform's core English customer service process to provide real-time translation for all languages. & \\
\hline\hline
\end{tabular}
\end{table}

These channels are not mutually exclusive, and their relative importance is likely to differ across applications. This is especially true across the seven workflows we study. For instance, Search Query Refinement and Live Chat Translation are most naturally expected to operate through friction reduction, because they improve matching or communication without materially changing the persuasive content seen by consumers. Chargeback Defense is similarly best understood as reducing seller-side frictions in a complex, language-intensive process. By contrast, Pre-sale Service Chatbot, Product Description, Marketing Push Message, and Google Advertising Title may affect outcomes through either informational or persuasive channels, since these applications directly shape the content consumers see. In these settings, GenAI may reduce search costs and information asymmetries by providing richer, more relevant, or more structured information, while also influencing behavior by making content more prominent, well-crafted, or compelling.\\

Importantly, these channels are suggestive interpretations rather than well-identified mechanisms. Our empirical setting does not allow us to separately and precisely identify the informational and persuasive channels through which GenAI may affect outcomes. Instead, our goal is to quantify their combined effect on sales when GenAI is added to an existing workflow. In each experiment, the treatment condition differs from the control condition only through the integration of GenAI, while prices, other inputs and the broader workflow environment remain unchanged. Although GenAI itself requires development, integration, and inference resources, there were no concurrent changes to the workflows other than the addition of GenAI that would confound the comparison. We therefore interpret our estimates as the short-run impact of GenAI on workflow-level sales performance, holding other aspects of the workflow as fixed as possible.

\subsection{Experimental Design}
\label{sec:empirical framework}
To test the sales productivity gains of the seven GenAI applications, the firm conducted a series of large-scale, randomized field experiments. Six of these experiments were executed at the consumer level, with participating consumers randomly assigned to either treatment or control groups. The only exception was the Google Advertising Title, which was conducted at the product level, where a subset of products selected for Google ads was randomly divided into treatment or control groups. Consumer overlap across experiments was minimal (below 1\%). In all cases, for each consumer or product, the treatment status remained fixed throughout the experimental period. The key distinction between treatment and control was that the treatment group was exposed to workflows re-engineered with GenAI, whereas the control group continued under the platform’s standard practices without GenAI integration.\\

A potential concern is that treatment exposure for some users may affect the experiences or behavior of others, creating possible SUTVA violations. In our setting, such spillovers could arise if treatment-induced demand shifts alter marketplace conditions, such as the composition of search results, or if treated customers share information with control customers. We expect these concerns to be limited in our context. Although we do not observe the exact share of platform users exposed in each workflow, our back-of-the-envelope calculations suggest that treated users represented only a small fraction of the platform’s relevant user base.\footnote{Because we do not observe the total number of users in each workflow across the entire platform, we approximate exposure using the number of treated purchasers and publicly available information on the platform’s annual active shoppers. These calculations suggest that treated purchasers account for less than 3\% of the platform’s active shoppers within the corresponding experimental time window across workflows.} Moreover, most experiments were short, and, except for Pre-sale Service Chatbot, AI operated mostly in the background and affected private user experiences, limiting awareness and diffusion.\\

The total size of the subject pool varied greatly between experiments, with the smallest experiment having 30 thousand subjects while the largest containing up to 13 million subjects. Most of the experiments featured an equal distribution between the treatment and control subjects, with each group comprising approximately half of the total sample. The exceptions are Pre-sale Service Chatbot and Live Chat Translation, where the treatment group consumers comprised two-thirds of the total sample. Below, we provide details on the experiments related to all seven business workflows, with a summary of key features presented in Table \ref{table:summary descriptions of the experiments}. Appendix \ref{appendix:interfaces} presents illustrative user interfaces and examples for each workflow.

\begin{sidewaystable}
\centering
\footnotesize
\caption{Summary Descriptions of the Experiments}
\label{table:summary descriptions of the experiments}
\begin{threeparttable}
\begin{tabular}{p{0.5em} >{\raggedright}p{3cm} >{\raggedright}p{3.5cm} >{\raggedright}p{3cm} >{\raggedright}p{3.5cm} >{\raggedright}p{3.7cm} >{\raggedright}p{1.7cm} p{2cm}}
\hline\hline
& Business Workflow  & Time Frame  & Sample Size  & Control  & Treatment  & Data \newline Availability & Product \newline Sold By \\
\hline
1 \newline & Pre-sale Service Chatbot & Two months from \newline Sep. to Oct. 2023 & 44,614  \newline consumers & Pre-programmed auto \newline response indicating \newline no customer service  & GenAI Agent  & Yes \newline & Platform \newline \\
 &   &   &   &   &  &  & \\
2 \newline \newline & Search Query Refinement \newline & Three nine-day sub-experiments from \newline May. to Jun. 2024 & 1,849,382 \newline consumers \newline & Basic query translation \newline with no semantic \newline comprehension & GenAI-translated \newline queries with semantic \newline comprehension  & Yes \newline  & Sellers \& \newline Platform \newline  \\
 &   &   &   &   &  & & \\
3 \newline \newline & Product \newline Description \newline & Five one-week sub-experiments in \newline Dec. 2023 & 4,772,937 \newline consumers \newline & Human-generated descriptions \newline  & AI-generated descriptions on top of those created by humans & Yes \newline \newline & Platform \newline \newline \\
 &   &   &   &   &  &  & \\
4 \newline & Marketing Push Message & One day in \newline Dec. 2023 \newline & 13,715,528 \newline consumers & Human-generated standardized messages & GenAI generated a large and diverse set of messages &  Yes \newline & Sellers \& \newline Platform \newline \\
 &   &   &   &   &  &  & \\
5 \newline & Google Advertising Title & Twelve days in Jan. 2024 \newline & 1,244,016 \newline products & Human-generated \newline ad titles & GenAI-optimized \newline ad titles  & Yes \newline & Sellers \& \newline Platform \newline \\
 &   &   &   &   &  & & \\
6 \newline & Chargeback \newline Defense  & Two months from \newline Oct. to Dec. 2023 & About 30 thousand \newline consumers & Human agent \newline & GenAI agent \newline & No \newline & Sellers \& \newline Platform \newline \\
 &   &   &   &   &  & & \\
7 \newline  & Live Chat Translation & One month in \newline Oct. 2023 \newline & About 0.2 million \newline consumers & Filipino agent without translation assistance & Filipino agent with GenAI real-time translation assistance & No \newline & Sellers \& \newline Platform \newline \\
\hline\hline
\end{tabular}
\begin{tablenotes}
\scriptsize
\item[1] In column ``Product Sold By", ``Platform" denotes products procured and sold directly by the platform---i.e., the platform's self-sold products. ``Sellers \& Platform" indicates that the experiment included products sold both by the platform itself and by third-party sellers on the platform.
\end{tablenotes}
\end{threeparttable}
\end{sidewaystable}

\paragraph{Pre-sale Service Chatbot} The experiment, conducted over a two-month period from September to October 2023, included a random sample of 44 thousand consumers who initiated pre-sale customer service requests for the platform's self-sold products during the experimental period. These consumers were randomly divided into treatment and control groups. Consumers in the control group received the platform’s automated response service, which delivered a pre-programmed standardized notification indicating that customer service was unavailable. This auto-response condition reflects the platform’s standard operating practice of prioritizing a limited number of human agents for post-sale rather than pre-sale support for self-sold products, given that pre-sale inquiries are generally less urgent. This setup also mirrors the constraints faced by many third-party sellers on the platform, particularly smaller-scale sellers, who lack the capacity to provide real-time multilingual customer service. By contrast, consumers in the treatment group were supported purely by GenAI-powered chatbots.\footnote{The platform did not provide chat transcripts or message-level content because of confidentiality and consumer-privacy restrictions. Our analysis therefore uses treatment assignment and transaction outcomes rather than the content of chatbot conversations.}

\paragraph{Search Query Refinement} The experiment comprised three sub-experiments, each targeting consumers using different languages: Arabic, Japanese, and Polish. These languages were chosen because they are less commonly used on the platform and have historically been underserved by the platform’s traditional translation of search queries.\footnote{On our focal platform, the search algorithm initially translates multilingual queries into English to facilitate matching with product and seller information stored in English.} The sub-experiments were launched at different points between May and June 2024, each lasting nine days. During each period, a random subset of consumers conducting searches was assigned to the experiment. These consumers were randomly divided into two groups, yielding a total sample of approximately 2 million consumers across all sub-experiments. In the control group, consumer search queries were subject only to basic translation without semantic comprehension. In the treatment group, GenAI was deployed to translate queries by comprehending their underlying intent and refining them to improve semantic accuracy and clarity.

\paragraph{Product Description} The experiment comprised five sub-experiments, each involving consumers who spoke English, Spanish, French, Portuguese, or Korean, which are among the most widely used languages on the platform. All sub-experiments ran for one week in December 2023, with staggered start dates. GenAI was employed to create multilingual, textual product descriptions for a predetermined product set of approximately 45,000 randomly selected platform self-sold products spanning a broad range of categories.\footnote{On this platform, product descriptions refer to the text content in the description module of product detail pages that summarizes key features and selling points.} According to our partner company, self-sold products are primarily sourced from Chinese vendors, who typically provide image-based product presentations with limited Chinese text embedded in the images. While such image-based content aligns with Chinese consumer preferences, global consumers are more accustomed to text-based descriptive bullet points, such as the “About this item” section on Amazon. Consequently, nearly half of the self-sold products either lack textual descriptions or contain only minimal textual information. During each sub-experiment period, a random subset of consumers who clicked into the product detail pages of the selected products entered the experiment and were evenly split into treatment and control groups, resulting in a total of approximately 5 million participants. Control group consumers viewed the original human-generated descriptions, whereas treatment group consumers were shown the GenAI-created descriptions on top of the original, human-created descriptions.\footnote{For products without existing human-generated descriptions, control group consumers saw no description—as was historically the case for such products, and treatment group consumers saw the AI-generated descriptions only.}

\paragraph{Marketing Push Message} The experiment took place over the course of approximately one month in December 2023. A random subset of consumers who received push notifications on their mobile devices entered the experiment and were randomly assigned to either control or treatment groups. Given the large scale of this experiment, we restricted our analysis to the first day, which contained 13 million consumers. On our partner platform, push messages were traditionally created by staff, requiring 1-2 employees several hours each month to produce a few dozen messages. Given the platform’s hundreds of millions of consumers, this limited volume meant that many consumers received identical content, constraining the potential for personalized marketing. Accordingly, in the control group of our experiment, consumers primarily received uniform, human-generated marketing content, totaling roughly 2,000 distinct messages. By contrast, during the platform's initial risk-managed rollout, approximately 40\% of consumers in the treatment group were exposed to AI-generated messages, while the remaining treatment-group consumers continued to receive human-generated messages. This partial exposure generated nearly 2.7 million unique AI-generated message variants and thus far greater differentiation across individuals.

\paragraph{Google Advertising Title} The experiment was conducted over twelve days in January 2024, with randomization occurring at the product level. The sample included 1.2 million products selected by the retail platform for advertising in the sponsored section of Google Shopping, representing a diverse set of categories. For Google ads, the quality of the advertisement title is critical: a well-crafted title not only increases product discovery by aligning with user search keywords but also enhances user clicks by incorporating appealing terms that drive consumer interest.\footnote{Many e-commerce platforms maintain libraries of such buzzwords which, based on historical data, are known to boost consumer click-through and conversion rates.} In our experiment, the control group retained the original product titles created by the sellers, while in the treatment group, titles used in the ads were optimized by GenAI based on seller titles.\footnote{When promoting products on Google Shopping, the platform also used the pricing and image information provided by sellers, and these factors remained unchanged across treatment and control groups in our experiment.} A key distinction of this experiment is that the GenAI model was not fine-tuned specifically for the advertising context within the e-commerce domain. As a result, the generated titles may fail to emphasize product attributes most relevant to consumer search and purchase decisions, leading us to adopt a more agnostic stance regarding the expected treatment effect in this setting.

\paragraph{Chargeback Defense} The experiment, conducted from late October to late December 2023, included over 30 thousand consumers. During this period, a random subset of consumers who initiated chargeback requests was assigned to the experiment and then randomly divided into two groups. Contesting chargeback disputes requires a broad skill set, including claim analysis, evidence collection, and persuasive defense writing, which is especially challenging in cross-border contexts characterized by language barriers and complex regulations and customs. As a result, more than half of chargeback disputes on the focal platform were left unaddressed by sellers. In the control group, consumer claims were initially addressed by sellers. If no action was taken, approximately 3-5 outsourced workers then intervened to resolve the claims. However, these employees could only handle a small fraction of cases elaborately, while most were processed using generalized templates that proved far less effective. In contrast, claims in the treatment group were initially managed by sellers and subsequently supported by GenAI agents.

\paragraph{Live Chat Translation} The experiment was conducted over one month in October 2023 and involved approximately 0.2 million non-English-speaking consumers who contacted the platform’s customer service for issues such as clarifying details of platform-level promotions or resolving disputes with sellers when no agreement was reached. Due to cost constraints, a large share of such requests is handled by customer service agents from the Philippines who provide assistance in English, as employing native agents for each market is roughly three times more expensive. During the experiment, non-English-speaking consumers who initiated inquiries to the platform's customer service were randomly split between treatment and control conditions. In the treatment group, consumers interacted with Filipino agents with real-time bidirectional GenAI translation support, while those in the control group engaged with Filipino agents without GenAI translation assistance.

\subsection{Data and Estimation}
\label{sec:estimation}
We obtained comprehensive granular consumer- or product-level data for the first five of the seven experiments, allowing for in-depth analysis of sales gains. For the remaining two experiments---Chargeback Defense and Live Chat Translation---the platform could not provide granular data. In these cases, we rely on analyses conducted by the platform's internal data science team. These estimates complement our direct observations, offering a broader perspective on the impact of GenAI across various business areas (see ``Data Availability'' in Table \ref{table:summary descriptions of the experiments}). \\

For experiments conducted at the consumer level, we record each consumer’s treatment status and observe their complete set of activities, including the number of product views (Views), product clicks (Clicks), product orders (Orders), and total expenditure on those orders (Sales).\footnote{In Search Query Refinement, product views represent the number of products a consumer browses on the search results page, which displays a summarized collection of products immediately after a query search. In Google Advertising Title, product views refer to the number of views of advertised products within the Google Shopping tab. In Search Query Refinement and Product Descriptions, product clicks capture the number of times consumers clicked into product detail pages. In Marketing Push Message, product clicks reflect consumer clicks on push notifications, while in  Google Advertising Title, they indicate clicks on advertised products.} For comparability across workflows, our main analysis focuses on sales, measured as total consumer expenditure. To shed light on the mechanisms underlying the sales effects, we additionally examine outcomes along the extensive and intensive margins. In particular, we analyze conversion rates, defined as a binary indicator for whether a consumer makes at least one purchase during the experimental period, and cart value, measured as average expenditure conditional on making a purchase. Conversion rates capture changes in purchase incidence and serve as a widely used proxy for consumer experience in online retail, while cart value reflects adjustments along the intensive margin. When available, we also analyze intermediate engagement metrics, such as click-through rates. For the product-level experiment, we collect the same outcomes at the product level.  Table \ref{table:summary statistics} reports summary statistics for these key variables across the five experiments for which granular data are available.

\begin{table}[htbp]
\centering
\small
\caption{Summary Statistics of Main Outcomes}
\label{table:summary statistics}
\begin{threeparttable}
\begin{tabular}{lccccc}
\hline\hline
                              & Mean      &Standard Dev. & Median  & Min    & Max        \\
                              \hline
\textbf{Pre-sale Service Chatbot}      &           &              &         &        &            \\
\hspace{2mm}Sales              & 1.86      & 9.75    & 0       & 0      & 517    \\
\hspace{2mm}Conversion Rate  & 0.07      & 0.25       & 0       & 0      & 1          \\
\hspace{2mm}Cart Value       &27.05      &26.41       &18.90    &1.11    & 517     \\\hline
\textbf{Search Query Refinement}      &           &              &         &        &            \\
\hspace{2mm}Views              & 313.36    & 615.02       & 125     & 1      & 105,883    \\
\hspace{2mm}Clicks             & 8.23      & 16.99        & 3       & 0      & 2,024       \\
\hspace{2mm}Orders             & 0.16      & 0.73         & 0       & 0      & 85        \\
\hspace{2mm}Sales               & 2.24      & 21.41        & 0       & 0      & 4,960   \\
\hspace{2mm}Conversion Rate   & 0.09      & 0.28         & 0       & 0      & 1          \\
\hspace{2mm}Cart Value          & 25.39    & 67.83         & 10.42  & 0.01       & 4,960    \\\hline
\textbf{Product Description}           &           &              &         &        &            \\
\hspace{2mm}Clicks             & 1.98      & 2.06         & 1       & 1      & 173        \\
\hspace{2mm}Orders             & 0.06      & 0.30         & 0       & 0      & 23         \\
\hspace{2mm}Sales               & 0.51      & 4.56         & 0       & 0      & 2,942   \\
\hspace{2mm}Conversion Rate   & 0.04      & 0.21         & 0       & 0      & 1          \\
\hspace{2mm}Cart Value          & 11.62     & 18.51        & 7.06    & 0.59      & 2,942    \\\hline
\textbf{Marketing Push Message}        &           &              &         &        &            \\
\hspace{2mm}Clicks             & 0.02     & 0.13        & 0       & 0      & 1          \\
\hspace{2mm}Orders             & 0.002    & 0.05        & 0       & 0      & 6           \\
\hspace{2mm}Sales               & 0.03     & 1.51        & 0       & 0      & 501     \\
\hspace{2mm}Conversion Rate   & 0.002    & 0.04        & 0       & 0      & 1          \\
\hspace{2mm}Cart Value        &15.52       &33.97        &7.81     &0.01     & 501    \\\hline
\textbf{Google Advertising Title}  &           &           &         &        &             \\
\hspace{2mm}Views              & 19.36      & 82.38       & 5       & 2      & 12,033      \\
\hspace{2mm}Clicks             & 0.22     & 1.69          & 0       & 0      & 627          \\
\hspace{2mm}Sales               & 0.13     & 2.97         & 0       & 0      & 322  \\
\hspace{2mm}Conversion Rate     & 0.004    & 0.06      & 0       & 0      & 1            \\
\hspace{2mm}Cart Value           &33.39     &34.39        &21.42    &0.54    &322     \\
\hline\hline
\end{tabular}
\begin{tablenotes}
\scriptsize
\item[1] ``Views" refers to the number of product views. ``Clicks" denotes the number of product clicks. ``Orders" is the number of product orders. ``Sales" represents the total expenditure on product orders. ``Conversion rate" measures consumers’ likelihood of making a purchase. It is a binary indicator for purchase, which equals 1 if a consumer makes at least one order during the experiment period, and 0 otherwise. ``Cart value" is the expenditure per consumer, conditional on the consumer making a purchase.
\item[2] In the Google Advertising Title experiment, the unit of observation is the product. The conversion rate is 1 if a product is purchased at least once during the experimental period, and 0 otherwise, while cart value refers to sales per product, conditional on the product being purchased.
\end{tablenotes}
\end{threeparttable}
\end{table}

To compare mean sales, conversion rates, and cart values between the treatment and control groups, we use the following general empirical specification, adapted as needed for each experiment:
\begin{equation}
\label{eq:main}
y_{i} =\beta \times Treat_{i} + \alpha_{c(i)} + \epsilon_{i},
\end{equation}
where $i$ denotes the randomized unit (consumer or product), and $y_{i}$ is the outcome. $Treat_{i}$ is the treatment indicator, which equals one if the consumer or product belongs to the treatment group and zero otherwise. $\alpha_{c(i)}$ denotes the cohort fixed effects. Specifically, in the Pre-sale Service Chatbot and Google Advertising Title experiments, consumers or products entered the experiments on different days, we therefore control for entry-day cohort fixed effects. In the Search Query Refinement and Product Description experiments, multiple sub-experiments were conducted across different languages at varying times, we thus include entry-day-by-language cohort fixed effects. For the Marketing Push Message experiment, the sample spans only a single day, so no cohort fixed effects are included. Details on the model specification for each experiment are provided in Appendix \ref{appendix:detailed main results}. \\

We estimate Equation \eqref{eq:main} via OLS, adjusting the standard errors for heteroskedasticity. Under random assignment, $\beta$ recovers the average treatment effect of GenAI adoption, expressed as the absolute lift in outcomes. For all detailed experiments estimated with this equation, except for Marketing Push Message, assignment to the treatment group coincides with exposure to the GenAI-enabled workflow, so this estimate is also the treatment-on-the-treated (TOT) effect for the corresponding workflow. Marketing Push Message differs because only approximately 40\% of consumers assigned to the treatment group were exposed to AI-generated messages during the platform's initial risk-managed rollout. For that experiment, $\beta$ is therefore an intent-to-treat (ITT) effect of assignment to the treatment group; in the aggregate-gains calculation, we report this ITT estimate as a lower bound and an exposure-adjusted upper bound under the assumption that exposure within the treatment group was random. We also report results in percent lift, rescaling $\beta$ by the control group mean. For sales, we use levels to address concerns regarding log transformations with zero outcomes \citep{ChenQJE2024}. For conversions (a binary outcome), we primarily estimate a linear probability model and confirm that the findings are robust to logit specifications. We also estimate the model using pre-experiment covariates as controls, and the results remain consistent. Consumer overlap across experiments was minimal (less than 1\%), and our findings are robust to excluding overlapping observations. This design allows treatment effects to be solely attributed to individual workflows, though it does not capture potential complementarities across GenAI applications.\\

To enrich our analysis, we obtained pre-experiment seller data for products included in the experiments (e.g., seller size measured by annual sales, operational years, and the number of sub-accounts linked to a seller's online store). These data enable us to examine seller-level heterogeneity in the three experiments involving products sold by both third-party sellers and the platform (see column ``Product Sold By" in Table \ref{table:summary descriptions of the experiments}). This analysis is not applicable to the Pre-sale Service Chatbot and Product Description experiments, as both involve platform self-sold products sold by a limited number of platform-operated sellers, resulting in insufficient variation in seller characteristics for meaningful heterogeneity analysis. Additionally, we collect product characteristics, such as the concentration level of the associated category, price, and annual sales quantity, to explore product-level heterogeneity.\\

\begin{figure}[t]
    \centering
       \caption{P-Values for Covariate Balance Checks Across Experiments} \includegraphics[width=1\textwidth]{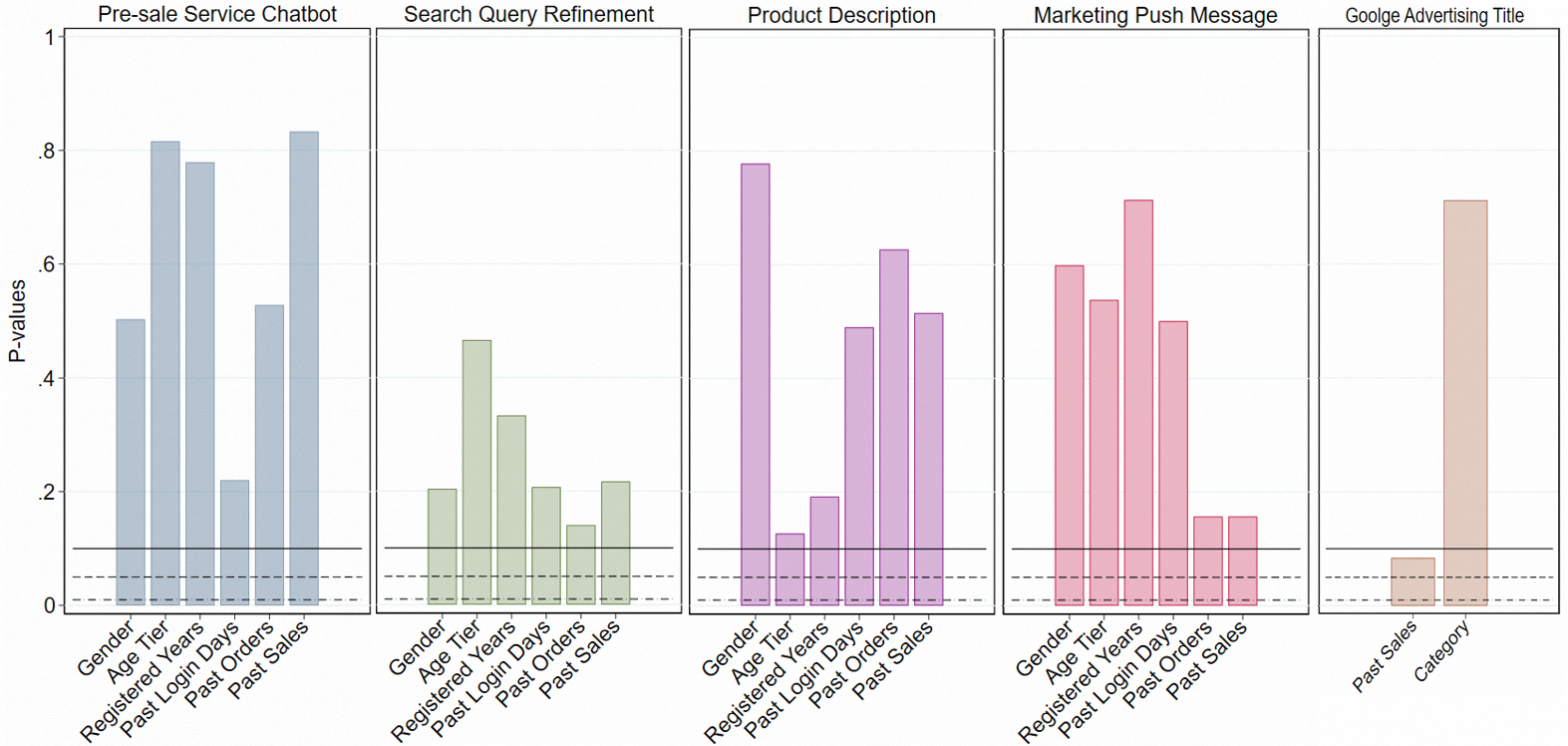}
    \label{fig:pvalues}
    \begin{threeparttable}
    \begin{tablenotes}
\scriptsize
\item[1] This figure presents the p-values for the covariate balance checks across all experiments. For the first four experiments, which are conducted at the consumer level, six consumers' demographic and behavioral variables are examined. For the single experiment conducted at the product level (Google Advertising Title), the focus is on two key measurements of products, which are the product’s historical sales and the distribution of product categories.
\item[2] The bar presents the p-values. The solid, dashed, and dash-dot lines indicate p-values of 0.1, 0.05, and 0.01, respectively.
\item[3] ``Gender" is predicted by the platform based on consumers' past shopping behaviors. ``Age Tier" is a 7-point scale from 1 (youngest) to 7 (oldest). ``Registered Years" indicates the duration from the year of consumer registration to the year of experiment. ``Past Login Days" represents the number of days consumers have logged into the platform in the 30 days prior to the experiment. ``Past Orders" is the number of product orders in the 30 days prior to the experiment. ``Past Sales" represents the total expenditure on product orders in the 30 days prior to the experiment. ``Category" is the category associated with a product.
\end{tablenotes}
\end{threeparttable}
\end{figure}

We also augmented the data with consumer demographics and pre-experiment shopping history (years of registration, activity level, purchase volume, etc.), which supported both the analysis of buyer-level heterogeneity and the verification of random group assignment in the experiments.\footnote{According to our research agreement with the partner platform, all consumers in our data are anonymous to ensure consumer privacy. We identify consumers by hashed IDs instead of knowing their actual names.} As confirmed by the covariate balance checks reported in Figure \ref{fig:pvalues}, we found no systematic significant pre-experiment differences across consumers between the control and treatment groups. Thus, the randomization process was effective at allocating comparable consumers and products into the two groups. Details on the covariate balance checks are presented in Appendix \ref{appendix:balance checks}.

\section{Main Results}
\label{sec:main results}

\subsection{Sales Impact by Workflow}
\label{subsection: productivity impact by workflow}

Table \ref{table:summarized main results} reports the estimated effects on sales, our primary outcome. Column (1) presents the absolute treatment effect, while column (2) reports the corresponding percentage change relative to the control group. As discussed in Section \ref{sec:estimation}, these estimates correspond to TOT effects for workflows in which treatment assignment coincides with GenAI exposure, and to an ITT effect for Marketing Push Message because of partial exposure within the assigned treatment group. The results presented here do not account for differences in the frequency of consumer interactions with the workflows or in the duration of the corresponding experiments. For example, search affects nearly all consumers on a daily basis, whereas chatbot interactions occur less frequently. In Section \ref{sec:aggregate_gains}, we provide adjusted measures that account for these differences. Additional results and robustness analyses are reported in Appendix \ref{appendix:detailed main results}.

\begin{table}[H]
\centering
\footnotesize
\caption{Average Treatment Effects of GenAI Adoption on Sales Across Workflows}
\label{table:summarized main results}
\begin{threeparttable}
\begin{tabular}{m{4.5cm} >{\raggedright}m{2.2cm} >{\raggedright}m{1.8cm} m{1.8cm} m{1.8cm}}
\hline\hline
& \multicolumn{4}{l}{Sales Productivity Impact (US \$)}  \\\cline{2-5}
& (1) & (2) & (3) & (4) \\
Business Workflow & Coefficient  & \% Change & Obs & Unit of Obs \\
\hline
\textbf{Pre-sale Service Chatbot} \newline
& 0.274*** \newline (0.0970)
& 16.3\% \newline
& 44,614 \newline
& Consumer \newline \\
&   &   &   &   \\
\textbf{Search Query Refinement} \newline
& 0.0648** \newline (0.0314)
& 2.9\% \newline
& 1,849,382 \newline
& Consumer \newline \\
&   &   &   &   \\
\textbf{Product Description} \newline
& 0.0104** \newline (0.00417)
& 2.1\% \newline
& 4,772,937 \newline
& Consumer \newline \\
&   &   &   &   \\
\textbf{Marketing Push Message} \newline
& 0.000402 \newline (0.000816)
& 1.6\% \newline
& 13,715,528 \newline
& Consumer \newline \\
&   &   &   &   \\
\textbf{Google Advertising Title} \newline
& -0.00602 \newline (0.00534)
& -4.5\% \newline
& 1,244,016 \newline
& Product \newline \\
&   &   &   &   \\\hdashline
\textbf{Chargeback Defense}\textsuperscript{†}
&  \multicolumn{4}{c}{15\% defense success rate increase}     \\
&   &   &   &   \\
\textbf{Live Chat Translation}\textsuperscript{†}
&  \multicolumn{4}{c}{5.2\% consumer satisfaction increase}    \\
\hline\hline
\end{tabular}
\begin{tablenotes}
\scriptsize
\item[1] ``Sales" represents the total expenditure on product orders, in USD.
\item[2]  Column (1) reports the estimated coefficients, with standard errors in parentheses. Column (2) reports \% Change, calculated as the treatment effect divided by the control group mean. Column (3) reports the number of observations. *** p$<$0.01, ** p$<$0.05, * p$<$0.1.
\item[3] \textsuperscript{†} For these experiments data are not available: we report findings estimated by the platform's internal data science team.
\end{tablenotes}
\end{threeparttable}
\end{table}

 The table shows that most GenAI deployments are associated with sales gains, with substantial heterogeneity across workflows. The largest effect is observed in the pre-sale service chatbot workflow, where GenAI increases sales by 16.3\% relative to the control group ($p<0.01$; Column 2). In this setting, consumers in the treatment group interact with a GenAI-powered chatbot, whereas consumers in the control group receive an automated message indicating that real-time support is unavailable. This control condition reflects the platform’s standard operating practice: human agents are prioritized for post-sale inquiries, leaving a substantial share of pre-sale inquiries for self-sold products without live assistance due to capacity constraints. This setting also mirrors the operating reality faced by many third-party sellers on the platform, particularly smaller-scale sellers who lack the capacity to provide real-time, multilingual customer support. Taken together, these features suggest that GenAI improves sales performance by performing tasks that are difficult for human agents to fulfill due to talent shortages or resource constraints.\\

One potential concern is that consumers in the control group may have been frustrated by the absence of assistance, mechanically reducing their likelihood of purchase and thus inflating the estimated treatment effects. To address this concern, we conducted additional experiments on the GenAI-powered chatbot. Appendix Table \ref{table:Results of Pre-sale Service Chatbot} reports a series of comparisons, including benchmarks against human agents and hybrid configurations that combine GenAI assistance with human escalation when needed. The results indicate that the GenAI chatbot delivers service quality comparable to human customer support (Column 2). Moreover, integrating GenAI with human agents yields substantially larger effects: relative to the no pre-sale service condition, sales increase by 25\% when GenAI assistance is combined with human escalation ($p<0.01$; Column 3), pointing to strong complementarities between GenAI and human labor. Most importantly, comparing consumers who receive GenAI-assisted service with human escalation to those served exclusively by human agents shows that the former spend 11.5\% more ($p<0.1$; Column 4). We interpret this comparison as a conservative lower bound on GenAI’s sales productivity impact in pre-sale customer support.\\

The remaining four workflows with detailed data reported in Table \ref{table:summarized main results} exhibit more modest effects on sales, ranging from a negative and statistically insignificant estimate to gains of up to 3\%. In particular, the Search Query Refinement application increases sales by 2.9\% ($p<0.05$), while automated Product Description generation raises sales by 2.1\% ($p<0.05$). Although smaller in magnitude, these effects are economically meaningful for a platform of this scale and maturity, especially given the near-universal exposure of consumers to search and product descriptions.\\

The Marketing Push Message workflow shows a positive yet not statistically significant improvement in sales (1.6\%). This likely reflects the combination of a very low baseline conversion rate (only 0.16\% of consumers make a purchase) and high variance in expenditures among converters, suggesting that broader implementation could provide sufficient power to detect treatment effects. Notably, however, this case exhibits significant increases in conversion rate and number of orders, which we discuss in detail later. It is also important to note that, for risk-mitigation purposes, only a subset of the treatment group was exposed to AI-generated messages in this experiment, which may attenuate the observed treatment effects under partial exposure.\footnote{The estimated effect is likely a lower-bound estimate. Because approximately 40\% of treatment customers were exposed to AI-generated messages during the initial rollout, under the assumption that exposure within the treatment group was random, rescaling the effect by the exposure rate gives \$0.000402 / 0.40 = \$0.001005 per exposed customer, implying a 4\% lift relative to the control group.}\\

By contrast, the Google Advertising Title workflow exhibits a statistically insignificant negative effect. This pattern is consistent with the GenAI model not being fine-tuned to the advertising context within e-commerce, leading it to omit commercially salient keywords commonly used in ad titles. As a result, these ads may receive lower quality scores from Google, reducing their probability of winning auctions at a given bid. In addition, Google’s advertising algorithm may identify and deprioritize AI-generated titles, further lowering their visibility. Consistent with this interpretation, Table \ref{table:Results of google advertising} in the Appendix shows that ads with AI-generated titles receive fewer views and clicks than those with human-generated titles. Taken together, these results underscore the importance of domain-specific fine-tuning or retraining of foundation models when deploying GenAI in industry-specific tasks that require specialized knowledge \citep{Deloitte2023}.\\

For the final two business processes studied—Chargeback Defense and Live Chat Translation—we do not observe granular transaction-level data and instead rely on the platform’s internal metrics and analyses, which do not report statistical significance. While these outcomes are not directly comparable to the sales effects discussed above, they nevertheless point to substantial improvements in two operational dimensions. Specifically, Chargeback Defense is associated with a 15\% increase in defense success rates, and Live Chat Translation increases consumer satisfaction by 5.2\%. Both effects point directionally to successful implementation of the GenAI applications.\footnote{In the case of Chargeback Defense, additional gains not captured in our calculations may arise from cost reductions, as the GenAI-enabled workflow eliminates the need for manual intervention.}\\

These results provide evidence on the potential of GenAI to improve front-office sales performance in online retail. A key distinction of our study is its focus on downstream revenue outcomes rather than worker-level gains, which makes our estimates not directly comparable to those from studies emphasizing labor productivity. Among the seven deployments we examine, most deliver measurable performance gains, showing that GenAI can generate substantial improvements in retail outcomes under real-world operating conditions. Overall, the evidence points to a positive but heterogeneous impact of GenAI along the customer journey.\\

The variation in effects across workflows is unlikely to reflect differences in implementation quality, which was comparable across applications. Instead, it appears to arise from differences in the marginal contribution of GenAI relative to baseline conditions. In each case, the control group reflects the firm’s standard operating practices prior to GenAI adoption. For example, in the Pre-sale Service Chatbot workflow, the control group received no pre-sale customer support; in Search Query Refinement, it relied on standard machine-learning-based search algorithms with lexical translation into English; and in most other workflows, the benchmark was human input. The results therefore point to genuine heterogeneity in where GenAI is most effective. Customer-support applications, such as Pre-sale Service Chatbots, generate the largest improvements; search and product-discovery tasks yield more modest gains; and advertising-related applications exhibit no statistically significant effects.\footnote{In the Marketing Push Message workflow, features of the experimental design may also affect measured impacts, as only a subset of treated users was exposed to GenAI, likely attenuating estimated effects.} Taken together, these findings highlight both the role of baseline conditions in shaping treatment effects and the differential effectiveness of GenAI across functional areas.

\subsection{Post-purchase Outcomes}

A natural concern with deploying GenAI in customer-facing processes is that, while it may provide richer, more relevant pre-purchase information and improve ex ante decision quality, it may also introduce hallucinations or provide overly persuasive content that misleads consumers, thereby reducing ex post consumer satisfaction. In our case, increases in revealed purchase behavior may not map cleanly into welfare gains, as higher sales could reflect distorted choices rather than improved consumer surplus. To assess this possibility, we examine two post-purchase outcomes that capture consumers’ evaluations of their purchase decisions. The first is the product return rate, defined as the share of orders that are returned. The second is the positive review rate, defined as the share of rated orders that receive four- or five-star ratings on a five-point scale.\footnote{The platform does not use auto-filled or pre-populated ratings; ratings are voluntarily submitted by consumers after purchase. We measure the positive review rate conditional on rated orders. For returns, we track purchases over a one-year post-purchase horizon, which is long relative to the platform's typical return window. Nevertheless, returns may still understate dissatisfaction if high transaction costs lead some consumers not to return products.}

\begin{figure}[h]
\begin{adjustwidth}{-1.2cm}{-1.2cm} 
    \centering
    \footnotesize
    \caption{Average Treatment Effects of GenAI Adoption on Product Returns and Reviews}
    \label{figure:main returns and reviews}
   \hspace{-0.3cm} \includegraphics[width=0.43\paperwidth]{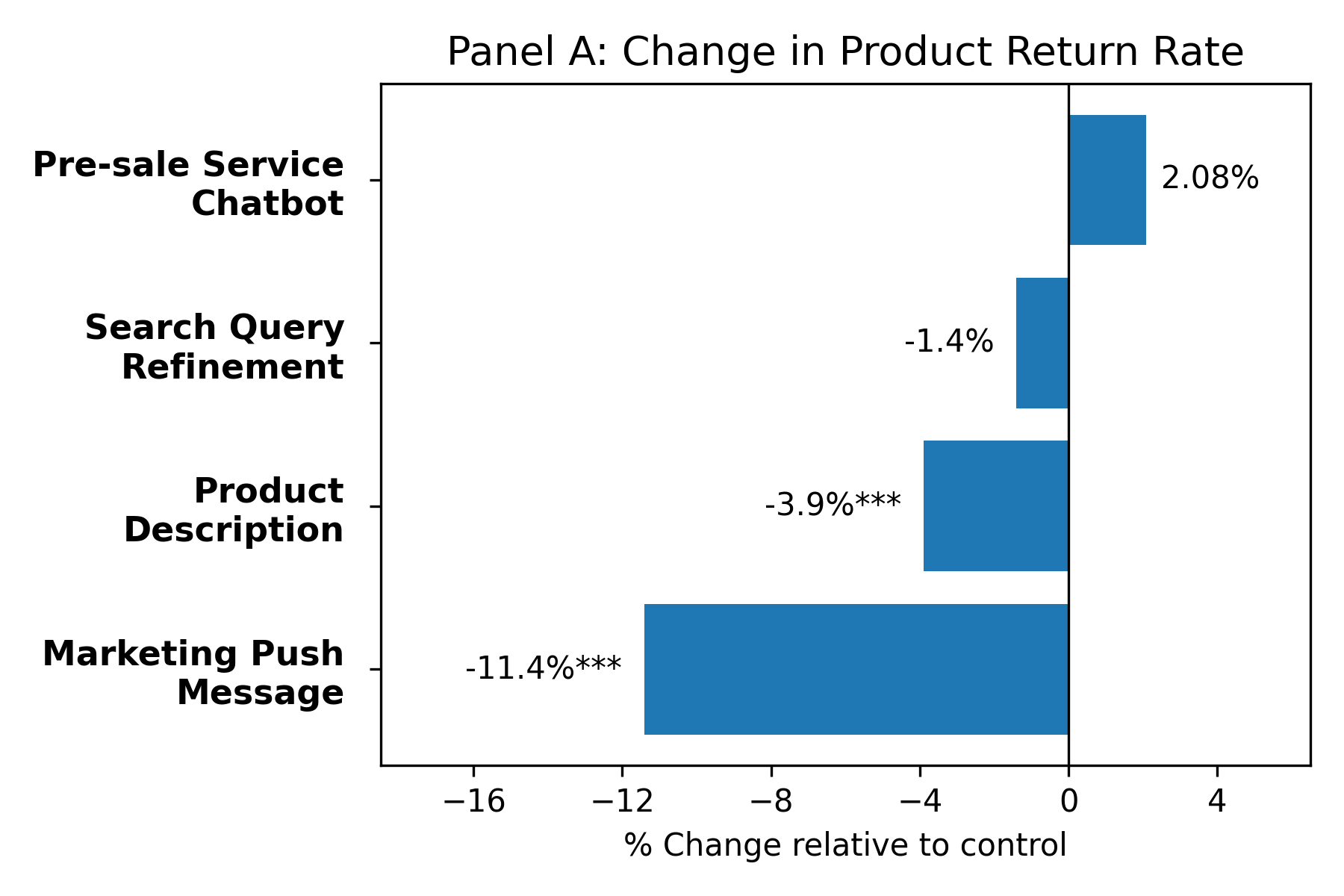}
    \includegraphics[width=0.43\paperwidth]{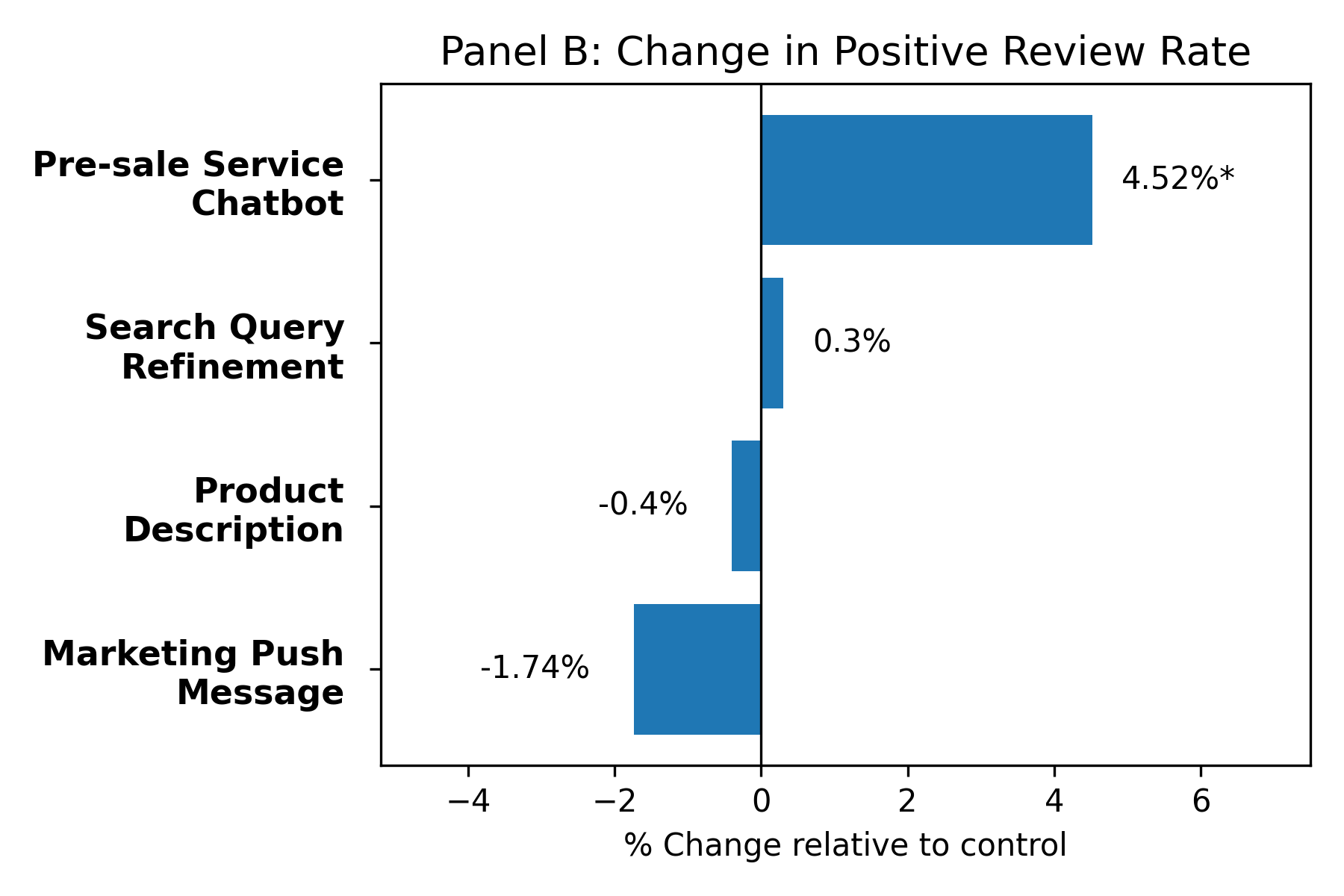}
\end{adjustwidth}
   \begin{threeparttable}
    \vspace{-0.5cm}\begin{tablenotes}
\scriptsize
\item[1] ``Return Rate" is defined as the share of orders that are returned within a year after the purchase. ``Positive Review Rate" is defined as the share of rated orders that receive four- or five-star ratings on a five-point scale. The estimation of Return Rate is conditional on consumers with orders, while the estimation of Positive Review Rate is conditional on consumers who provide ratings. Such data are not available for the Google Advertising Title workflow.
\item[2] Full regression results are reported in Table \ref{table:main returns and reviews}. *** p$<$0.01, ** p$<$0.05, * p$<$0.1.
\end{tablenotes}
\end{threeparttable}
\end{figure}

As shown in Figure \ref{figure:main returns and reviews} (full regression results are provided in Table \ref{table:main returns and reviews}), we find no evidence that the observed sales gains come at the expense of measured ex post consumer satisfaction. Product return rates and customer ratings do not deteriorate following GenAI adoption. Notably, return rates decline significantly for Product Descriptions and Marketing Push Messages (by 3.9\% and 11.4\%, respectively; $p<0.01$), consistent with improvements in both ex ante decision quality and ex post consumer satisfaction in these workflows. We also observe a positive effect on customer ratings for the Pre-sale Service Chatbot (a 4.5\% increase; $p<0.10$). In all other cases, the estimated effects are statistically insignificant. We therefore interpret these post-purchase outcomes as useful but imperfect evidence that the sales gains are not accompanied by lower revealed consumer satisfaction.

\subsection{Potential Mechanisms}

We investigate two potential drivers behind the observed impact on sales. The first is an increase in purchase incidence, captured by an increase in the probability that a consumer completes a purchase at the extensive margin. We measure this pattern using conversion rates, a widely used industry metric that reflects consumers’ revealed satisfaction and perceived friction during the ex ante search and evaluation process in e-commerce. The second pattern is a change in the behavior of consumers who make a purchase, reflected in adjustments to spending intensity and order composition. In this case, GenAI may influence consumers’ willingness to pay, alter the products that consumers see, or shift demand toward products at different price points. We assess this pattern using cart value, defined as expenditure per purchasing consumer, and order intensity, defined as the number of orders per purchasing consumer, both measured conditional on consumers making at least one purchase.

\begin{table}[H]
\centering
\footnotesize
\caption{Average Treatment Effects of GenAI Adoption on Conversion Rate and Cart Value}
\label{table:main conversion and cart value}
\begin{threeparttable}
\begin{tabular}{m{4.5cm} >{\raggedright}m{1.6cm} >{\raggedright}m{1.5cm} >{\raggedright}m{1.2cm} >{\raggedright}m{0.1em} >{\raggedright}m{1.5cm} >{\raggedright}m{1.5cm} m{1.2cm}}
\hline\hline
& \multicolumn{3}{l}{Conversion Rate} && \multicolumn{3}{l}{Cart Value (\$)}  \\\cline{2-4}\cline{6-8}
& (1) & (2) & (3) && (4) & (5) & (6)   \\
Business Workflow & Coefficient  & \% Change & Obs && Coefficient  & \% Change & Obs \\
\hline
\textbf{Pre-sale Service Chatbot} \newline &0.0131*** \newline (0.00252) & 21.7\% \newline &44,614 \newline && -1.264 \newline (1.036) & -4.5\% \newline & 3,076 \newline \\
& & & & & & & \\
\textbf{Search Query Refinement} \newline &0.00101** \newline (0.000411) &  1.2\% \newline &1,849,382 \newline && 0.370 \newline (0.334) & 1.5\% \newline & 163,381 \newline \\
& & & & & & & \\
\textbf{Product Description} \newline &0.000554*** \newline (0.000187) & 1.3\% \newline &4,772,937 \newline && 0.0944 \newline (0.0805) & 0.8\% \newline & 210,155  \newline \\
& & & & & & & \\
\textbf{Marketing Push Message} \newline &0.000048** \newline (0.0000218) & 3.0\% \newline &13,715,528 \newline && -0.206 \newline (0.454) & -1.3\% \newline & 22,425 \newline \\
& & & & & & & \\
\textbf{Google Advertising Title} \newline &-0.000090 \newline (0.000111)&  -2.3\% \newline  &1,244,016 \newline && -0.784 \newline (0.992) & -2.3\% \newline & 4,811 \newline \\
\hline\hline
\end{tabular}
\begin{tablenotes}
\scriptsize
\item[1] For workflows 1-4, ``Conversion Rate" measures consumers' likelihood of making purchases. It is a binary indicator for purchase, which equals 1 if a consumer makes at least one order during the experiment period, and 0 otherwise. ``Cart Value'' refers to the expenditure per consumer, conditional on the consumer making a purchase. For workflow 5, ``Conversion Rate" is 1 if a product is purchased at least once during the experimental period, and 0 otherwise. ``Cart Value'' refers to the sales per product, conditional on the product being purchased.
\item[2] Columns (1) and (4) report the estimated coefficients, with standard errors in parentheses. Columns (2) and (5) report \% Change, calculated as the treatment effect divided by the control group mean. Columns (3) and (6) report the number of observations. *** p$<$0.01, ** p$<$0.05, * p$<$0.1.
\end{tablenotes}
\end{threeparttable}
\end{table}

Table \ref{table:main conversion and cart value} reports the results on conversion rates and cart value, with additional evidence on order intensity provided in Appendix \ref{appendix:detailed main results}. Across workflows, we document economically and statistically significant increases in conversion rates ranging from 1\% to 22\% (Column 2), which translate directly into higher output as measured by sales. By contrast, we find no evidence of effects along the intensive margin. Columns 4 and 5 show that cart value, as measured by the average order value among consumers who make at least one purchase (or among products purchased at least once, in the case of Google Advertising Title), remains unchanged following GenAI adoption. We likewise find no significant changes in order intensity, defined as the number of orders placed conditional on making at least one purchase (Appendix \ref{appendix:detailed main results}). Taken together, these results indicate that GenAI primarily drives sales gains by increasing purchase incidence rather than by altering spending patterns among existing buyers.\\

The evidence on increased conversion rates in Table \ref{table:main conversion and cart value} suggests that GenAI has the potential to reduce market frictions, both by enabling new services and by improving existing ones, thereby enhancing the shopping experience and increasing the probability of purchase. For example, GenAI can mitigate search costs and information asymmetries by providing relevant and timely assistance through a pre-sale chatbot (a 21.7\% increase in conversion rates, $p<0.01$) and by generating more comprehensive and structured product descriptions (a 1.3\% increase, $p<0.01$). It can also reduce search frictions and improve match quality by enhancing the translation and semantic understanding of consumer queries (a 1.2\% increase, $p<0.05$). In addition, GenAI enables personalization of marketing content at scale by generating customized messages across a broad product portfolio (a 3\% increase, $p<0.05$). Finally, evidence from Chargeback Defense and Live Chat Translation reported in Table \ref{table:summarized main results} points to improvements consistent with enhanced experiences for sellers and consumers, respectively, including higher dispute success rates and greater customer satisfaction.\\

In Appendix \ref{appendix:detailed main results}, we report additional evidence on how GenAI reduces frictions along the customer journey, where such data are available. For example, Table \ref{table:more exploration of search query refinement} shows that GenAI-refined search queries increase the likelihood of a product click by 0.3\% ($p<0.01$; Column 1), consistent with improved search performance in combination with the previously documented increase in purchase conversion. Columns 2 and 3 further indicate that treated consumers view fewer products prior to clicking or purchasing (by 0.5\% and 1.5\%, respectively; $p<0.1$), suggesting a reduction in search intensity. These patterns align with prior literature showing that improvements in search quality increase conversion while reducing search effort \citep{Yang2013, Zhou2025}. In addition, we observe a statistically significant 2.0\% increase in the click-through rate ($p<0.01$; Column 4), defined as the ratio of product clicks to product views, while the click-to-order conversion rate (Column 5), defined as the ratio of orders to clicks, remains statistically insignificant. Together, these findings suggest that GenAI-refined queries primarily affect the composition of products retrieved at the search stage, rather than the information displayed on product detail pages, reinforcing the proposed query-refinement mechanism.\\

Table \ref{table:Results of product description} shows that AI-generated product descriptions increase the number of orders placed by 1.1\% ($p<0.05$; Column 2). Moreover, when we stratify products by the length of their original, human-generated descriptions, we find substantial heterogeneity in effects. Products with no or insufficient textual descriptions (fewer than 50 words) experience a 6.5\% increase in sales following augmentation with AI-generated content ($p<0.05$), whereas products with sufficiently detailed descriptions (more than 50 words) exhibit no significant increase (Columns 1 and 2 of Table \ref{table:more exploration of product description}).\footnote{Based on internal research and expert surveys conducted by our partner platform, descriptions containing fewer than 50 words are classified as providing insufficient or minimal textual information. This analysis is restricted to the English-language sub-experiment because description-length data are available only for English-language content.} This pattern indicates that the descriptions augmented by GenAI are most effective for products with limited baseline information.\\

In the marketing push workflow, GenAI enables large-scale variation in customer-facing content. The treatment group includes approximately 2.7 million unique message variants, compared with only roughly 2,000 variants in the control group, which relies exclusively on human-generated content. Consistent with the expanded scope for personalization, AI-generated marketing messages increase clicks by 3.1\% ($p<0.01$) and orders by 2.8\% ($p<0.10$; columns 1 and 2 of Table \ref{table:Results of marketing push}). Taken together, these results show that GenAI improves intermediate engagement outcomes, such as clicks, orders, and click-through rates, and are consistent with a smoother and more informative shopping experience that may help explain the higher conversion and sales gains documented above.\\

Overall, the evidence in this section is best interpreted as suggestive evidence on channels, rather than as definitive evidence that separately identifies the mechanisms underlying the treatment effects. Consistent with the conceptual discussion in Section \ref{subsec:study-setting}, the increases in conversion, clicks, and orders, together with the absence of systematic changes in cart value or order intensity, are consistent with GenAI reducing search, information, communication, and personalization frictions along the customer journey. At the same time, the experiments identify the combined effect of adding GenAI to each workflow and do not separately isolate informational channels from persuasion, attention, trust, or novelty. The evidence therefore clarifies where the sales gains arise in consumer behavior---primarily at the purchase-incidence margin---while remaining cautious about why they arise.

\subsection{Aggregate Gains Across Workflows \label{sec:aggregate_gains}}
In this section, we aggregate the sales gains across workflows to estimate the AI-driven incremental value per consumer observed in our experiments, explicitly accounting for differences in the frequency of consumer interactions across workflows. Specifically, we focus on the four workflows with positive treatment effects in sales, excluding the Google Advertising Title experiment, which serves as a testable bad case and can be readily improved in future iterations.\\

Table \ref{table:back of envelope} summarizes the key variables used in our calculations. Column 1 reports the AI-driven incremental value per consumer for each experimental workflow, reflecting the estimated average treatment effects (the absolute sales lift per consumer) reported in Column 1 of Table \ref{table:summarized main results}.
Because each experiment had a different duration, Column 2 presents the time multiplier used to extrapolate these effects to an annual horizon.\footnote{We assume that each experiment’s duration reflects the typical interval between treatment opportunities for a representative consumer on the platform. For example, a representative consumer is assumed to interact with the Pre-sale Service Chatbot about once every two months, whereas they could receive a Marketing Push Message daily.} For instance, the Pre-sale Service Chatbot experiment spans two months, yielding a time multiplier of six (i.e., $12/2$).
For each workflow, we obtain the annualized AI-driven incremental value per consumer in Column 3 by multiplying the estimated effect in Column 1 by the corresponding time multiplier in Column 2. This calculation assumes that the treatment effects observed during the experimental period remain constant over time, abstracting from potential amplification (e.g., greater engagement and purchases as AI-generated content enhances consumer satisfaction and platform loyalty) or attenuation (e.g., consumer dissatisfaction and product returns due to potential mismatches between AI-generated content and actual product characteristics).\\

\begin{table}[t]
\centering
\footnotesize
\caption{Aggregate Gains Across Workflows}
\label{table:back of envelope}
\begin{threeparttable}
\hspace*{-0.1cm}\begin{tabular}{p{4.2cm} >{\raggedright\arraybackslash}p{3.5cm} >{\raggedright\arraybackslash}p{2cm}  >{\raggedright\arraybackslash}p{3.5cm}}
\hline\hline
& {\footnotesize(1)} & {\footnotesize(2)} & {\footnotesize(3)}    \\
{\footnotesize Business Workflow} & {\footnotesize Incremental Value \newline Per Consumer (\$)} & {\footnotesize Time Multiplier} & {\footnotesize Annualized Incremental \newline Value  Per Consumer (\$)}  \\
\hline
\textbf{Pre-sale Service Chatbot} & 0.274 (upper-bound)  \newline  0.218 (lower-bound)  & 6.0 \newline 6.0 & 1.64 \newline 1.31  \\
&  &  &      \\
\textbf{Search Query Refinement} & 0.0648 & 40.6 & 2.63  \\
&  &  &      \\
\textbf{Product Description} & 0.0104 & 52.1 & 0.54   \\
&  &  &      \\
\textbf{Marketing Push Message} & 0.001005  (upper-bound) \newline  0.000402 (lower-bound) & 365.0 \newline 365.0 & 0.37  \newline 0.15 \\\hline
\textbf{Total (linear additivity)} &  &  & \textbf{5.18} (upper-bound) \newline \textbf{4.63} (lower-bound)   \\
\hline\hline
\end{tabular}
\begin{tablenotes}
\scriptsize
\item[1] Column (1) reports the absolute lifts in sales (treatment effects of GenAI) for each workflow from Table \ref{table:summarized main results}. Column (2) shows the factor used to annualize the workflow-specific estimates. Column (3) reports the annualized values for each workflow.
\end{tablenotes}
\end{threeparttable}
\end{table}

Finally, we aggregate the annualized estimates across the four workflows to obtain the total annual incremental value per consumer attributable to GenAI (see ``Total'' in Column 3). This aggregation assumes that effects across workflows are linearly additive and abstracts from potential cross-workflow interactions, such as cannibalization across touchpoints or expansion through synergies. Based on the four GenAI applications with positive sales effects, we estimate an annual incremental value of up to \$5.18 per consumer when using the upper-bound estimates from the Marketing Push experiment (implied full-exposure effect) and the Pre-sale Service Chatbot (no service vs. GenAI assistance), and \$4.63 per consumer when using the lower-bound estimates from these experiments (Pre-sale Service Chatbot: human agents vs. GenAI assistance equipped with human escalation; Marketing Push: ITT effect with partial exposure). These effects represent roughly 5.5--6.2\% of the per-user revenue growth observed in global e-commerce between 2023 and 2024 \citep{statista2025}, highlighting the economic significance of these gains relative to broader industry trends.\\

It is worth noting that these estimates capture only a partial and early-stage view of the firm’s efforts to scale up GenAI across workflows. While in 2023 the platform applied GenAI to only a handful of workflows, by 2024 it was deployed in more than 40 applications and by 2025 in over 60. This rapid expansion is also reflected in the growth of API calls to the platform's GenAI tools: in mid-2024, AI-related API requests averaged over 50 million per day, rising to more than 1 billion per day by mid-2025, representing a twentyfold increase. Hence, our point estimates should be interpreted with caution and in light of these rapid adoption trends, which nonetheless indicate that the firm anticipates substantial value from GenAI.\\

Taken together, these results reveal sizable gains in targeted workflows and measurable contributions to overall platform sales, with substantial potential as GenAI applications diffuse across use cases and models are further refined for domain-specific tasks. This pattern aligns with the view that the aggregate productivity effects of new general-purpose technologies materialize gradually as complementary investments and organizational adaptations accumulate (e.g., \cite{acemoglu2025simple}).

\section{Heterogeneous Treatment Effects}
\label{sec:hte}
A central question in online retail is which platform participants benefit most from GenAI-driven improvements. To address this, we examine heterogeneity in treatment effects across buyers, sellers, and products. If GenAI primarily reduces frictions on both the demand and supply sides, one would expect relatively larger gains among participants with lower baseline capabilities—namely, less experienced buyers with limited platform engagement, smaller and newer sellers, and products in the long tail of the sales distribution or in less concentrated categories.\\

For each dimension, we classify observations into “high” and “low” groups based on pre-experiment characteristics that capture experience, scale, or market position. For buyers, we use measures of prior online shopping experience and purchasing intensity. For sellers, we classify firms based on size and tenure on the platform. For products, we rely on indicators of category concentration, sales volume, and relative price. These groupings allow us to assess whether GenAI adoption disproportionately benefits less sophisticated buyers, smaller or newer sellers, and products in the long tail, thereby shedding light on how GenAI reshapes outcomes across segments of the marketplace.\\

We find consistent evidence of heterogeneity on the demand side: buyers with lower prior spending and less platform experience benefit more from GenAI-powered workflows. On the supply side, the estimated effects point to larger gains for smaller and newer sellers, although these estimates are less precise. Finally, heterogeneity across product groups is more context-dependent. We note that, due to data limitations and workflow-specific features, for some workflows we could not conduct all heterogeneity analyses.

\subsection{Heterogeneous Effects Across Consumers}

We classify consumers into high- and low-experience groups based on three pre-experiment indicators of online shopping experience: total expenditure over the 30-day period preceding the experiment (Past Purchases), number of login days in the 30 days prior to the experiment (Past Login Days), and years since registration on the platform (Registered Years). For each indicator, consumers in the low group are defined as relatively inexperienced if they fall below the median of the corresponding distribution. We estimate treatment effects separately for high- and low-experience consumers and test for differences in percentage treatment effects across groups using Wald tests implemented via seemingly unrelated estimation.\\

\begin{figure}[h!]
\centering
\caption{Heterogeneous Treatment Effects on Sales Across Consumers}
\label{fig:summarized_hte_customer}

\begin{tabular}{c c}

\raisebox{1.1\height}{\rotatebox{90}{\footnotesize Chatbot}} &
\includegraphics[width=0.6\textwidth]{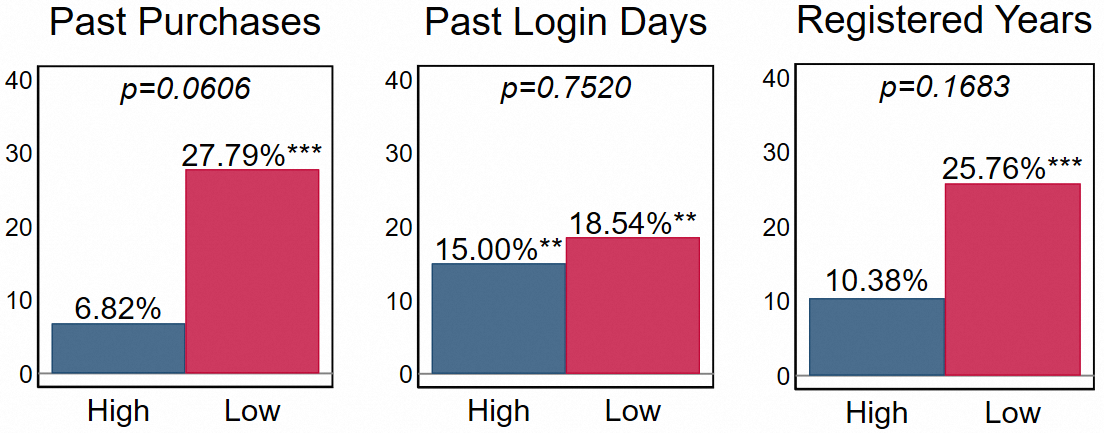} \\

\raisebox{1.5\height}{\rotatebox{90}{\footnotesize Search}} &
\includegraphics[width=0.6\textwidth]{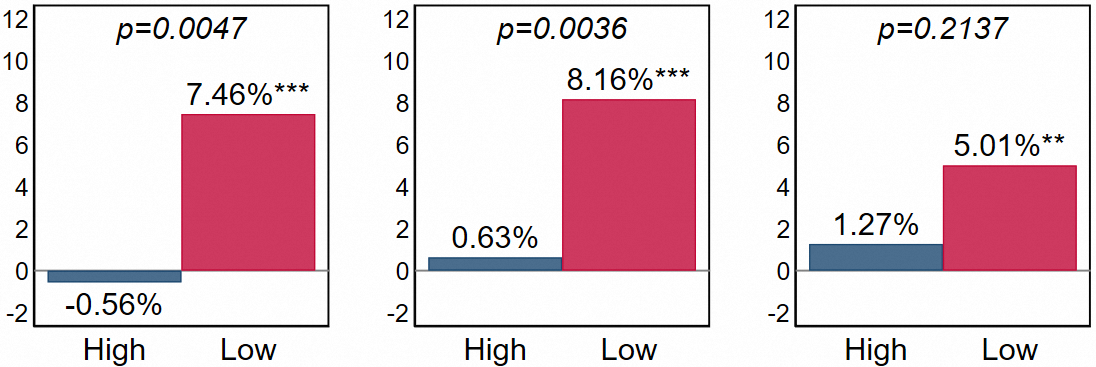} \\

\raisebox{0.7\height}{\rotatebox{90}{\footnotesize Description}} &
\includegraphics[width=0.6\textwidth]{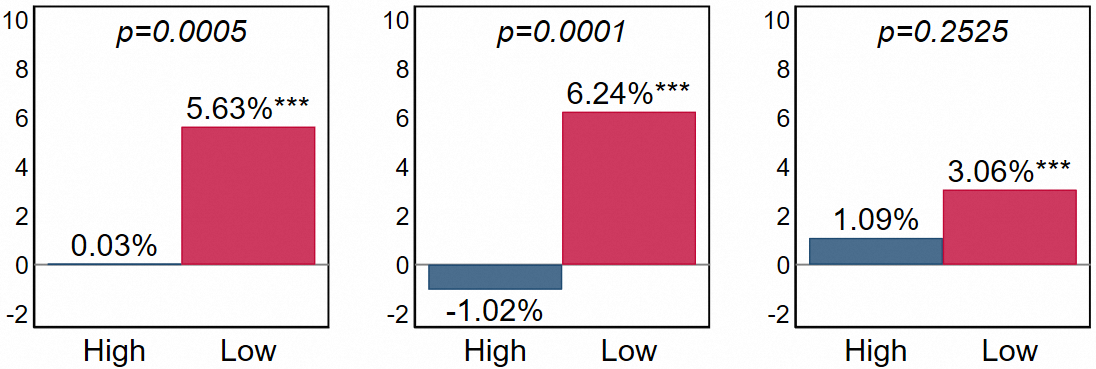} \\

\raisebox{2.1\height}{\rotatebox{90}{\footnotesize Push}} &
\includegraphics[width=0.612\textwidth]{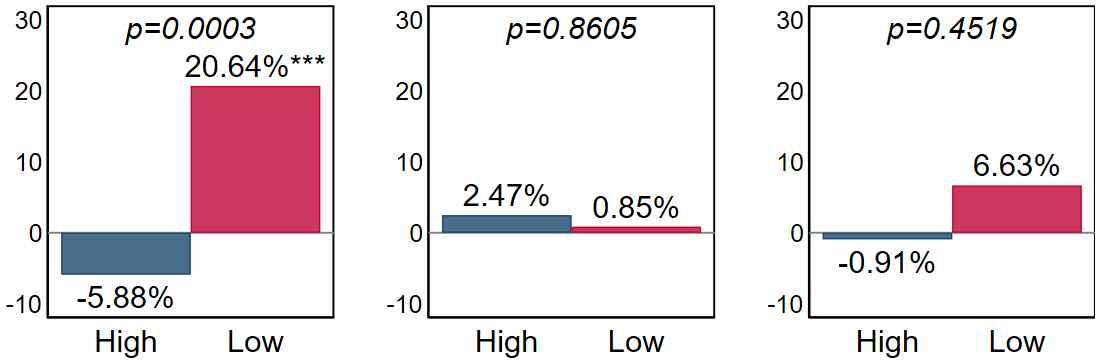}
\end{tabular}

\par\medskip
\begin{minipage}{0.9\textwidth}
\scriptsize
\textit{Notes:} The bars indicate the \% change in treatment effects for the high (blue) and low (red) groups. The p-values are from seemingly unrelated estimations testing the equality of \% treatment effects across the two groups.
\end{minipage}
\end{figure}

Results on sales are summarized in Figure \ref{fig:summarized_hte_customer}, with regression results and additional evidence on conversion rates and cart values reported in Appendix \ref{appendix:detailed_hte_customer}.\footnote{The Google Advertising Title experiment is not suitable for consumer-level heterogeneity analysis, as it was conducted off-platform at the product level, preventing access to detailed consumer characteristics.} Overall, we find statistically significant heterogeneity in GenAI’s effects across consumer segments. We begin by analyzing heterogeneity by past purchases, our preferred proxy for consumers’ prior experience on the platform. Across workflows, GenAI-driven gains are significant and systematically larger for consumers with fewer past purchases, while the effects for consumers with more purchase experience are smaller and insignificant. Further analysis confirms that the differences between these two consumer groups are statistically significant. We find qualitatively similar patterns when stratifying consumers by past login activity and platform tenure, although the differences along these dimensions are generally less pronounced.\\

Our results indicate that the technology disproportionately benefits less experienced and less sophisticated users, aligning with prior evidence that GenAI can disproportionately benefit vulnerable groups, such as low-skilled workers \citep{Bjorn2025, Noy2023, Cui2026}, and extend this insight to the consumer domain. The results are also consistent with earlier work demonstrating that improvements in e-commerce technologies, such as enhanced information provision, search refinement, and personalization, generate greater benefits for consumers who are newer to the platform or have lower purchasing power \citep{sun2024value}. \\

A plausible explanation for the observed buyer heterogeneity is that less experienced consumers face greater informational and search frictions throughout the purchase journey. With limited domain knowledge and fewer alternative information sources, they rely more heavily on customer service and product descriptions to acquire relevant information. They also exhibit weaker search skills and less familiarity with online search environments, making GenAI-enhanced search particularly valuable for articulating their needs and identifying relevant products. In addition, GenAI-driven message personalization increases the perceived relevance of marketing content, which is especially effective in supporting the purchase decisions of less experienced consumers.

\subsection{Heterogeneous Effects Across Sellers}

We examine heterogeneity in GenAI’s effects across sellers by focusing on differences in seller size and sophistication. Sellers are classified into high and low groups based on three pre-experiment characteristics: annual sales volume (Annual Sales), years of operation on the platform (Operation Years), and the number of sub-accounts associated with the seller’s store (\# of Sub-Accounts). Sellers in the low group correspond to smaller and less established firms, defined by lower sales, shorter platform tenure, or fewer sub-accounts.\footnote{Sellers are classified into the low group if they (i) account for the bottom 50\% of cumulative annual sales, (ii) have operated on the platform for fewer than five years (the platform treats five years as a key performance benchmark), or (iii) maintain fewer than three sub-accounts (following platform guidance, small stores are typically individually operated or mom-and-pop businesses, whereas stores with more sub-accounts generally employ additional staff). Platform-operated sellers are classified as large.} We then decompose consumer purchases into transactions involving high- and low-group sellers and estimate treatment effects separately for each group. Differences in percentage treatment effects are formally tested using Wald tests implemented via seemingly unrelated estimation.\\

\begin{figure}[tb]
\centering
\caption{Heterogeneous Treatment Effects on Sales Across Sellers}
\label{fig:summarized_hte_seller}

\begin{tabular}{c c}

\raisebox{1.6\height}{\rotatebox{90}{\footnotesize Search}} &
\includegraphics[width=0.6\textwidth]{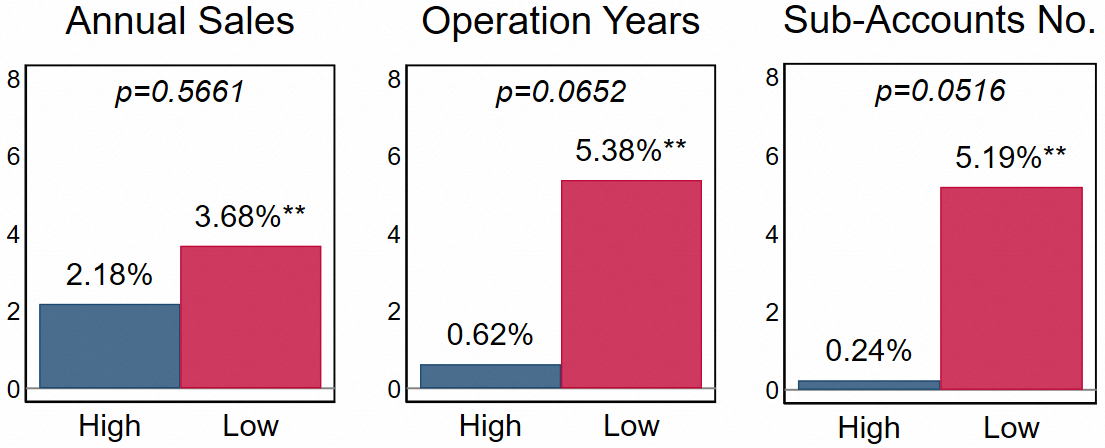} \\

\raisebox{2.1\height}{\rotatebox{90}{\footnotesize Push}} &
\includegraphics[width=0.605\textwidth]{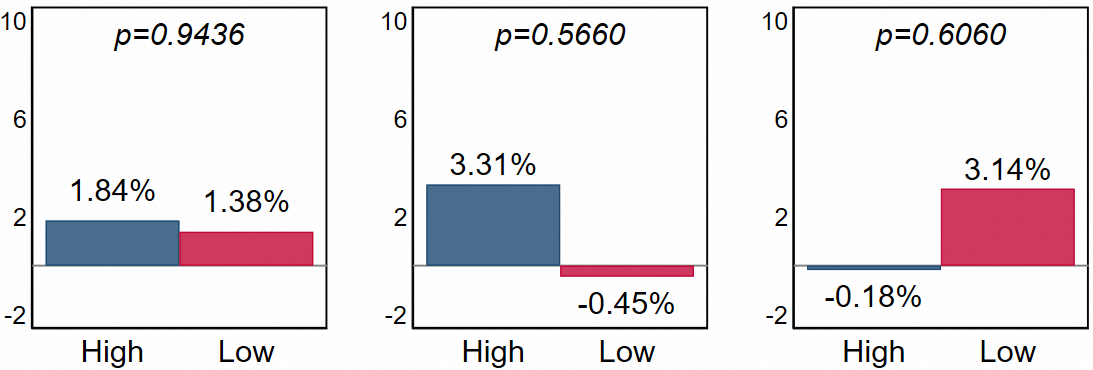} \\

\raisebox{1.4\height}{\rotatebox{90}{\footnotesize Google}} &
\includegraphics[width=0.615\textwidth]{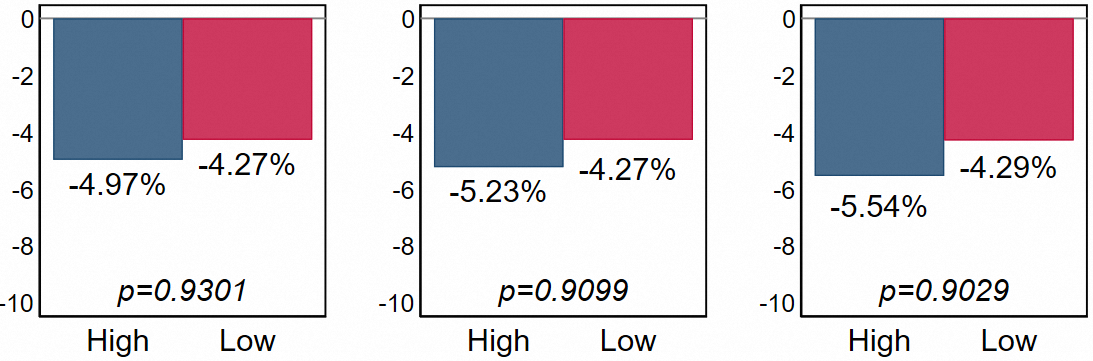}
\end{tabular}

\par\medskip
\begin{minipage}{0.9\textwidth}
\scriptsize
\textit{Notes:} The bars indicate the \% change in treatment effects for the high (blue) and low (red) groups. The p-values are from seemingly unrelated estimations testing the equality of \% treatment effects across the two groups.
\end{minipage}

\end{figure}

Summary results are reported in Figure \ref{fig:summarized_hte_seller}, with regression results and additional details on conversion rates and cart values provided in Appendix \ref{appendix:detailed_hte_seller}. Overall, we find limited statistical evidence of differential effects across seller types. While point estimates often suggest larger gains for smaller and less established sellers, most differences between high- and low-group sellers are statistically insignificant, reflecting imprecise estimates and limited power. This pattern holds across the workflows for which seller-level heterogeneity can be assessed.\footnote{We cannot assess seller heterogeneity in the Pre-sale Service Chatbot and Product Description experiments, as both were conducted exclusively on platform self-sold products with only a handful of platform-operated sellers, leaving insufficient variation for meaningful analysis.}

\subsection{Heterogeneous Effects Across Products}

\begin{figure}[tb]
\centering
\caption{Heterogeneous Treatment Effects on Sales Across Products}
\label{fig:summarized_hte_product}

\begin{tabular}{c c}

\raisebox{1.0\height}{\rotatebox{90}{\footnotesize Chatbot}} &
\includegraphics[width=0.6\textwidth]{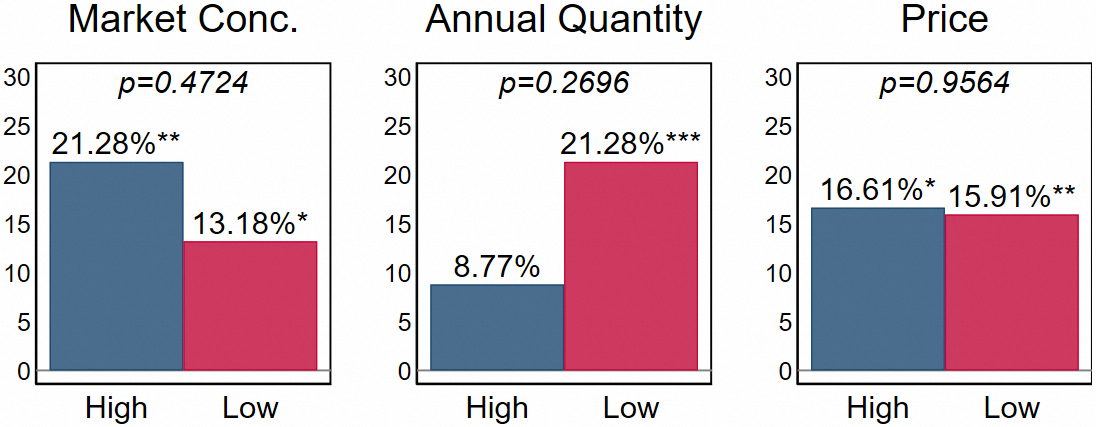} \\

\raisebox{1.6\height}{\rotatebox{90}{\footnotesize Search}} &
\includegraphics[width=0.6\textwidth]{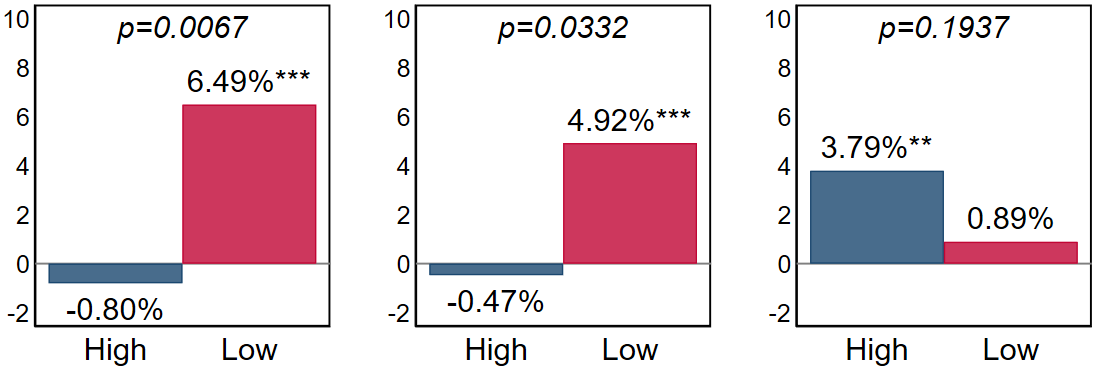} \\

\raisebox{0.7\height}{\rotatebox{90}{\footnotesize Description}} &
\includegraphics[width=0.6\textwidth]{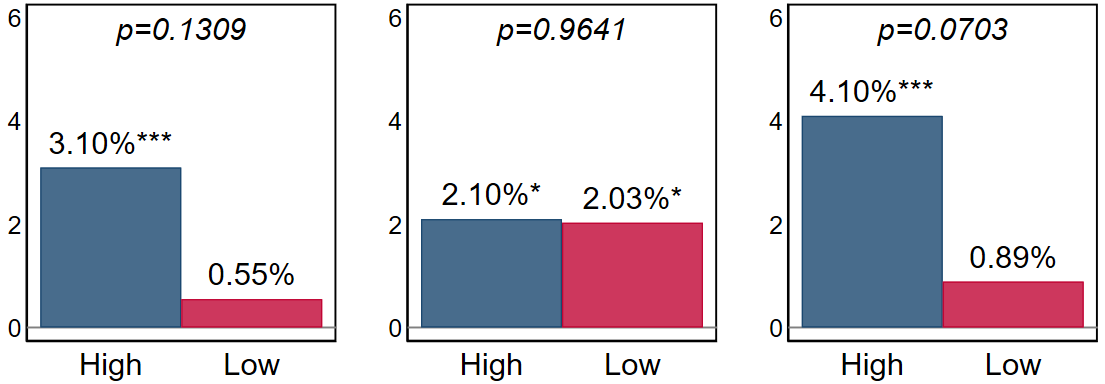}  \\

\raisebox{2.3\height}{\rotatebox{90}{\footnotesize Push}} &
\includegraphics[width=0.6\textwidth]{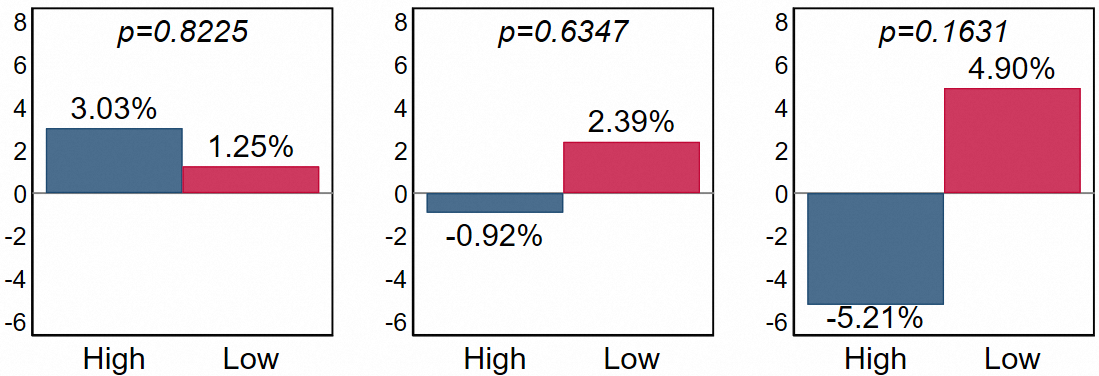}  \\

\raisebox{1.5\height}{\rotatebox{90}{\footnotesize Google}} &
\includegraphics[width=0.6\textwidth]{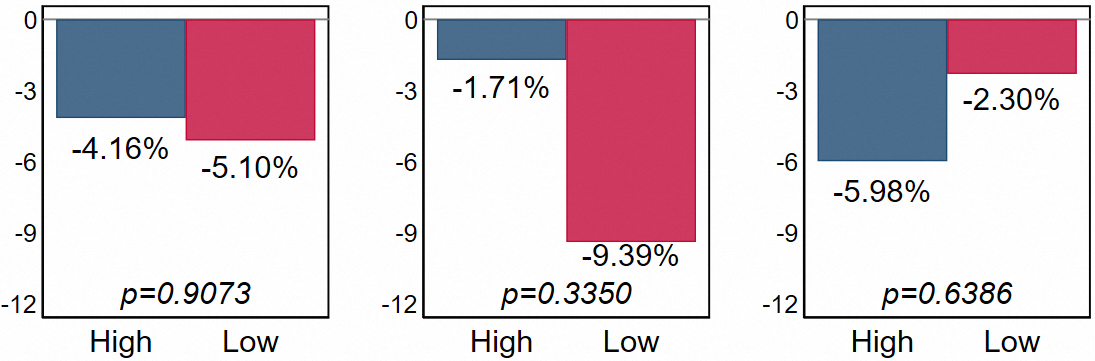}
\end{tabular}

\par\medskip
\begin{minipage}{0.9\textwidth}
\scriptsize
\textit{Notes:} The bars indicate the \% change in treatment effects for the high (blue) and low (red) groups. The p-values are from seemingly unrelated estimations testing the equality of \% treatment effects across the two groups.
\end{minipage}

\end{figure}

Finally, we study heterogeneity in GenAI’s effects across products by classifying them into high and low groups along three pre-experiment dimensions. First, category market concentration, measured by the sales share of the top 1\% of products within each category, captures the degree of product standardization and demand homogeneity. Second, annual sales quantity within category distinguishes head from tail products. Third, product price within category proxies for purchase stakes and potential information asymmetries. For each dimension, products in the low group correspond to less concentrated categories, lower-sales (tail) products, or lower-priced items.\footnote{Products are classified into the low group if they (i) belong to categories with concentration below the platform average, (ii) account for the bottom 50\% of cumulative sales quantity within their category, or (iii) are priced below the category median.} We decompose consumer purchases into transactions involving high- and low-group products and estimate treatment effects separately by group, testing for differences in percentage treatment effects using seemingly unrelated estimation. \\

Summary results are reported in Figure \ref{fig:summarized_hte_product}, with regression results and additional details provided in Appendix \ref{appendix:detailed_hte_product}. Overall, we find no uniform or sharply defined pattern of product-level heterogeneity across workflows. While several point estimates suggest larger GenAI-driven gains for products in less concentrated categories, tail products or high-priced items—depending on the specific application—most differences between high- and low-group products are statistically insignificant. The direction of the effects is often consistent with GenAI reducing search and information frictions in settings with greater product differentiation, limited sales history or higher decision stakes, but these patterns are not robust across all workflows. Overall, the evidence points to limited and context-dependent product-level heterogeneity in the effects of GenAI.

\section{Discussion and Conclusions}

The rapid advances in GenAI have generated widespread expectations among investors and business leaders, driving unprecedented investment in infrastructure and applications. Yet important questions remain about where GenAI creates business value, how large those gains are in real-world enterprise settings, and whether they extend beyond improvements in narrowly defined worker-level tasks. This paper provides large-scale experimental evidence on these questions in online retail, studying the deployment of GenAI across multiple consumer-facing workflows within a single global platform. The setting allows us to examine how GenAI affects front-office sales and related business outcomes under a common organizational environment.\\

Our findings yield three main insights. First, GenAI can deliver measurable short-run improvements in sales across several business workflows, holding other observed inputs and prices fixed. Although the magnitude of these gains varies widely, from negligible effects to double-digit increases, the evidence shows that GenAI can generate economically meaningful value when deployed in consumer-facing retail processes. Second, the pattern of results is consistent with GenAI improving the consumer experience by reducing marketplace frictions along the customer journey. Across workflows, we observe higher conversion rates but little evidence of changes in spending intensity, and post-purchase outcomes such as return rates and customer ratings do not deteriorate. By enriching pre-sale communication, refining search queries, generating richer product descriptions, and personalizing marketing messages, GenAI appears to improve matching efficiency and mitigate information asymmetries, although our data do not allow us to uniquely separate these channels from others such as persuasion, trust, or novelty. Third, the benefits of GenAI adoption are heterogeneous across market participants. On the demand side, less experienced consumers derive disproportionately larger gains. On the supply side, point estimates suggest larger benefits for smaller and newer sellers, though these effects are less precisely estimated. Heterogeneity across products is more context-dependent. Taken together, these patterns suggest that GenAI may be especially valuable for participants facing greater baseline frictions.\\

Back-of-the-envelope calculations provide a sense of the economic magnitude of these short-run sales impacts. Annualizing workflow-specific gains and assuming linear additivity, the four GenAI applications with positive sales effects generate an estimated annual incremental value of approximately \$4.6--\$5.2 per consumer. These effects correspond to roughly 5.5--6.2\% of global per-user e-commerce revenue growth in 2023--2024. These figures should be interpreted cautiously: they assume that gains from different workflows can be summed without accounting for synergies or overlap, reflect the platform’s partial adoption of GenAI in selected workflows during a short experimental period, and capture top-line sales gains rather than bottom-line profitability or full return on investment. Nevertheless, they indicate that even a small number of GenAI applications can generate substantial value for a large, mature retailer. By 2025, the partner platform had deployed GenAI in more than 60 workflows, with usage rising twentyfold as API calls to its proprietary GenAI tools increased between 2024 and 2025. Although our data do not allow us to estimate ROI directly---since we do not observe the full capital, engineering, inference, and organizational costs of GenAI deployment---this accelerating expansion is consistent with the platform internally realizing returns that more than offset these costs at scale.\\

Our study has several limitations that inform the interpretation of the results and point to avenues for future research. First, the adoption horizon in our experiments was short, spanning days to months. Therefore, our analysis reflects only the immediate, short-run effects of GenAI. The impacts of sustained GenAI use may differ as sellers and consumers adapt their behavior over time and as platforms refine model deployment. Relatedly, we lack data to assess longer-term outcomes such as consumer retention and repeat purchasing. For example, while our results show no deterioration in post-purchase outcomes, as measured by return rates and customer ratings, they do not speak to longer-run consumer responses, such as changes in trust, loyalty, or engagement with the platform.\\

Second, our analysis is limited to seven workflows that were selected by the platform based on managerial assessments of technical feasibility, organizational costs, and expected sales productivity improvements, rather than representing the full spectrum of business processes where the technology could be deployed. Other business processes, including logistics, inventory management, or dynamic pricing, remain unexplored and could yield distinct productivity effects.\\

Third, our estimates do not account for potential changes in labor and capital inputs. Many of the studied processes, including customer service, content creation, and chargeback defense, are currently staffed or supported by human labor. Over time, GenAI adoption could displace or augment these functions, yielding additional labor efficiency improvements not captured in our current analysis.\\

Fourth, we cannot uniquely attribute the observed sales gains to market expansion, because some gains may reflect intertemporal or cross-category substitution. While we lack the data to assess intertemporal substitution directly, we examined cross-category substitution within the platform through additional, non-tabulated analyses of consumers' other purchases outside the focal experimental context. For example, in Search Query Refinement, we separately examine purchases made outside the search context, while in Product Description, we examine purchases of products not included in the description experiment. The estimated effects on these other purchases are statistically insignificant across experiments, suggesting limited cross-category cannibalization during the experimental period.\footnote{We cannot report the detailed tables because these analyses use purchase data outside the focal workflows or product categories, which are commercially sensitive.}\\

Other limitations concern external validity and general-equilibrium effects. Our experiments were conducted on a single, albeit large, global retail platform and may not generalize to other marketplaces. Moreover, as GenAI adoption spreads and competing platforms deploy similar technologies, the estimated effects may be different. For instance, our experiments abstract from strategic responses by competitors, such as changes in pricing or advertising strategy, which could either amplify or attenuate realized sales gains. However, if GenAI improves consumer experience and reduces frictions more broadly, there may still be scope for market expansion even in a more competitive environment.\\

Taken together, our results show both the current scope and the future potential of GenAI adoption in online retail. In the short run, GenAI delivers measurable gains by creating demand-side value within selected workflows, partly through reductions in market frictions and improvements in the consumer shopping experience. These gains are economically meaningful given the scale and maturity of the partner platform. Over the longer run, GenAI’s impact may expand as firms move beyond early use cases, capture cost-reduction opportunities, and adapt organizational structures to integrate the technology more effectively. Continued advances in computational speed, accuracy, and domain coverage may further amplify these effects. At the aggregate level, however, widespread adoption raises open questions about equilibrium effects, competitive dynamics, and the persistence of observed gains. Our study provides a first step by offering causal evidence on how GenAI can reshape core retail workflows to improve business outcomes, while pointing to important directions for future research on general-equilibrium effects, cost-side adjustments, and long-term productivity growth.

\printbibliography

\clearpage

\appendix

\section*{Appendix}
\section{Illustrative User Interfaces and Examples of Each Experiment}
\setcounter{figure}{0}
\renewcommand{\thefigure}{A\arabic{figure}}
\renewcommand{\theHfigure}{A\arabic{figure}} 
\setcounter{table}{0}
\renewcommand{\thetable}{A\arabic{table}}
\renewcommand{\theHtable}{A\arabic{table}} 

\label{appendix:interfaces}

\begin{figure}[h!]
    \begin{center}
    \caption{Illustration of Pre-sale Service Chatbot}
    \label{fig:presale_chatbot}	
        \begin{subfigure}[t]{0.3\linewidth}
            \centering
            \includegraphics[width=\textwidth]{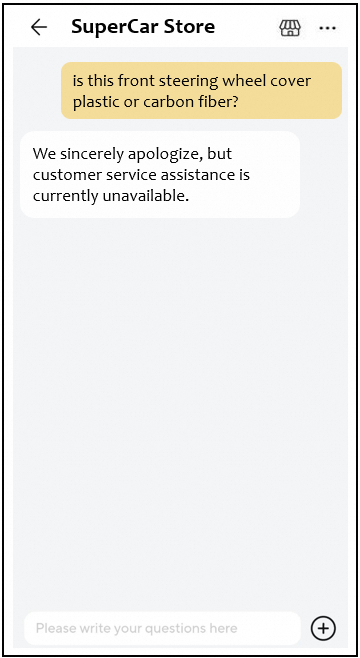}
            \caption{Auto-Response (No Service)}
        \end{subfigure}
         \hspace{0.05\textwidth}
        \begin{subfigure}[t]{0.3\linewidth}
            \centering
            \includegraphics[width=\textwidth]{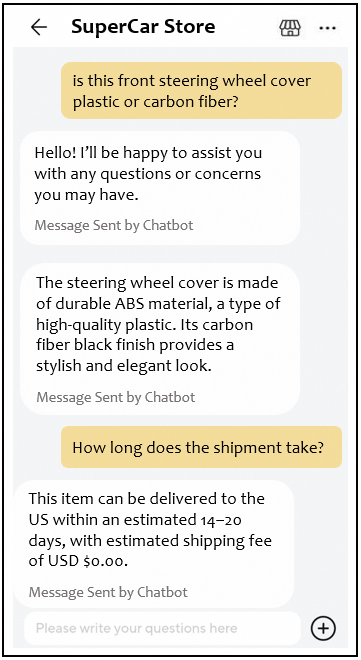}
            \caption{GenAI-Driven Chatbot}
        \end{subfigure}
    \end{center}
\end{figure}

\paragraph{Pre-sale Service Chatbot} In Figure \ref{fig:presale_chatbot}, Panel (a) depicts the control condition, where a pre-programmed auto response indicates that no service is available, while Panel (b) presents the treatment condition, featuring support from a GenAI-driven chatbot. The chatbot acts as a virtual sales assistant, available 24/7 to provide instant responses to pre-sale inquiries, covering product features, pricing, availability, and delivery options across multiple languages. \\

\begin{figure}[h!]
    \begin{center}
    \caption{Illustration of Search Query Refinement}
    \label{fig:search_query_refinement}	
        \begin{subfigure}[t]{0.405\linewidth}
            \centering
            \includegraphics[width=\textwidth]{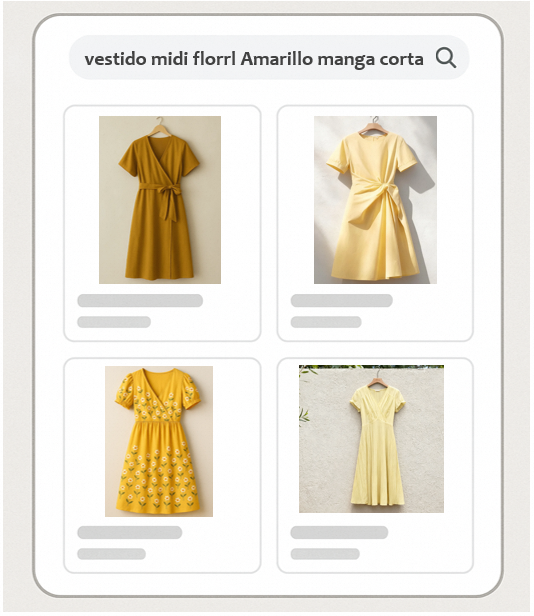}
            \caption{Search Results of Consumer Query}
        \end{subfigure}
         \hspace{0.05\textwidth}
        \begin{subfigure}[t]{0.4\linewidth}
            \centering
            \includegraphics[width=\textwidth]{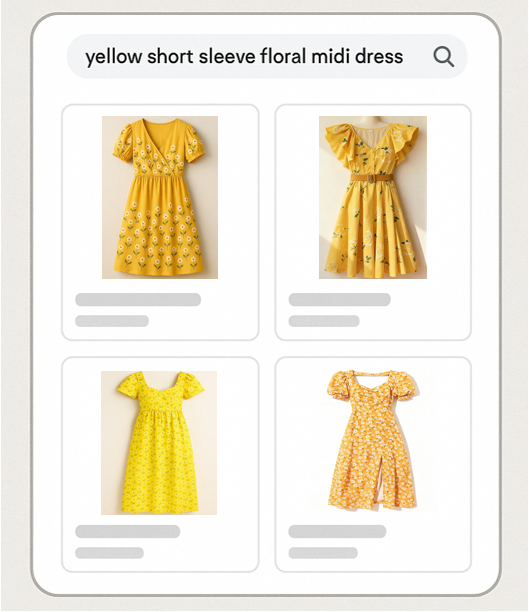}
            \caption{Search Results of GenAI-Refined Query}
        \end{subfigure}
    \end{center}
\end{figure}

\paragraph{Search Query Refinement} In Figure \ref{fig:search_query_refinement}, Panel (a) displays the consumer’s original search query in Spanish, along with the corresponding results, while Panel (b) presents the structured English query translated and refined by GenAI, as well as the corresponding search results retrieved. The GenAI-powered query refinement can improve the expression of consumer demand and facilitate the matching efficiency of search engines.

\begin{figure}[h!]
    \begin{center}
    \caption{Illustration of Product Description}
    \label{fig:product_description}	
        \begin{subfigure}[t]{0.48\linewidth}
            \centering
            \includegraphics[width=\textwidth]{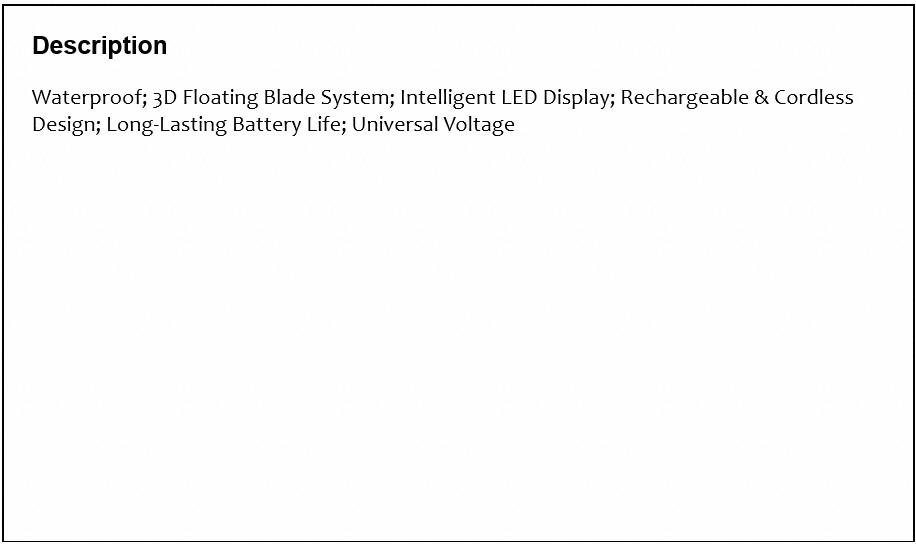}
            \caption{Human-Created Description}
        \end{subfigure}
        \begin{subfigure}[t]{0.48\linewidth}
            \centering
            \includegraphics[width=\textwidth]{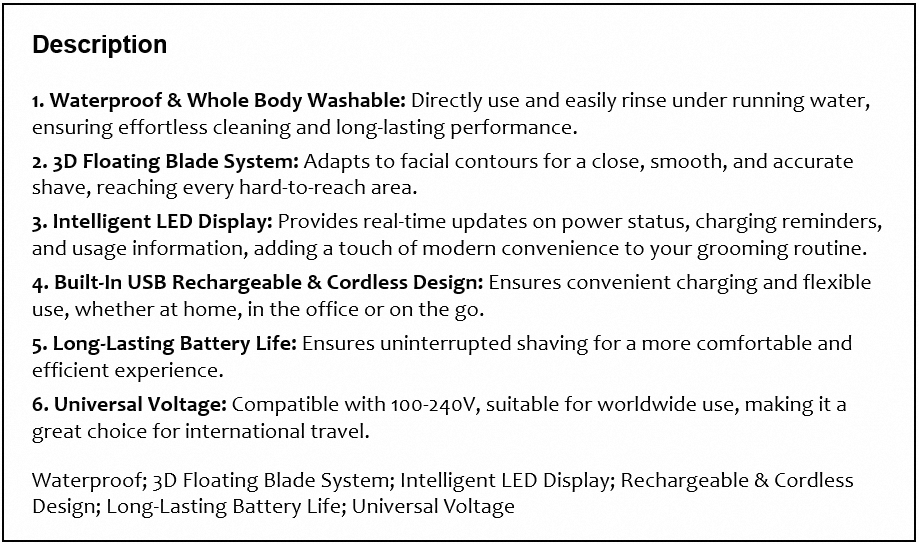}
            \caption{GenAI-Created Description Above Original Human Input}
        \end{subfigure}
    \end{center}
\end{figure}

\paragraph{Product Description} In Figure \ref{fig:product_description}, Panel (a) illustrates the control condition with human-generated descriptions, while Panel (b) shows the treatment condition, where GenAI-created descriptions are layered on top of those written by humans. The AI-generated content provides more comprehensive and structured bullet-point-style descriptions that highlight product features, benefits, and typical use cases.

\begin{figure}[h!]
    \begin{center}
    \caption{Illustration of Marketing Push Message}
    \label{fig:marketing_push}
        \begin{subfigure}[t]{0.48\linewidth}
            \centering
            \includegraphics[width=\textwidth]{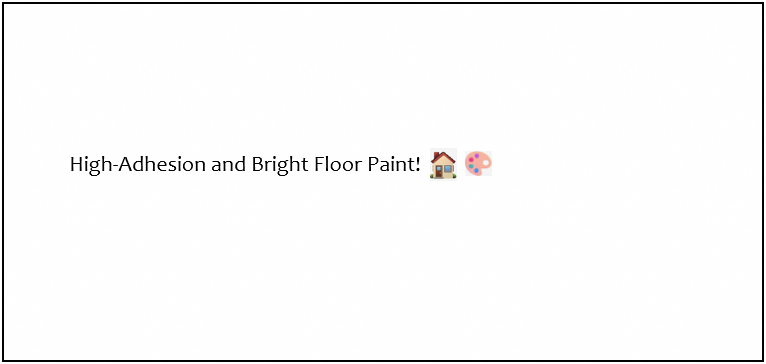}
            \caption{Human-Created Push Message}
        \end{subfigure}
        \begin{subfigure}[t]{0.48\linewidth}
            \centering
            \includegraphics[width=\textwidth]{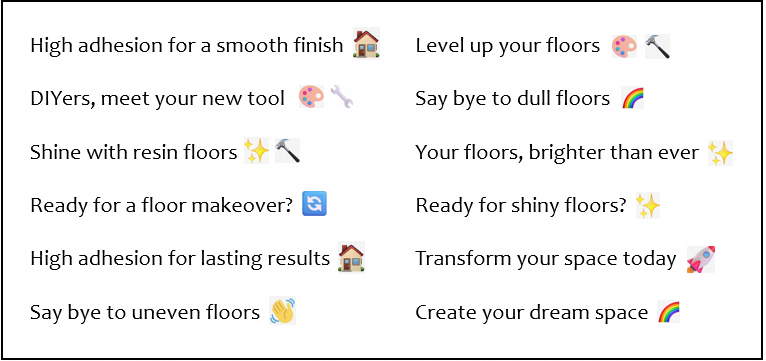}
            \caption{GenAI-Created Push Messages}
        \end{subfigure}
    \end{center}	
\end{figure}

\paragraph{Marketing Push Message} In Figure \ref{fig:marketing_push}, Panel (a) illustrates a human-generated marketing push message, whereas Panel (b) displays multiple messages produced by GenAI for the same product. Generative AI enables large-scale creation of diverse marketing content, increasing the likelihood that consumers encounter differentiated messages and thereby allowing platforms to leverage the benefits of personalized marketing.

\begin{figure}[!htp]
    \begin{center}
    \caption{Illustration of Google Shopping Interface}
    \label{fig:google_ad_title}
    \includegraphics[width=0.49\paperwidth]{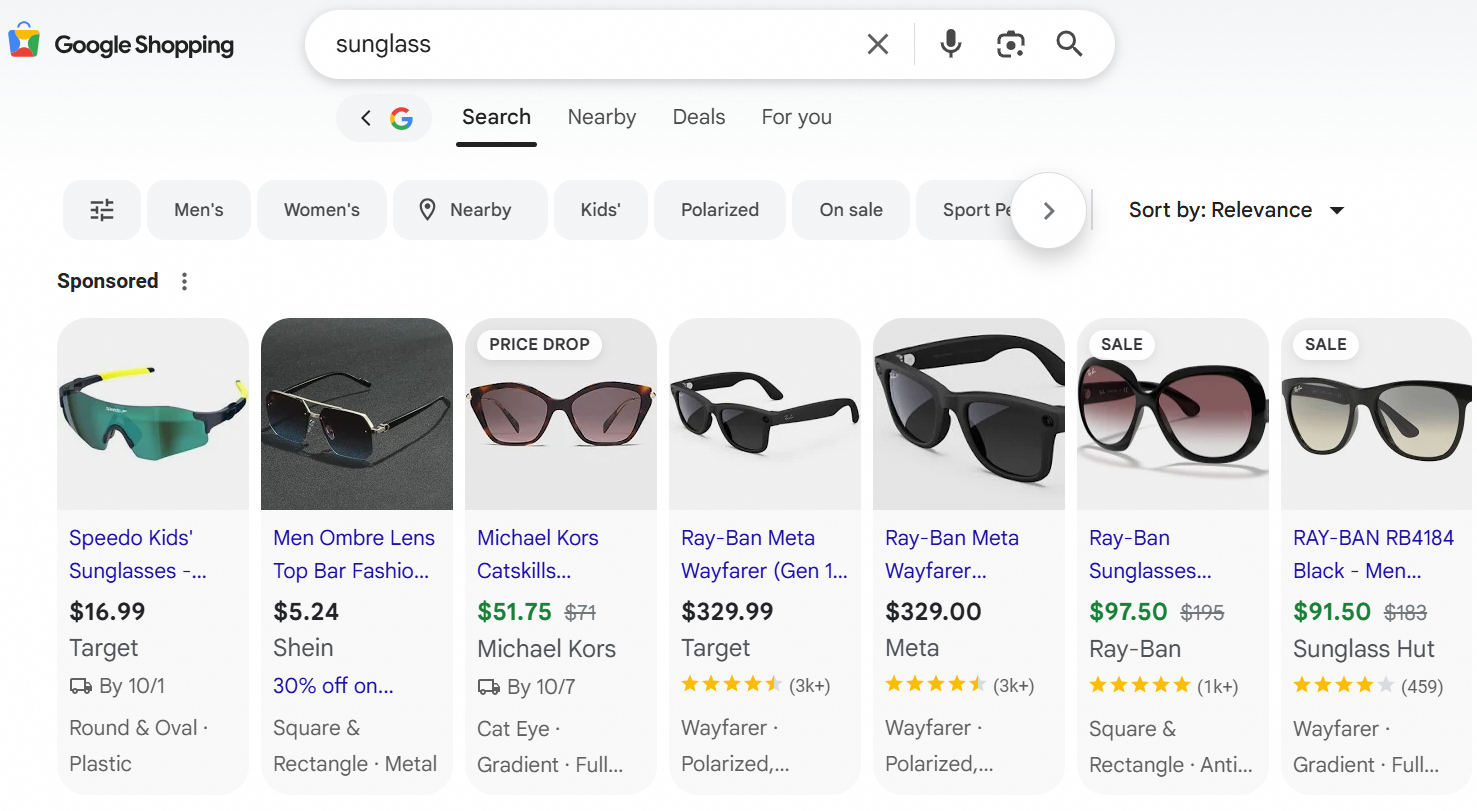}
    \end{center}
\end{figure}

\paragraph{Google Advertising Title} In this case, GenAI is applied to optimize human-generated product titles for Google advertising. Since the model was not fine-tuned with e-commerce domain knowledge, the performance of GenAI-optimized titles is lower than that of human-generated titles. For example, the original human-generated title for a pair of sunglasses is: ``2024 New Arrival Polarized Pitboss 2 Sunglasses Men Cycling Eyewear Goggles Bicycle Glasses". The GenAI-optimized version is: ``Men's Polarized Pitboss 2 Sunglasses - Polycarbonate Frame for Cycling, Sports, Bike Goggles Bicycle". On Google Shopping, the first few words of a product title are the most prominent, as shown in Figure \ref{fig:google_ad_title}. Thus, by removing popular keywords such as “New Arrival,” GenAI may reduce consumer attention.

\begin{figure}[!htp]
    \begin{center}
    \caption{Illustration of Chargeback Defense}
    \label{fig:chargeback_defense}
    \includegraphics[width=0.50\paperwidth]{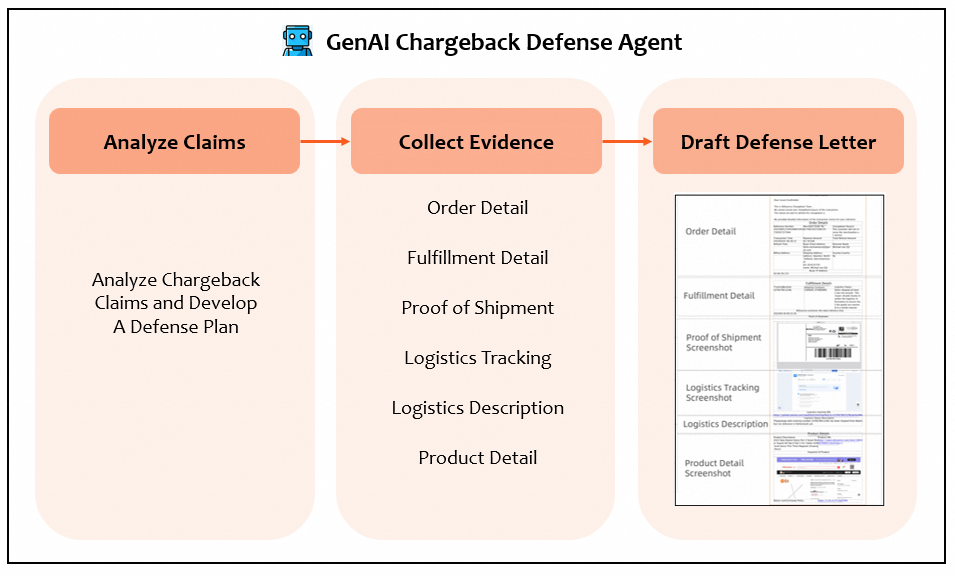}
    \end{center}
\end{figure}

\paragraph{Chargeback Defense} Figure \ref{fig:chargeback_defense} illustrates the process by which the chargeback defense agent interprets and analyzes consumer claims, gathers relevant evidence—including transaction records, product details, and shipping information—and drafts persuasive defense letters. By automating this complex workflow, GenAI enables sellers to respond to chargeback claims more quickly and consistently, thereby improving win rates while reducing compliance burdens and financial losses.

\begin{figure}[h!]
    \begin{center}
    \caption{Illustration of Live Chat Translation}
    \label{fig:live_chat_translation}	
        \begin{subfigure}[t]{0.23\linewidth}
            \centering
            \includegraphics[width=\textwidth]{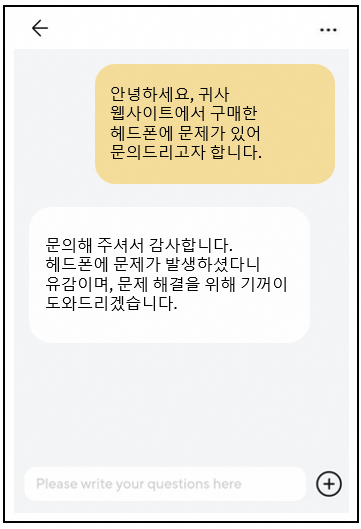}
            \caption{Consumer Interface}
        \end{subfigure}
         \hspace{0.01\textwidth}
        \begin{subfigure}[t]{0.55\linewidth}
            \centering
            \includegraphics[width=\textwidth]{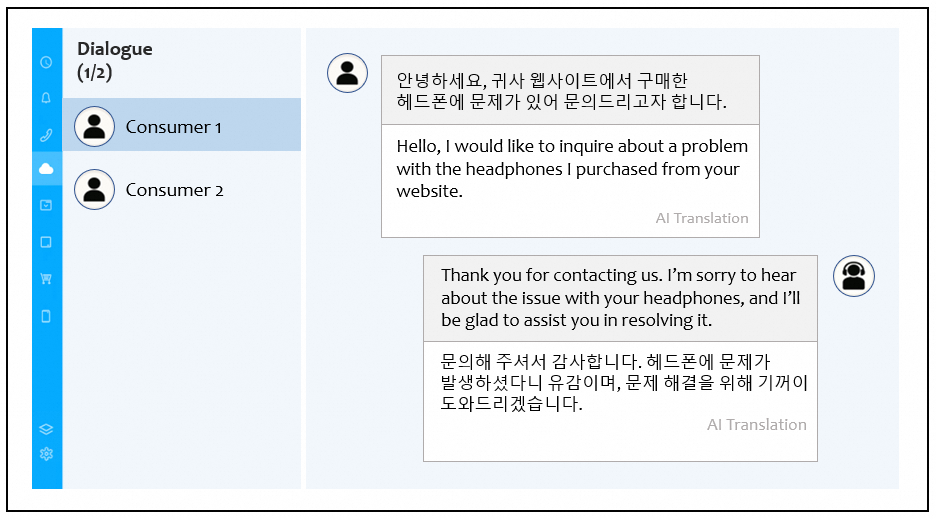}
            \caption{Customer Service Agent Interface}
        \end{subfigure}
    \end{center}
\end{figure}

\paragraph{Live Chat Translation} Figure \ref{fig:live_chat_translation} depicts the live chat translation system. Panel (a) shows the consumer interface, where a Korean consumer submits a query. Panel (b) presents the service agent interface, where a Filipino agent receives the query translated in real time from Korean to English by GenAI. The system supports bidirectional translation: the agent’s English response is simultaneously translated into Korean, allowing the consumer to receive the reply in their native language. This functionality enables English-speaking Filipino agents to communicate seamlessly with consumers across multiple languages on the platform.

\clearpage

\section{Covariate Balance Checks for the Field Experiments}
\label{appendix:balance checks}
\setcounter{figure}{0}
\renewcommand{\thefigure}{B\arabic{figure}}
\renewcommand{\theHfigure}{B\arabic{figure}} 
\setcounter{table}{0}
\renewcommand{\thetable}{B\arabic{table}}
\renewcommand{\theHtable}{B\arabic{table}} 
In this section, we present detailed covariate balance checks for each of the five experiments for which granular data are available.

\begin{table}[!htpb]
\centering
\small
\caption{Covariate Balance and Randomization Checks} 
\label{table:balance check}
\begin{threeparttable} 
\begin{tabular}{lcccc}
\hline\hline
		                       & Control & Treatment & \textit{p-value} (C=T) \\ \hline
\textbf{Pre-sale Service Chatbot}      &&& \\
\hspace{2mm} Gender		      & 1.000 (0.600)	  & 1.004 (0.598)	  & 0.503	\\	
\hspace{2mm} Age Tier		  & 1.000 (0.280)	  & 1.001 (0.277)	  & 0.816	\\	
\hspace{2mm} Registered	Years & 1.000 (0.988)   &	0.997 (0.957)	  &	0.779	\\		
\hspace{2mm} Past Login	Days  & 1.000 (1.071)	  &	1.014 (1.131)	  &	0.220	\\	
\hspace{2mm} Past Orders	  & 1.000 (3.108)   &	0.981 (2.863)	  &	0.528	\\	
\hspace{2mm} Past Sales		  & 1.000 (6.441)	  &	1.017 (9.097)	  &	0.833	\\	
\hspace{2mm} N. of Consumers              & 15,457  & 29,157    &   \\ \hline
\textbf{Search Query Refinement}       &&& \\
\hspace{2mm} Gender           & 1.000 (0.799)   & 1.001 (0.799)     & 0.204 \\
\hspace{2mm} Age Tier         & 1.000 (0.289)   & 1.000 (0.289)     & 0.466 \\
\hspace{2mm} Registered Years & 1.000 (0.860)   & 1.001 (0.861)     & 0.333 \\
\hspace{2mm} Past Login Days  & 1.000 (1.051)   & 1.002 (1.054)     & 0.207  \\
\hspace{2mm} Past Orders      & 1.000 (3.214)   & 1.007 (3.468)     & 0.140 \\
\hspace{2mm} Past Sales         & 1.000 (6.850)   & 1.014 (8.206)     & 0.217  \\ 
\hspace{2mm} N. of Consumers              & 929,188 & 920,194   &   \\ \hline  
\textbf{Product Description}           &&& \\
\hspace{2mm} Gender           & 1.000 (0.912)   & 1.000 (0.912)     & 0.778 \\
\hspace{2mm} Age Tier         & 1.000 (0.283)   & 1.000 (0.283)     & 0.127 \\
\hspace{2mm} Registered Years & 1.000 (0.942)   & 1.001 (0.942)     & 0.192 \\
\hspace{2mm} Past Login Days  & 1.000 (0.942)   & 0.999 (0.944)     & 0.490  \\
\hspace{2mm} Past Orders      & 1.000 (2.437)   & 1.001 (2.516)     & 0.627 \\
\hspace{2mm} Past Sales         & 1.000 (3.613)   & 0.998 (3.613)     & 0.515  \\
\hspace{2mm} N. of Consumers              & 2,392,803  & 2,380,134 &             \\ \hline  
\textbf{Marketing Push Message}        &&& \\
\hspace{2mm} Gender           & 1.000 (0.894)  & 1.000 (0.894)    & 0.599 \\
\hspace{2mm} Age Tier         & 1.000 (0.275)  & 1.000 (0.277)    & 0.538 \\
\hspace{2mm} Registered Years & 1.000 (1.009)  & 1.000 (1.008)    & 0.714 \\
\hspace{2mm} Past Login Days  & 1.000 (2.386)  & 0.999 (2.371)    & 0.501  \\
\hspace{2mm} Past Orders      & 1.000 (3.304)  & 1.003 (3.062)    & 0.157 \\
\hspace{2mm} Past Sales         & 1.000 (6.115)  & 1.004 (5.278)    & 0.157  \\ 
\hspace{2mm} N. of Consumers              & 6,869,558  & 6,845,970 &        \\ \hline 
\textbf{Google Advertising Title}      &&& \\
\hspace{2mm} Past Sales         & 1.000 (2.261)  & 0.993 (2.238)    & 0.084       \\
\hspace{2mm} Industry ID      & 1.000 (0.414)  & 1.000 (0.416)    & 0.712      \\ 
\hspace{2mm} N. of Products              & 621,133 & 622,883 &     \\ \hline\hline
\end{tabular}
\begin{tablenotes}   
\scriptsize  
\item[1] Mean (Std. Dev.) are shown with all values normalized by the corresponding variable's control group mean. For the definitions of each variable, refer to the notes in Figure \ref{fig:pvalues}. 
\item[2] The first four experiments are conducted at the consumer level and thus the unit of observation is the consumer. The Google Advertising Title experiment is conducted at the product level and thus the unit of observation is the product. 
\end{tablenotes}
\end{threeparttable} 
\end{table}

\pagebreak
\section{Main Results: Model, Estimation, and Additional Outcomes}
\label{appendix:detailed main results}

\setcounter{figure}{0}
\renewcommand{\thefigure}{C\arabic{figure}}
\renewcommand{\theHfigure}{C\arabic{figure}} 
\setcounter{table}{0}
\renewcommand{\thetable}{C\arabic{table}}
\renewcommand{\theHtable}{C\arabic{table}}

\paragraph{Pre-sale Service Chatbot} Pre-sale inquiries regarding product and seller information (e.g., product attributes, promotions, and logistics) play a critical role in shaping consumer purchase decisions. To reduce search costs, support decision-making, improve customer service, and enhance the overall consumer experience, the platform introduced a GenAI-powered chatbot capable of delivering accurate, content-rich responses tailored to a diverse consumer base and available around the clock. \\

We conduct our analysis using the following regression model:
\begin{align}\label{eq:reg-chatbot}
     y_{i} &= \beta \times Treat_{i}  + \alpha_{c(i)} + \epsilon_{i}, 
\end{align}
where $i$ indicates the consumer, $y_{i}$ stands for a consumer’s outcome (e.g., conversion rate or sales), $Treat_{i}$ is an indicator for whether the consumer is assigned to the treatment group. Since consumers entered the experiments on different
days, we control for their entry-day cohort fixed effects using $\alpha_{c(i)}$. \\

In addition to the main experiment comparing an auto-response indicating service unavailability (``No Service") with a GenAI chatbot service (``GenAI Reply"), we also studied three supplementary experiments: (1) ``No Service" versus ``GenAI+Human Reply", where consumers initially interacted with a GenAI chatbot and unresolved issues were escalated to human agents; (2) ``Human Reply", where consumers were exclusively served by human agents, versus ``GenAI Reply"; (3) ``Human Reply" versus ``GenAI+Human Reply". \\

Table \ref{table:Results of Pre-sale Service Chatbot} presents the impact of the GenAI chatbot on sales, while Table \ref{table:IM of Pre-sale Service Chatbot} reports its effects on conversion rates and cart value. In both tables, Columns (1)–(2) report effects when the treatment is GenAI Reply, and Columns (3)–(4) report effects for GenAI+Human Reply. Within each set, Columns (1) and (3) use No Service as the control, while Columns (2) and (4) use Human Reply as the control. \\

Focusing on the comparison between the No Service control and the GenAI Reply treatment (Column 1), we continue to find sizable sales productivity improvements: sales increase by 16.3\% and conversion rate rises by 21.7\% (both significant at the 1\% level). Using Human Reply as the control (Column 2), the GenAI Reply treatment shows no statistically significant differences in either sales or conversion, suggesting that the GenAI chatbot matches the quality of human service but does not outperform it. When the No Service control is compared with the GenAI+Human Reply treatment (Column 3), the gains are even larger—sales improve by 25.0\% and conversion by 29.0\%—indicating complementarities between GenAI and human agents. By contrast, relative to the Human Reply control (Column 4), the GenAI+Human Reply treatment yields a marginally significant 11.5\% increase in sales, with no statistically significant change in conversion (4.8\%), implying that the hybrid approach can enhance revenue even when benchmarked against human agents. Finally, Table \ref{table:IM of Pre-sale Service Chatbot} shows that all estimated effects on cart value are statistically insignificant.

\begin{table}[!htpb]
\centering
\small
\caption{Effects of Pre-sale Service Chatbot on Sales Across Experiments} 
\label{table:Results of Pre-sale Service Chatbot}
\begin{threeparttable}
\hspace*{-0.4cm}
\begin{tabular}{lcccc}
\hline\hline
Treatment: & \multicolumn{1}{c}{GenAI Reply}             
           & \multicolumn{1}{c}{GenAI Reply}  
           & \multicolumn{1}{c}{GenAI+Human Reply}             
           & \multicolumn{1}{c}{GenAI+Human Reply}  \\
Control:   & \multicolumn{1}{c}{No Service}             
           & \multicolumn{1}{c}{Human Reply}  
           & \multicolumn{1}{c}{No Service}             
           & \multicolumn{1}{c}{Human Reply}  \\   
    & (1)& (2)& (3)& (4)         \\ \hline
    &    &    &    &    \\
Treat   
& 0.274***  
& 0.0701    
& 0.422***  
& 0.218*    \\

& (0.0970)  
& (0.0967)  
& (0.115)   
& (0.114)  \\
    &    &    &    &    \\
\%Change  
& 16.25\%  
& 3.71\%   
& 25.03\%  
& 11.53\%    \\
    &    &    &    &    \\
Observations  
& 44,614  
& 44,736   
& 30,345  
& 30,467 \\
R-squared     
& 0.000  
& 0.000  
& 0.000  
& 0.000  \\ 
\hline\hline
\end{tabular}

\begin{tablenotes}   
\scriptsize  
\item[1] ``Sales" represents the total expenditure on product orders. 
\item[2] Standard errors are in parentheses. \% Change is calculated by dividing the treatment effect by the control group mean. *** p$<$0.01, ** p$<$0.05, * p$<$0.1.
\end{tablenotes}
\end{threeparttable}
\end{table}

\begin{table}[!htpb]
\centering
\small
\caption{Conversion Rate and Cart Value in Pre-sale Service Chatbot Experiments} 
\label{table:IM of Pre-sale Service Chatbot}
\begin{threeparttable}
\hspace*{-0.4cm}
\begin{tabular}{lcccc}
\hline\hline
Treatment: & \multicolumn{1}{c}{GenAI Reply}             & \multicolumn{1}{c}{GenAI Reply}  & \multicolumn{1}{c}{GenAI+Human Reply}             & \multicolumn{1}{c}{GenAI+Human Reply}  \\
Control:     & \multicolumn{1}{c}{No Service}             & \multicolumn{1}{c}{Human Reply}  & \multicolumn{1}{c}{No Service}             & \multicolumn{1}{c}{Human Reply}  \\  
    & (1)& (2)& (3)& (4)        \\ \hline
    &    &  &    &    \\
\multicolumn{5}{l}{\textit{Extensive margin: Conversion Rate}} \\
Treat  & 0.0131*** & -0.000768 & 0.0175*** & 0.00358 \\
  & (0.00252) & (0.00259) & (0.00291) & (0.00304) \\
  \%Change & 21.70\% & -1.03\% & 28.99\% & 4.82\% \\
  Observations & 44,614 & 44,736 & 30,345 & 30,467 \\
  R-squared & 0.001 & 0.000 & 0.001 & 0.000 \\
  &&&& \\
\multicolumn{5}{l}{\textit{Intensive margin: Cart Value (conditional on purchase)}} \\
Treat  & -1.264 & 1.220 & -0.859 & 1.624 \\
  & (1.036) & (0.929) & (1.203) & (1.078) \\
\%Change & -4.52\% & 4.79\%  & -3.08\% & 6.38\%\\
Observations & 3,076 & 3,300 & 2,092 & 2,316  \\
R-squared & 0.000 & 0.001 & 0.000 & 0.001 \\  \hline\hline
\end{tabular}

\begin{tablenotes}   
\scriptsize  
\item[1] The first block reports the \textit{Conversion Rate} (binary indicator equal to 1 if consumer makes at least one order during the experimental period). The second block reports \textit{Cart Value}, expenditure per consumer conditional on making a purchase. 
\item[2] Standard errors are in parentheses. \% Change is calculated by dividing the treatment effect by the control group mean. *** p$<$0.01, ** p$<$0.05, * p$<$0.1.
\end{tablenotes}
\end{threeparttable}
\end{table}

\paragraph{Search Query Refinement} Consumers arrive at e-commerce platforms with diverse needs. The search engine serves as the primary channel to facilitate consumers' discovery of desired products, allowing them to express preferences through search queries. Our focal platform seeks to accurately decode the latent demands behind consumers' multilingual queries, translate the queries, and retrieve products that align with their underlying needs. The effectiveness of this process is crucial in determining match quality, which in turn impacts consumer purchase decisions and platform revenues. GenAI is well-positioned to improve the search algorithm’s capabilities in translating consumer queries based on semantic understanding and refinement. \\

We conduct our analysis using the following regression model:
\begin{align}\label{eq:reg-search}
     y_{i} &= \beta \times Treat_{i} + \alpha_{cl(i)} + \epsilon_{i}, 
\end{align}
where $i$ denotes a consumer, $y_{i}$ stands for a consumer’s outcome variables, $Treat_{i}$ is an indicator of a consumer's treatment status. Consumers are classified into various cohorts based on their language groups and their first day of entry in the experiment. As multiple sub-experiments were conducted across consumers in different languages at varying dates, we include entry-day-by-language cohort fixed effects, $\alpha_{cl(i)}$.  \\

The results are summarized in Table \ref{table:Results of search query refinement}. Column 1 indicates no significant differences in product views between the two groups. However, treatment group consumers generated 1.1\% more clicks (Column 2), spent 2.9\% more (Column 4), and were 1.2\% more likely to make a purchase (Column 5). We find no significant impact on cart value and order intensity (Columns 6 and 7). \\

We also explore further mechanism analysis of the Search Query Refinement application. In Table \ref{table:more exploration of search query refinement}, we find that the likelihood of a product click increases by 0.3\% ($p<0.01$; Column 1), consistent with improved search performance in combination with the documented purchase conversion increase in our main effect. Columns 2 and 3 further indicate that treated consumers view fewer products prior to clicking or purchasing, suggesting a reduction in search intensity. In addition, we observe a statistically significant 2.0\% increase in the click-through rate ($p<0.01$; Column 4), defined as the ratio of product clicks to product views, suggesting that consumers found the exposed products more appealing and chose to seek additional details after viewing the summarized search results. The click-to-order conversion rate (Column 5), defined as the ratio of orders to clicks, remains statistically insignificant. This pattern echoes the fact that query refinement influenced only the composition of products retrieved immediately after a query search, not the information displayed on product detail pages. Overall, GenAI-facilitated query refinement enhanced the search algorithm's ability to satisfy consumers' demand, resulting in more effective matching and improved consumer purchases.

\begin{table}[!htpb]
\caption{Main Effect of Search Query Refinement} 
\label{table:Results of search query refinement}
\centering
\footnotesize
\begin{threeparttable}
\resizebox{\columnwidth}{!}{
\begin{tabular}{lccccccc}
\hline\hline
        & (1)   & (2)    & (3)    & (4)    & (5)    & (6) &     (7)           \\
        &    &     &      &    & Conversion  & Cart &  Order  \\
        & Views    & Clicks   & Orders   & Sales  & Rate   & Value  &  Intensity   \\ \hline
        &         &           &           &          &            &               &    \\
Treat   & -0.549  & 0.0901*** & 0.00154   & 0.0648** & 0.00101**  & 0.370 & -0.00315   \\
        & (0.887) & (0.0245)  & (0.00107) & (0.0314) & (0.000411) & (0.334) & (0.00842)     \\
        &         &           &           &          &            &               &    \\
\%Change & -0.18\% & 1.10\%    & 0.94\%    & 2.93\%   & 1.15\%     & 1.47\% & -0.17\%       \\
         &     &     &    &     &     &     &    \\
Observations & 1,849,382 & 1,849,382 & 1,849,382  & 1,849,382 & 1,849,382  & 163,381 & 163,381 \\
R-squared & 0.038   & 0.041     & 0.019    & 0.004    & 0.030   & 0.010  & 0.004       \\ \hline\hline
\end{tabular}
}
\begin{tablenotes}   
\scriptsize  
\item[1] ``Views" stands for the number of product views in the summarized search results pages. ``Clicks" stands for the number of product clicks into product detail pages. ``Orders" is the number of product orders. ``Sales" represents the total expenditure on product orders. ``Conversion Rate" measures consumers' likelihood of making purchases. It is a binary indicator for purchase, which equals 1 if a consumer makes at least one order during the experimental period, and 0 otherwise. ``Cart Value" refers to the expenditure per consumer, conditional on making a purchase. ``Order Intensity" refers to the number of orders per consumer, conditional on making a purchase. 
\item[2] Standard errors are in parentheses. \% Change is calculated by dividing the treatment effect by the control group mean. *** p$<$0.01, ** p$<$0.05, * p$<$0.1.
\end{tablenotes}
\end{threeparttable} 
\end{table}

\begin{table}[!htpb]
\caption{More Exploration of Search Query Refinement} 
\label{table:more exploration of search query refinement}
\centering
\footnotesize
\begin{threeparttable}
\resizebox{\columnwidth}{!}{
\begin{tabular}{lccccc}
\hline\hline
         & (1)         & (2)             & (3)         & (4)             & (5)             \\
         &             & Views            & Views        & Click-          & Click-to-Order   \\
         &             & Conditional on  & Conditional & through         & Conversion       \\
         & Is Click    & Click           & Order       & Rate            &  Rate            \\ \hline
         &             &                 &             &                 &                  \\
Treat    & 0.00267***  & -1.728*       & -10.34*       & 0.000767***     & -0.000088         \\
         & (0.00057)   & (1.034)      & (5.362)     & (0.000079)      & (0.000182)        \\
         &             &                 &             &                 &                   \\
\%Change & 0.33\%   & -0.46\%     & -1.48\%   & 2.02\%        & -0.35\%  \\
         &     &     &     &    & \\
Observations & 1,849,382 & 1,508,873  & 163,381 & 1,849,382 & 1,508,873 \\
R-squared    & 0.013    & 0.039   & 0.069  & 0.003      & 0.014   \\ \hline\hline
\end{tabular}
}
\begin{tablenotes}   
\scriptsize  
\item[1] ``Is Click" measures consumers' likelihood of making a click. It is a binary indicator for click, which equals 1 if a consumer makes at least one click during the experimental period, and 0 otherwise. ``Views Conditional on Click" refers to the product views per consumer, conditional on making a click. ``Views Conditional on Order" refers to the product views per consumer, conditional on making a purchase. ``Click-through Rate" is the ratio of the number of product clicks to the number of product views, measuring the degree of conversion from views to clicks. ``Click-to-Order Conversion Rate" stands for the ratio of the number of product orders to the number of product clicks, measuring the degree of conversion from clicks to purchases. 
\item[2] Standard errors are in parentheses. \% Change is calculated by dividing the treatment effect by the control group mean. *** p$<$0.01, ** p$<$0.05, * p$<$0.1.
\end{tablenotes}
\end{threeparttable} 
\end{table}

\paragraph{Product Description} Well-crafted product descriptions are essential for informing consumers about product features, benefits, and uses, thereby reducing information asymmetry, facilitating consumer decision-making and driving platform sales. Despite its importance, our studied platform shows that nearly half of the self-sold products either lack a textual description or contain only a minimal description. GenAI's strengths in content recognition, comprehension, and generation can offer an effective solution to create comprehensive and structured product descriptions for a global audience. \\

We conduct our analysis using the following regression model:
\begin{align}\label{eq:reg-description}
     y_{i} &=\beta \times Treat_{i} + \alpha_{cl(i)}  + \epsilon_{i}, 
\end{align}
where $i$ denotes a consumer, $y_{i}$ stands for a consumer’s outcome variables, $Treat_{i}$ is an indicator for whether the consumer is assigned to the treatment group. Because multiple sub-experiments were implemented across language groups on different start dates, we include entry-day-by-language cohort fixed effects, $\alpha_{cl(i)}$. \\

Column 1 of Table \ref{table:Results of product description} shows no statistically significant differences in product clicks between the control and treatment groups. However, conditional on consumers clicking through to product detail pages—where the descriptions are displayed—treated consumers place 1.1\% more orders and spend 2.1\% more on those orders (Columns 2 and 3). This improvement is also reflected in a 1.3\% increase in the conversion rate (Column 4), indicating a higher likelihood of purchase following exposure to AI-generated descriptions. By contrast, we detect no statistically significant changes in cart value or order intensity. Taken together, these results suggest that augmenting human-generated product descriptions with AI-generated content primarily affects purchase incidence rather than spending intensity, leading to higher overall sales. \\

Moreover, we stratify products based on the length of their original, human-generated descriptions, distinguishing between those with no or insufficient text (fewer than 50 words) and those with sufficiently detailed descriptions (more than 50 words). Products in the former group experience a 6.5\% increase in sales following augmentation with AI-generated content ($p<0.05$), whereas products in the latter group show no significant effect (Columns 1 and 2 of Table \ref{table:more exploration of product description}), indicating that richer descriptions facilitate consumer decision-making.\footnote{Based on internal research and expert surveys conducted by our partner platform, descriptions containing fewer than 50 words are classified as providing insufficient textual information. This analysis is restricted to the English-language sub-experiment, as description-length data are available only for English-language content.} \\

\begin{table}[!htpb]
\caption{Main Effect of Product Description} 
\label{table:Results of product description}
\centering
\small
\begin{threeparttable}
\begin{tabular}{lcccccc}
\hline\hline
             & (1)       & (2)        & (3)        & (4)             & (5)        & (6)  \\
             & Clicks     & Orders      & Sales      & Conversion Rate & Cart Value & Order Intensity \\ \hline
             &           &            &            &                 &         & \\
Treat        & 0.00233    & 0.000600**   & 0.0104**   & 0.000554***       & 0.0944   & -0.00186 \\
             & (0.00183)  & (0.000278)   & (0.00417)   & (0.000187)        & (0.0805)   & (0.00320)   \\
             &           &            &            &                 &     &  \\
\%Change     & 0.12\%    & 1.08\%     & 2.05\%     & 1.27\%          &  0.82\%  & -0.15\%   \\
             &           &            &            &                 &     &  \\
Observations & 4,772,937 & 4,772,937  & 4,772,937  & 4,772,937       & 210,155  & 210,155  \\
R-squared    & 0.055     & 0.008      & 0.002      & 0.008           & 0.010    & 0.021   \\ \hline\hline
\end{tabular}
\begin{tablenotes}   
\scriptsize  
\item[1] ``Clicks" stands for the number of product clicks into product detail pages. ``Orders" is the number of product orders. ``Sales" represents the total expenditure on product orders. ``Conversion Rate" measures consumers' likelihood of making purchases. It is a binary indicator for purchase, which equals 1 if a consumer makes at least one order during the experimental period, and 0 otherwise. ``Cart Value" refers to the expenditure per consumer, conditional on making a purchase. ``Order Intensity" refers to the number of orders per consumer, conditional on making a purchase. 
\item[2] Standard errors are in parentheses. \% Change is calculated by dividing the treatment effect by the control group mean. *** p$<$0.01, ** p$<$0.05, * p$<$0.1. 
\end{tablenotes}
\end{threeparttable} 
\end{table}

\begin{table}[!htpb]
\caption{Product Description: Heterogeneity by Length of Original, Human-generated Description} 
\label{table:more exploration of product description}
\centering
\small
\begin{threeparttable}
\begin{tabular}{lcccccc}
\hline\hline
             & (1)       & (2)        & (3)        & (4)   & (5) & (6)          \\
             & Sales     & Sales      & Conversion Rate      & Conversion Rate & Cart Value & Cart Value \\ 
             &$\leq$50 words   &$>$50 words        &$\leq$50 words     & $>$50 words &$\leq$50 words     &$>$50 words \\ \hline
             &           &            &            &   &            &              \\
Treat        & 0.0115**    & 0.0001   & 0.00047**   & 0.00044* & 0.367 & -0.188    \\
             & (0.0050)  & (0.0064)   & (0.00022)   & (0.00026) & (0.235)  & (0.198)       \\
             &           &            &            &  &            &               \\
\%Change     & 6.50\%    & 0.03\%     & 2.43\%     & 1.60\%  & 3.99\% & -1.50\%        \\
P-value      & 0.0536    &            & 0.5490     &         & 0.0653 &    \\
             &           &            &            &         &            &        \\
Observations & 1,628,409 & 1,628,409  & 1,628,409  & 1,628,409  & 31,759 & 45,500     \\
R-squared    & 0.000     & 0.001      & 0.003      & 0.005   & 0.001  & 0.004        \\ \hline\hline
\end{tabular}
\begin{tablenotes}   
\scriptsize  
\item[1] ``Sales" represents the total expenditure on product orders. ``Conversion Rate" measures consumers' likelihood of making purchases. It is a binary indicator for purchase, which equals 1 if a consumer makes at least one order during the experimental period, and 0 otherwise. ``Cart Value" refers to the expenditure per consumer, conditional on making a purchase. 
\item[2] ``$\leq$50 words" refers to descriptions containing fewer than 50 words, while ``$>$50 words" refers to descriptions containing more than 50 words.
\item[3] Standard errors are in parentheses. \% Change is calculated by dividing the treatment effect by the control group mean. The p-values are from seemingly unrelated estimations testing the equality of \% treatment effects across the two groups. *** p$<$0.01, ** p$<$0.05, * p$<$0.1. 
\end{tablenotes}
\end{threeparttable} 
\end{table}

\paragraph{Marketing Push Message} The platform leverages marketing push notifications, direct messages sent to consumers via their platform app, to draw traffic and boost transactions. Because creating a large volume of diverse and targeted marketing messages manually was challenging, the number of marketing messages is far smaller than the hundreds of millions of consumers, often resulting in many consumers receiving identical content. With the introduction of GenAI, the platform can produce millions of distinct messages, enabling highly personalized marketing strategies through push notifications. \\ 

We conduct our analysis using the following regression model:
\begin{align}\label{eq:reg2}
y_i= \alpha + \beta\times Treat_i+ \epsilon_i
\end{align}
where $i$ denotes a consumer, $y_{i}$ stands for a consumer’s outcome variables, $Treat_{i}$ is an indicator for whether the consumer is assigned to the treatment group. \\

Table \ref{table:Results of marketing push} shows that the use of AI-generated marketing messages leads to increases in consumer engagement. Specifically, clicks increase by 3.1\%, orders by 2.8\%, and the point estimate for total purchase amount increases by 1.6\%, though it is not statistically significant (Columns 1--3). In addition, the probability that a consumer makes a purchase rises by 3\%, while cart value and order intensity remain statistically unchanged (Columns 4--6). Taken together, these results suggest that GenAI enhances the effectiveness of marketing content primarily by expanding consumer participation, consistent with its ability to unlock personalization at scale in settings where human-generated content is constrained by limited resources.

\begin{table}[htpb]
\centering
\small
\caption{Main Effect of Marketing Push} 
\label{table:Results of marketing push}
\begin{threeparttable}
\begin{tabular}{lccccccc}
\hline\hline
             & (1)         & (2)        & (3)     & (4)  & (5)     & (6)           \\
             &       &       &     &  Conversion     & Cart   & Order         \\
             & Clicks       & Orders      & Sales     &  Rate    &  Value  &  Intensity   \\ \hline
             &             &                &            &            &    &         \\
Treat   & 0.000529***  & 0.000052*       & 0.000402   & 0.000048** &  -0.206  &-0.00138    \\
             & (0.000071)  & (0.000028)      & (0.000816) &    (0.0000218) & (0.454)  & (0.00854)   \\
             &             &                &            &            &     &        \\
\%Change     & 3.08\%       & 2.83\%          & 1.60\%      & 2.95\%      &    -1.32\%  &   -0.12\%         \\
             &             &                &            &            &     &        \\
Observations & 13,715,528  & 13,715,528     & 13,715,528 & 13,715,528 &  22,425  &22,425   \\
R-squared    & 0.000       & 0.000          & 0.000      & 0.000      &  0.000   &0.000   \\ \hline\hline
\end{tabular}
\begin{tablenotes}   
\scriptsize  
\item[1] ``Clicks" stands for the number of clicks on the marketing messages. ``Orders" is the number of product orders. ``Sales" represents the total expenditure on product orders. ``Conversion Rate" measures consumers' likelihood of making purchases. It is a binary indicator for purchase, which equals 1 if a consumer makes at least one order during the experimental period, and 0 otherwise. ``Cart Value" refers to the expenditure per consumer, conditional on making a purchase. ``Order Intensity" refers to the number of orders per consumer, conditional on making a purchase. 
\item[2] Standard errors are in parentheses. \% Change is calculated by dividing the treatment effect by the control group mean. *** p$<$0.01, ** p$<$0.05, * p$<$0.1. 
\end{tablenotes}
\end{threeparttable}
\end{table}

\paragraph{Google Advertising Title} The platform buys advertising slots in the sponsored section of Google Shopping to promote its products on Google and attract traffic to its site. To maximize purchases derived from Google advertisements, a critical operational decision for the platform is how to design product titles to increase both product discoverability and the likelihood of user clicks. The platform leveraged GenAI to refine titles based on the original seller-created titles, optimizing them for better visibility and engagement. \\

We conduct our analysis employing the following regression model:
\begin{align}\label{eq:reg-google-ads}
     y_{i} &= \beta \times Treat_{i}  + \alpha_{c(i)} + \epsilon_{i},
\end{align}
where $i$ denotes a product, $y_{i}$ stands for the outcome variables for a product, $Treat_{i}$ is an indicator for whether the product is assigned to the treatment group. We control for product entry-day cohort fixed effect $\alpha_{c(i)}$. \\

Table \ref{table:Results of google advertising} presents the main findings. Columns 1 and 2 indicate a 7.7\% decrease in ad views and a 10.3\% decrease in ad clicks for the treatment group, respectively. Columns 3, 4, and 5 report a non-significant reduction in sales, conversion rate, and cart value, respectively. As we discussed in Section \ref{subsection: productivity impact by workflow}, the null effect on sales can be attributed to the lack of fine-tuning using e-commerce domain knowledge when setting up the GenAI model.

\begin{table}[htpb]
\centering
\small 
\caption{Main Effect of Google Advertising} 
\label{table:Results of google advertising}
\begin{threeparttable}
\begin{tabular}{lccccc}
\hline\hline
             & (1)         & (2)         & (3)         & (4)         & (5)    \\
             & Views        & Clicks       & Sales       & Conversion Rate& Cart Value         \\ \hline
             &             &             &             &           &    \\
    Treat & -1.547*** & -0.0247***  & -0.00602 & -0.000090   &  -0.784\\
          & (0.148) & (0.00304) & (0.00534) & (0.000111)   & (0.992)\\
          &       &       &       &  \\
\%Change     & -7.68\%      & -10.26\%        & -4.56\%    & -2.29\%    &  -2.32\% \\
             &             &             &             &          &     \\
    Observations & 1,244,016 & 1,244,016 & 1,244,016 & 1,244,016  &  4,811 \\
    R-squared & 0.000 & 0.000 & 0.000 & 0.000   & 0.000  \\ \hline\hline
\end{tabular}
\begin{tablenotes}   
\scriptsize  
\item[1] ``Views" represents the number of times the product advertisement is viewed on Google. ``Clicks" refers to the number of times the product advertisement is clicked on Google. ``Sales" represents the total expenditure on product orders. ``Conversion Rate" is 1 if a product is purchased at least once during the experiment period, and 0 otherwise. ``Cart Value" refers to the expenditure per product, conditional on the product being purchased. 
\item[2] Standard errors are in parentheses. \% Change is calculated by dividing the treatment effect by the control group mean. *** p$<$0.01, ** p$<$0.05, * p$<$0.1. 
\end{tablenotes}
\end{threeparttable} 
\end{table}

\paragraph{Chargeback Defense} Online sellers often struggle to defend against chargebacks due to reasons such as non-receipt of goods. Chargebacks can lead to significant financial losses and jeopardize the long-term sustainability of sellers' businesses on the platform. The whole process of contesting chargeback disputes includes analyzing claims, collecting necessary evidence (e.g., order details, fulfillment records, proof of shipment and logistics tracking through ways like interfacing with diverse APIs), and crafting compelling chargeback defense letters. The rapid advancement of GenAI enabled the platform to develop a chargeback defense agent that offers a one-stop solution to streamline the intricacies of chargeback disputes. \\

As the data was not available for us to review,  we report findings estimated by the platform's internal data science team. Their estimates indicate that the adoption of the GenAI agent helps increase the success rate of chargeback defense by 15\%.

\paragraph{Live Chat Translation} E-commerce platforms must provide robust consumer services for consumers seeking consultation or negotiation with the platform, such as addressing inquiries about the platform's promotional details and resolving disputes when consensus with sellers is not reached. For our focal platform, delivering native-language consumer services to a diverse, multilingual consumer base incurs significant costs. Thus, a significant portion of non-English consumer inquiries were supported by Filipino consumer service agents in English. A straightforward application of GenAI allows the platform to equip low-cost Filipino agents with robust real-time translation support, aiming to provide more effective communication between consumers and agents. \\

Similarly, due to the unavailability of raw data, the effect estimate for this process is from the platform's internal data science team and could not be independently verified. During the experiment, consumers were queried whether they were satisfied or not with the service via a survey question immediately after service completion. As a result, there was a 5.2\% increase in consumer satisfaction, suggesting that GenAI helped Filipino agents to better serve non-English-speaking consumers.

\paragraph{Post-purchase Outcomes} In Table \ref{table:main returns and reviews}, we provide evidence on two important post-purchase measures of consumer satisfaction with the products, namely product return rates and customer ratings. We do not find evidence that GenAI applications deteriorate such outcomes.

\begin{table}[h!]
\centering
\footnotesize
\caption{Average Treatment Effects of GenAI Adoption on Product Returns and Reviews}
\label{table:main returns and reviews}
\begin{threeparttable}
\begin{tabular}{m{4.5cm} >{\raggedright}m{1.6cm} >{\raggedright}m{1.4cm} >{\raggedright}m{1.3cm} >{\raggedright}m{0.1em} >{\raggedright}m{1.6cm} >{\raggedright}m{1.4cm} m{1.3cm}}
\hline\hline
& \multicolumn{3}{l}{Product Return Rate} & & \multicolumn{3}{l}{Positive Review Rate}  \\\cline{2-4}\cline{6-8}
& (1) & (2) & (3) && (4) & (5) & (6)   \\
Business Workflow & Coefficient  & \% Change & Obs && Coefficient  & \% Change & Obs \\ 
\hline
\textbf{Pre-sale Service Chatbot} \newline &    0.0208     \newline (0.0582)  &     2.08\%     \newline &    3,076      \newline &&        0.0452*  \newline (0.0247) &      4.52\%  \newline & 1,008        \newline \\
&   &   &   &   &  &  & \\
\textbf{Search Query Refinement} \newline &-0.0141 \newline (0.0152) & -1.41\% \newline &  163,381 \newline && 0.0031 \newline (0.0040) &  0.31\% \newline &  26,209 \newline \\
&   &   &   &   &  &  & \\
\textbf{Product Description} \newline &-0.0386*** \newline (0.0142) & -3.86\% \newline &  210,155 \newline && -0.0042 \newline (0.0034) & -0.42\% \newline &  31,916 \newline \\
&   &   &   &   &  &  & \\
\textbf{Marketing Push Message} \newline &    -0.114***     \newline (0.0394) &        -11.40\% \newline &       22,425   \newline && -0.0174   \newline (0.0122) &          -1.74\% \newline &     2,853   \newline \\
\hline\hline
\end{tabular}
\begin{tablenotes}   
\scriptsize
\item[1] ``Return Rate" is defined as the share of orders that are returned within a year after the purchase. ``Positive Review Rate" is defined as the share of rated orders that receive four- or five-star ratings on a five-point scale. The estimation of Return Rate is conditional on consumers with orders, while the estimation of Positive Review Rate is conditional on consumers with rated orders. Such data are not available for the Google Advertising Title workflow. To protect proprietary business information, both outcome variables are normalized by the control-group mean. Consequently, coefficients can be interpreted as percentage changes relative to the control condition.
\item[2] Columns (1) and (4) report the estimated coefficients, with standard errors in parentheses. Columns (2) and (5) report \% Change, calculated as the treatment effect divided by the control group mean. Columns (3) and (6) report the number of observations. *** p$<$0.01, ** p$<$0.05, * p$<$0.1.
\end{tablenotes}
\end{threeparttable} 
\end{table}

\pagebreak 
\section{Details on Consumer Treatment Effect Heterogeneity}
\label{appendix:detailed_hte_customer}
\setcounter{figure}{0}
\renewcommand{\thefigure}{D\arabic{figure}}
\renewcommand{\theHfigure}{D\arabic{figure}} 
\setcounter{table}{0}
\renewcommand{\thetable}{D\arabic{table}}
\renewcommand{\theHtable}{D\arabic{table}} 

\paragraph{Pre-sale Service Chatbot} In this experiment, we focus on the comparison between consumers who are served only by the GenAI chatbot and those who are served by an auto-response indicating service unavailability. Table \ref{table:customer_hte_pre_sale_service_chatbot} reports the heterogeneous treatment effects across consumer groups. Our analysis reveals that the gains in sales are more pronounced among newer consumers with shorter registration histories (Column 2), less active consumers with fewer login days (Column 4), and consumers with less past spending (Column 6). However, differences across consumer groups are generally not statistically significant, with the exception of sales across past purchases, for which the difference is significant at the 10\% level.

\paragraph{Search Query Refinement} Table \ref{table:customer_hte_search_query_refinement} reports the heterogeneous treatment effect of refining search query with GenAI on different consumer groups. For sales, the benefits are significant and larger for inexperienced consumers (Columns 2, 4, and 6), while the effects for more experienced consumers are smaller and insignificant (Columns 1, 3, and 5). The differences between these two consumer groups are statistically significant when consumers are classified by past purchases and past login days. When focusing on conversion rates, we also find statistically significant differences between consumer groups defined by years since registration. Since consumers with limited online experience often struggle to effectively articulate their needs through query-based searches, they tend to benefit more from enhanced match quality achieved by applying GenAI to the semantic translation of consumer queries.

\begin{table}
\caption{Consumer HTE for Pre-sale Service Chatbot} 
\label{table:customer_hte_pre_sale_service_chatbot}
\centering
\scriptsize
\begin{threeparttable}
\begin{tabular}{m{2cm} >{\raggedright}m{1.5cm} >{\raggedright}m{1.8cm} >{\raggedright}m{0.1em} >{\raggedright}m{1.5cm} >{\raggedright}m{1.5cm} >{\raggedright}m{0.1em} >{\raggedright}m{1.5cm} m{1.5cm}}
\hline\hline
~ & (1) & (2) && (3) & (4) && (5) & (6)  \\\cline{2-3}\cline{5-6}\cline{8-9}
~ & \multicolumn{2}{l}{Registered Years} && \multicolumn{2}{l}{Past Login Days} && \multicolumn{2}{l}{Past Purchases}   \\ 
~ & High & Low && High & Low && High & Low  \\ \hline
\multicolumn{2}{l}{\textit{Panel A: Sales}} & ~ & ~ & ~ & ~ & & &   \\
Treat    & 0.217 & 0.329*** &  & 0.318** & 0.230** &  & 0.159   & 0.343*** \\
         & (0.153) & (0.118) &  & (0.156) & (0.114) &  & (0.180) & (0.105)  \\
\%Change & 10.38\%  & 25.76\%  && 15.00\%  & 18.54\%  && 6.82\%   & 27.79\% \\
P-value &  0.1683   &&&  0.7520   &&&   0.0606  &  \\ 
~ & ~ & ~ & ~ & ~ & ~ & & &  \\ 
\multicolumn{2}{l}{\textit{Panel B: Conversion Rate}} & ~ & ~ & ~ & ~ & & &  \\
Treat    & 0.0131*** & 0.0131*** &  & 0.0128*** & 0.0135*** &  & 0.0128*** & 0.0130*** \\
         & (0.00377) & (0.00333) &  & (0.00386) & (0.00321) &  & (0.00434) & (0.00300) \\
\%Change & 18.77\% & 25.77\% && 17.11\% & 29.62\% && 16.41\% & 27.09\%   \\
P-value &   0.4000  &&&  0.1212   &&&   0.1955  &  \\ 
~ & ~ & ~ & ~ & ~ & ~ & & &  \\ 
\multicolumn{2}{l}{\textit{Panel C: Cart Value}} & ~ & ~ & ~ & ~ & & &  \\
Treat         & -2.130 & 0.0141 && -0.516 & -2.331 && -2.470  & 0.157 \\
 & (1.432) & (1.471) && (1.387) & (1.525) && (1.568) & (1.320) \\
\% Change    & -7.11\% & 0.06\% && -1.82\% & -8.56\% && -8.27\% & 0.61\%  \\
P-value & 0.3257    &&&  0.3702   &&& 0.2170    &  \\ 
~ & ~ & ~ & ~ & ~ & ~ & & &  \\ 
Observation (Panel A\&B) &  22,461  &22,153  &&22,570  &22,044  &&18,610  &26,004    \\ 
Observation (Panel C) & 1,761   &1,315  && 1,877 & 1,199 && 1,609 &  1,467  \\ 
\hline\hline
\end{tabular}
\begin{tablenotes}   
\scriptsize  
\item[1] We classify consumers into high and low groups and capture their online shopping experience based on 
registration duration, 30-day pre-experiment login days and 30-day pre-experiment purchase. For each indicator, consumers in the low group are defined as relatively inexperienced if they fall below the median of the corresponding distribution.
\item[2] ``Sales" represents the total expenditure on product orders. ``Conversion Rate" measures consumers' likelihood of making purchases. It is a binary indicator for purchase, which equals 1 if a consumer makes at least one order during the experimental period, and 0 otherwise. ``Cart Value" refers to the expenditure per consumer, conditional on making a purchase.
\item[3] Standard errors are in parentheses. \% Change is calculated by dividing the treatment effect by the control group mean. *** p$<$0.01, ** p$<$0.05, * p$<$0.1.
\end{tablenotes}
\end{threeparttable} 
\end{table}

\begin{table}
\caption{Consumer HTE for Search Query Refinement} 
\label{table:customer_hte_search_query_refinement}
\centering
\scriptsize
\begin{threeparttable}
\begin{tabular}{m{2cm} >{\raggedright}m{1.5cm} >{\raggedright}m{1.8cm} >{\raggedright}m{0.1em} >{\raggedright}m{1.5cm} >{\raggedright}m{1.5cm} >{\raggedright}m{0.1em} >{\raggedright}m{1.5cm} m{1.5cm}}
\hline\hline
~ & (1) & (2) && (3) & (4) && (5) & (6)  \\\cline{2-3}\cline{5-6}\cline{8-9}
~ & \multicolumn{2}{l}{Registered Years} && \multicolumn{2}{l}{Past Login Days} && \multicolumn{2}{l}{Past Purchases}   \\ 
~ & High & Low && High & Low && High & Low  \\ \hline
\multicolumn{2}{l}{\textit{Panel A: Sales}} & ~ & ~ & ~ & ~ & & &   \\
Treat    &  0.0360  & 0.0867** && 0.0203   & 0.104*** && -0.0216  & 0.106***   \\ 
         & (0.0526) & (0.0380) && (0.0566) & (0.0303) && (0.0757) & (0.0291)   \\ 
\%Change &  1.27\%  & 5.01\%   && 0.63\%   & 8.16\%   && -0.56\%   & 7.46\%   \\ 
P-value & 0.2137 &&& 0.0036 &&& 0.0047 &  \\ 
~ & ~ & ~ & ~ & ~ & ~ & & &  \\ 
\multicolumn{2}{l}{\textit{Panel B: Conversion Rate}} & ~ & ~ & ~ & ~ & & &  \\
Treat    &  -0.000104   & 0.00186*** && 0.000500   & 0.00141***  && 0.000651   & 0.00115***   \\ 
         & (0.000697) & (0.000487) && (0.000675) & (0.000482) && (0.000907) & (0.000421)   \\
\%Change & -0.09\%     & 2.79\%     && 0.43\%     & 2.32\%     && 0.44\%     & 1.93\%       \\ 
P-value & 0.0023 &&& 0.0421 &&& 0.1122 &  \\ 
~ & ~ & ~ & ~ & ~ & ~ & & &  \\ 
\multicolumn{2}{l}{\textit{Panel C: Cart Value}} & ~ & ~ & ~ & ~ & & &  \\
Treat &  0.329 & 0.440 && -0.0145 & 1.126** && -0.355 & 1.224***   \\ 
~ & (0.431) & (0.526) && (0.452) & (0.460) && (0.484) & (0.450)   \\
\%Change & 1.34\% & 1.69\%  && -0.05\% & 5.36\% && -1.35\% & 5.13\%  \\ 
P-value & 0.8965 &&& 0.0363 &&& 0.0139 &  \\ 
~ & ~ & ~ & ~ & ~ & ~ & & &  \\ 
Observation (Panel A\&B) & 813,317 & 1,036,065 && 890,044 & 959,338 && 596,721 & 1,252,661  \\
Observation (Panel C) & 93,362 & 70,019 && 104,531 & 58,850 && 88,135 & 75,246  \\\hline\hline
\end{tabular}
\begin{tablenotes}   
\scriptsize  
\item[1] Refer to Table \ref{table:customer_hte_pre_sale_service_chatbot} for detailed notes.
\end{tablenotes}
\end{threeparttable} 
\end{table}

\paragraph{Product Description} In Table \ref{table:customer_hte_product_description}, we observe a similar pattern to Search Query Refinement. There is a significant and more pronounced sales increase for inexperienced consumers (Columns 2, 4, and 6). The differences between high- and low-group consumers are statistically significant for consumers classified by past purchases and past login days. Augmenting product descriptions with AI-generated content substantially enhances the sufficiency and clarity of product information, thereby motivating consumer purchase decisions, particularly among less experienced consumers.

\begin{table}
\caption{Consumer HTE for Product Description} 
\label{table:customer_hte_product_description}
\centering
\scriptsize
\begin{threeparttable}
\begin{tabular}{m{2cm} >{\raggedright}m{1.5cm} >{\raggedright}m{1.8cm} >{\raggedright}m{0.1em} >{\raggedright}m{1.5cm} >{\raggedright}m{1.5cm} >{\raggedright}m{0.1em} >{\raggedright}m{1.5cm} m{1.5cm}}
\hline\hline
~ & (1) & (2) && (3) & (4) && (5) & (6)  \\\cline{2-3}\cline{5-6}\cline{8-9}
~ & \multicolumn{2}{l}{Registered Years} && \multicolumn{2}{l}{Past Login Days} && \multicolumn{2}{l}{Past Purchases}   \\ 
~ & High & Low && High & Low && High & Low  \\ \hline
\multicolumn{2}{l}{\textit{Panel A: Sales}} & ~ & ~ & ~ & ~ & & &   \\
Treat    &  0.00671  & 0.0130*** && -0.00606   & 0.0263*** && 0.000224  & 0.0206***   \\ 
         & (0.00761) & (0.00455) && (0.00655) & (0.00522) && (0.00718) & (0.00423)   \\ 
\%Change &  1.09\%  & 3.06\%   && -1.02\%   & 6.24\%   && 0.03\%   & 5.63\%   \\ 
P-value & 0.2525 &&& 0.0001 &&& 0.0005 &  \\ 
~ & ~ & ~ & ~ & ~ & ~ & & &  \\ 
\multicolumn{2}{l}{\textit{Panel B: Conversion Rate}} & ~ & ~ & ~ & ~ & & &  \\
Treat    &  0.000575*   & 0.000520** && 0.000325   & 0.000785***  && 0.000568*   & 0.000533**   \\ 
         & (0.000312) & (0.000229) && (0.000284) & (0.000245) && (0.000292) & (0.000233)   \\
\%Change & 1.10\%     & 1.39\%     && 0.65\%     & 2.06\%     && 1.05\%     & 1.59\%       \\ 
P-value & 0.7441 &&& 0.1194 &&& 0.5235 &  \\ 
~ & ~ & ~ & ~ & ~ & ~ & & &  \\ 
\multicolumn{2}{l}{\textit{Panel C: Cart Value}} & ~ & ~ & ~ & ~ & & &  \\
Treat &  0.003 & 0.184* && -0.180 & 0.438*** && -0.127 & 0.447***   \\ 
~ & (0.126) & (0.098) && (0.112) & (0.115) && (0.115) & (0.098)   \\
\%Change & 0.03\% & 1.62\%  && -1.50\% & 3.96\% && -1.06\% & 4.09\%  \\ 
P-value & 0.2496 &&& 0.0001 &&& 0.0001 &  \\ 
~ & ~ & ~ & ~ & ~ & ~ & & &  \\ 
Observation (Panel A\&B) & 2,035,278 & 2,737,659 && 2,322,437 & 2,450,500 && 2,386,336 & 2,386,601  \\ 
Observation (Panel C) & 107,098 & 103,057 && 115,708 & 94,447 && 129,462 & 80,693  \\\hline\hline
\end{tabular}
\begin{tablenotes}   
\scriptsize  
\item[1] Refer to Table \ref{table:customer_hte_pre_sale_service_chatbot} for detailed notes.
\end{tablenotes}
\end{threeparttable} 
\end{table}

\paragraph{Marketing Push Message} In Table \ref{table:customer_hte_marketing_push_message}, we find that the effects of AI-generated marketing push messages are concentrated among consumers with fewer past purchases. For this group, treatment significantly increases both sales and conversion rates (Column 6), while the corresponding estimates for high-purchase consumers are negative and statistically insignificant (Column 5); the high-low differences are statistically significant. For registered years and past login days, conversion-rate gains are larger for lower-experience consumers, but the differences across groups are not statistically significant. These patterns suggest that AI-generated push messages are particularly effective for consumers with less purchase history, likely by making promotional content more relevant or salient.

\begin{table}
\caption{Consumer HTE for Marketing Push Message} 
\label{table:customer_hte_marketing_push_message}
\centering
\scriptsize
\begin{threeparttable}
\begin{tabular}{m{2cm} >{\raggedright}m{1.5cm} >{\raggedright}m{1.8cm} >{\raggedright}m{0.1em} >{\raggedright}m{1.5cm} >{\raggedright}m{1.5cm} >{\raggedright}m{0.1em} >{\raggedright}m{1.5cm} m{1.5cm}}
\hline\hline
~ & (1) & (2) && (3) & (4) && (5) & (6)  \\\cline{2-3}\cline{5-6}\cline{8-9}
~ & \multicolumn{2}{l}{Registered Years} && \multicolumn{2}{l}{Past Login Days} && \multicolumn{2}{l}{Past Purchases}   \\ 
~ & High & Low && High & Low && High & Low  \\ \hline
\multicolumn{2}{l}{\textit{Panel A: Sales}} & ~ & ~ & ~ & ~ & & &   \\
Treat        & -0.000269  & 0.00128   &  & 0.000694   & 0.000195 &  & -0.00250  & 0.00238*** \\
             & (0.00115) & (0.00112) &  & (0.00136) & (0.00100) &  & (0.00161) & (0.000712) \\
\% Change    & -0.91\%     & 6.63\%     &  & 2.47\%      & 0.85\%    &  & -5.88\%    & 20.64\%     \\
P-value &  0.4519   &&&  0.8605   &&&   0.0003  &  \\ 
~ & ~ & ~ & ~ & ~ & ~ & & &  \\ 
\multicolumn{2}{l}{\textit{Panel B: Conversion Rate}} & ~ & ~ & ~ & ~ & & &  \\
Treat        & 0.0000314   & 0.0000681** &  & 0.0000122 & 0.0000737***  &  & -0.0000590  & 0.000111*** \\
             & (0.0000301) & (0.0000312) &  & (0.0000345)  & (0.0000281) &  & (0.0000416) & (0.0000210) \\
\% Change    & 1.76\%       & 4.98\%       &  & 0.71\%        & 4.82\%      &  & -2.22\%      & 14.17\%      \\
P-value &   0.4372  &&&  0.2810   &&&  0.0000   &  \\ 
             &             &             &  &              &             &  &             &             \\
   \\
\multicolumn{2}{l}{\textit{Panel C: Cart Value}} & ~ & ~ & ~ & ~ & & &  \\
Treat         & -0.432 & 0.225&&0.285  &-0.573  && -0.600 & 0.828\\
 & (0.578) & (0.731) && (0.714) & (0.583) && (0.558) & (0.761) \\
\% Change    & -2.62\% & 1.60\% && 1.75\% & -3.80\% && -3.76\% & 5.62\%  \\
P-value &   0.4956  &&&  0.3551   &&&  0.1286   &  \\ 
~ &  & ~ & ~ & ~ & ~ &  & &  \\
Observations (Panel A\&B) & 7,959,851   &5,755,677  &  & 5,796,995   & 7,918,533
   &  & 6,085,076   & 7,630,452 \\ 
Observations (Panel C) & 14,351   &8,074  &  & 10,026   & 12,399
   &  & 16,025   & 6,400 \\
\hline\hline
\end{tabular}
\begin{tablenotes}   
\scriptsize  
\item[1] Refer to Table \ref{table:customer_hte_pre_sale_service_chatbot} for detailed notes.
\end{tablenotes}
\end{threeparttable} 
\end{table}

\section{Details on Seller Treatment Effect Heterogeneity}
\label{appendix:detailed_hte_seller}
\setcounter{figure}{0}
\renewcommand{\thefigure}{E\arabic{figure}}
\renewcommand{\theHfigure}{E\arabic{figure}} 
\setcounter{table}{0}
\renewcommand{\thetable}{E\arabic{table}}
\renewcommand{\theHtable}{E\arabic{table}} 

\paragraph{Search Query Refinement} Table \ref{table:seller_hte_search_query_refinement} reports the heterogeneous impact on high- versus low-type sellers classified based on three metrics: annual past sales, tenure on the platform, and number of sub-accounts. Small sellers with lower transaction volumes (Column 2), shorter operational histories (Column 4), and fewer sub-accounts (Column 6) experience a significant and larger increase in sales from treated consumers. By contrast, sales on larger, tenured, and scaled sellers show no significant change. The differences between these two seller groups are statistically significant when sellers are divided by operation years and the number of sub-accounts. When using conversion rates, we also find statistically significant differences between seller groups defined by annual sales. Thus, the GenAI-powered search query refinement generates greater value for smaller sellers.

\paragraph{Marketing Push Message} In Table \ref{table:seller_hte_marketing_push_message}, we find limited evidence of heterogeneity across seller groups. Most high-low differences are statistically insignificant across sales, conversion rates, and cart values. The main exception is conversion rates by the number of sub-accounts: sellers with fewer sub-accounts experience a larger conversion-rate gain, and the high-low difference is marginally significant at the 10\% level.

\paragraph{Google Advertising Title} Table \ref{table:seller_hte_google_advertising_title} shows that the differences between high- and low-group sellers are all statistically insignificant.

\begin{table}
\caption{Seller HTE for Search Query Refinement} 
\label{table:seller_hte_search_query_refinement}
\centering
\scriptsize
\begin{threeparttable}
\begin{tabular}{m{2cm} >{\raggedright}m{1.5cm} >{\raggedright}m{1.8cm} >{\raggedright}m{0.1em} >{\raggedright}m{1.5cm} >{\raggedright}m{1.5cm} >{\raggedright}m{0.1em} >{\raggedright}m{1.5cm} m{1.5cm}}
\hline\hline
~ & (1) & (2) && (3) & (4) && (5) & (6)  \\\cline{2-3}\cline{5-6}\cline{8-9}
~ & \multicolumn{2}{l}{Annual Sales} && \multicolumn{2}{l}{Operation Years} && \multicolumn{2}{l}{\# of Sub-Accounts}   \\ 
~ & High & Low && High & Low && High & Low  \\ \hline
\multicolumn{2}{l}{\textit{Panel A: Sales}} & ~ & ~ & ~ & ~ & & &   \\
Treat &  0.0241 & 0.0407** && 0.0070 & 0.0578** && 0.00238 & 0.0624**   \\ 
~ & (0.0243) & (0.0180) && (0.0182) & (0.0237) && (0.0174) & (0.0246)   \\ 
\%Change &  2.18\% & 3.68\% && 0.62\% & 5.38\% && 0.24\% & 5.19\%   \\ 
P-value & 0.5661 &&& 0.0652 &&& 0.0516 &  \\ 
~ & ~ & ~ & ~ & ~ & ~ & & &  \\ 
\multicolumn{2}{l}{\textit{Panel B: Conversion Rate}} & ~ & ~ & ~ & ~ & & &  \\
Treat &  0.000085 & 0.00105*** && 0.000444 & 0.000673** && 0.000242 & 0.00100***   \\ 
~ & (0.000291) & (0.000352) && (0.000343) & (0.000306) && (0.000317) & (0.000330)   \\
\%Change & 0.21\% & 1.69\%  && 0.76\% & 1.47\% && 0.49\% & 1.85\%  \\ 
P-value & 0.0573 &&& 0.3493 &&& 0.0738 &  \\ 
~ & ~ & ~ & ~ & ~ & ~ & & &  \\ 
\multicolumn{2}{l}{\textit{Panel C: Cart Value}} & ~ & ~ & ~ & ~ & & &  \\
Treat &  0.550 & 0.238 && -0.042 & 0.773 && -0.054 & 0.613   \\ 
~ & (0.554) & (0.267) && (0.287) & (0.486) && (0.325) & (0.426)   \\
\%Change & 2.06\% & 1.34\%  && -0.22\% & 3.29\% && -0.26\% & 2.75\%  \\ 
P-value & 0.7711 &&& 0.1484 &&& 0.2069 &  \\ 
~ & ~ & ~ & ~ & ~ & ~ & & &  \\ 
Observation (Panel A\&B) & 1,849,382 & 1,849,382 && 1,849,382 & 1,849,382 && 1,849,382 & 1,849,382  \\
Observation (Panel C) & 76,536 & 115,804 && 108,790 & 85,195 && 91,774 & 100,643  \\ \hline\hline
\end{tabular}
\begin{tablenotes}   
\scriptsize  
\item[1] We classify sellers into high and low groups based on three different proxies. Low-group sellers are generally small sellers, as defined by meeting the following pre-experiment criteria: (1) accounting for the bottom 50\% cumulative share of total sales; (2) having operated on the platform for fewer than five years; or (3) maintaining fewer than three sub-accounts for their online store. Platform-operated sellers are classified as large.
\item[2] ``Sales" represents the total expenditure on product orders. ``Conversion Rate" measures consumers' likelihood of making purchases. It is a binary indicator for purchase, which equals 1 if a consumer makes at least one order during the experimental period, and 0 otherwise. ``Cart Value" refers to the expenditure per consumer, conditional on making a purchase.
\item[3] Standard errors are in parentheses. \% Change is calculated by dividing the treatment effect by the control group mean. *** p$<$0.01, ** p$<$0.05, * p$<$0.1.
\end{tablenotes}
\end{threeparttable} 
\end{table}

\begin{table}
\caption{Seller HTE for Marketing Push Message} 
\label{table:seller_hte_marketing_push_message}
\centering
\scriptsize
\begin{threeparttable}
\begin{tabular}{m{2cm} >{\raggedright}m{1.5cm} >{\raggedright}m{1.8cm} >{\raggedright}m{0.1em} >{\raggedright}m{1.5cm} >{\raggedright}m{1.5cm} >{\raggedright}m{0.1em} >{\raggedright}m{1.5cm} m{1.5cm}}
\hline\hline
~ & (1) & (2) && (3) & (4) && (5) & (6)  \\\cline{2-3}\cline{5-6}\cline{8-9}
~ &  \multicolumn{2}{l}{Annual Seller Sales} && \multicolumn{2}{l}{Operation Years} && \multicolumn{2}{l}{\# of Sub-Accounts}   \\ 
~ &  High & Low && High & Low && High & Low  \\ \hline
\multicolumn{2}{l}{\textit{Panel A: Sales}} & ~ & ~ & ~ & ~ &   &   &  \\
Treat         & 0.000216 & 0.000186 && 0.000453 & -0.0000514 && -0.0000206 & 0.000422 \\
 & (0.000594) & (0.000559) && (0.000579) & (0.000575) && (0.000517) & (0.000631) \\
\% Change    & 1.84\% & 1.38\% && 3.31\% & -0.45\% && -0.18\% & 3.14\%    \\
P-value &  0.9436   &&&  0.5660   &&&  0.6060   &  \\ 
~ & ~ & ~ & ~ & ~ & ~ & & &  \\ 
\multicolumn{2}{l}{\textit{Panel B: Conversion Rate}} & ~ & ~ & ~ & ~ & & &  \\
Treat         & 0.0000240 & 0.0000236 && 0.0000166 & 0.0000310** && 0.00000409 & 0.0000435*** \\
 & (0.0000147) & (0.0000161) && (0.0000170) & (0.0000137) && (0.0000157) & (0.0000152) \\
\% Change    & 3.28\% & 2.69\% && 1.68\% & 4.97\% && 0.48\% & 5.68\%  \\
P-value &   0.8299  &&&    0.2370  &&&  0.0557   &  \\ 
~ &  & ~ & ~ & ~ & ~ &  & &  \\ 
\multicolumn{2}{l}{\textit{Panel C: Cart Value}} & ~ & ~ & ~ & ~ & & &  \\
Treat &  -0.221 & -0.195 && 0.223 & -0.951 && -0.0912 & -0.420  \\
 & (0.734) & (0.564) && (0.530) & (0.813) && (0.553) & (0.729) \\
\% Change    & -1.38\% & -1.27\% && 1.61\% & -5.16\% && -0.66\% & -2.40\%  \\
P-value & 0.9852 &&& 0.2488 &&& 0.7633 & \\ 
~ &  & ~ & ~ & ~ & ~ &  & &  \\ 
Observation (Panel A\&B) & 13,715,528 & 13,715,528 && 13,715,528 & 13,715,528 && 13,715,528 & 13,715,528  \\ 
Observation (Panel C) & 10,212 & 12,213 && 13,649 & 8,776 && 11,618 & 10,807  \\ 
\hline\hline
\end{tabular}
\begin{tablenotes}   
\scriptsize  
\item[1] Refer to Table \ref{table:seller_hte_search_query_refinement} for detailed notes.
\end{tablenotes}
\end{threeparttable} 
\end{table}

\begin{table}
\caption{Seller HTE for Google Advertising Title} 
\label{table:seller_hte_google_advertising_title}
\centering
\scriptsize
\begin{threeparttable}
\begin{tabular}{m{2cm} >{\raggedright}m{1.5cm} >{\raggedright}m{1.8cm} >{\raggedright}m{0.1em} >{\raggedright}m{1.5cm} >{\raggedright}m{1.5cm} >{\raggedright}m{0.1em} >{\raggedright}m{1.5cm} m{1.5cm}}
\hline\hline
~ & (1) & (2) && (3) & (4) && (5) & (6)  \\\cline{2-3}\cline{5-6}\cline{8-9}
~ &  \multicolumn{2}{l}{Annual Seller Sales} && \multicolumn{2}{l}{Operation Years} && \multicolumn{2}{l}{\# of Sub-Accounts}   \\ 
~ &  High & Low && High & Low && High & Low \\ \hline
\multicolumn{2}{l}{\textit{Panel A: Sales}} & ~ & ~ & ~ & ~ &   &   &  \\
Treat     & -0.00605  & -0.00610  &  & -0.00656  & -0.00578  &  & -0.00759 & -0.00562  \\
          & (0.00711) & (0.00797) &  & (0.00935) & (0.00650) &  & (0.0125) & (0.00589) \\
\% Change & -4.97\%    & -4.27\%    &  & -5.23\%    & -4.27\%    &  & -5.54\%   & -4.29\%    \\
P-value &  0.9301    &&&  0.9099  &&&   0.9029  &  \\ 
~ & ~ & ~ & ~ & ~ & ~ & ~ & ~ &   \\
\multicolumn{2}{l}{\textit{Panel B: Conversion Rate}} & ~ & ~ & ~ & ~ & & & \\
Treat   & -0.000186  & 0.0000057  && -0.0000884    & -0.0000906   && -0.0000464   & -0.000100   \\
        & (0.000155) & (0.000159)  && (0.000189)    & (0.000137)  && (0.000232)  & (0.000127)    \\
\% Change & -4.82\% & 0.14\% &  & -2.46\% & -2.23\% &  & -1.33\% & -2.49\% \\
P-value &  0.4048   &&&  0.9713   &&&   0.8736  &  \\ 
~ &  & ~ & ~ & ~ & ~ &  & &  \\
\multicolumn{2}{l}{\textit{Panel C: Cart Value}} & ~ & ~ & ~ & ~ & & &  \\
Treat &  -0.0528 & -1.588 && -0.990 & -0.697 && -1.672 & -0.602 \\
 & (1.372) & (1.425) && (1.896) & (1.162) && (2.520) & (1.071) \\
\% Change    & -0.17\% & -4.41\% && -2.84\% & -2.09\% && -4.27\% & -1.85\%  \\
P-value & 0.4551 &&& 0.9014 &&& 0.7378 & \\ 
~ &  & ~ & ~ & ~ & ~ &  & &  \\ 
Observations (Panel A\&B) & 622,083 & 621,933 &  & 397,339 & 846,677 &  & 258,157 & 985,859 \\
Observation (Panel C) & 2,343 & 2,468 && 1,411 & 3,400 && 897 & 3,914  \\ 
\hline\hline
\end{tabular}
\begin{tablenotes}   
\scriptsize  
\item[1] Refer to Table \ref{table:seller_hte_search_query_refinement} for detailed notes.
\end{tablenotes}
\end{threeparttable} 
\end{table}

\section{Details on Product Treatment Effect Heterogeneity}\label{appendix:detailed_hte_product}
\setcounter{figure}{0}
\renewcommand{\thefigure}{F\arabic{figure}}
\renewcommand{\theHfigure}{F\arabic{figure}}
\setcounter{table}{0}
\renewcommand{\thetable}{F\arabic{table}}
\renewcommand{\theHtable}{F\arabic{table}}

\paragraph{Pre-sale Service Chatbot} Table \ref{table:product_hte_pre_sale_service_chatbot} reports the product-level heterogeneous treatment effects of the Pre-sale Service Chatbot experiment. When focusing on conversion rates, we detect statistically significant differences between product groups classified by annual sales quantity, indicating that the GenAI-driven chatbot service is more effective at converting consumers for tail products.

\begin{table}
\caption{Product HTE for Pre-sale Service Chatbot} 
\label{table:product_hte_pre_sale_service_chatbot}
\centering
\small
\scriptsize
\begin{threeparttable}
\begin{tabular}{m{2cm} >{\raggedright}m{1.5cm} >{\raggedright}m{1.5cm} >{\raggedright}m{0.1em} >{\raggedright}m{1.5cm} >{\raggedright}m{1.5cm} >{\raggedright}m{0.1em} >{\raggedright}m{1.5cm} m{1.5cm} }
\hline\hline
~ & (1) & (2) && (3) & (4) && (5) & (6)  \\\cline{2-3}\cline{5-6}\cline{8-9}
~ &  \multicolumn{2}{l}{Market Concentration} && \multicolumn{2}{l}{Annual Quantity}&& \multicolumn{2}{l}{Price}   \\ 
~ &  High & Low  && High & Low  && High & Low  \\ \hline
\multicolumn{2}{l}{\textit{Panel A: Sales}} & ~ & ~ & ~ & ~  & & & \\ 
Treat      & 0.136**  & 0.138*  && 0.0597   & 0.214***  && 0.137*  & 0.137**     \\
         & (0.0532) & (0.0821)   &&  (0.0564)   & (0.0799)   && (0.0782)  & (0.0589)  \\
\% Change  &  21.28\% &  13.18\% && 8.77\%  &  21.28\% &&  16.61\% &  15.91\%   \\ 
P-value &  0.4724   &&& 0.2696    &&&  0.9564   &  \\ 
~ & ~ & ~ & ~ & ~ & ~  &&& \\ 
\multicolumn{2}{l}{\textit{Panel B: Conversion Rate}}& ~  & & &  & & &   \\
Treat     & 0.00647*** & 0.00666*** && 0.00265   & 0.0105***  && 0.00404**  & 0.00910***  \\
          & (0.00166) & (0.00196)   && (0.00162) & (0.00198) && (0.00158)   & (0.00202) \\
\% Change  &  26.60\% &  18.48\% && 10.32\% &  30.28\%  && 17.44\% &  24.46\% \\
P-value &  0.3443  &&&   0.0182   &&& 0.4139    &  \\ 
\\
\multicolumn{2}{l}{\textit{Panel C: Cart Value}} & ~ & ~ & ~ & ~ & & &  \\
Treat   & -1.108 & -1.300 &&  -0.366 & -1.990 && -0.258  & -1.584* \\
 & (1.182) & (1.551)  && (1.370) & (1.472) && (2.182) & (0.926) \\
\% Change    & -4.22\% & -4.47\%  && -1.38\% & -6.86\% && -0.72\% & -6.84\%  \\
P-value &  0.9704   &&& 0.4498    &&& 0.3981    &  \\ 
~ & ~ & ~ & ~ & ~ & ~ & & &  \\ 
Observation (Panel A\&B) &  44,614  &44,614  &&44,614  &44,614  &&44,614  &44,614    \\ 
Observation (Panel C)& 1,274  & 1,802   && 1,223 & 1,853 &&1,151  &1,925    \\ 
\hline\hline
\end{tabular}
\begin{tablenotes}   
\scriptsize  
\item[1] We classify products into high and low groups based on three key pre-experiment dimensions. (1) Category market concentration, measured by the sales share of the top 1\% of products (ranked by annual sales) within each category. Products in the low group belong to categories with concentration levels below the platform average. (2) Annual sales quantity, also defined within the category. Products in the low group—referred to as tail products—are those comprising the bottom 50\% cumulative share of total sales quantity within each category. (3) Product price, defined within each category to control for category-level pricing variation. Low-priced products are those priced below the median of their respective category. 
\item[2] ``Sales" represents the total expenditure on product orders. ``Conversion Rate" measures consumers' likelihood of making purchases. It is a binary indicator for purchase, which equals 1 if a consumer makes at least one order during the experimental period, and 0 otherwise. ``Cart Value" refers to the expenditure per consumer, conditional on making a purchase.
\item[3] Standard errors are in parentheses. \% Change is calculated by dividing the treatment effect by the control group mean. *** p$<$0.01, ** p$<$0.05, * p$<$0.1.
\end{tablenotes}
\end{threeparttable} 
\end{table}

\paragraph{Search Query Refinement} In Table \ref{table:product_hte_search_query_refinement}, we find that for sales, the gains are larger and statistically significant only for products in low-concentration categories (Column 2), products with lower sales volume (Column 4), and products with high price level (Column 5). The differences between these two product groups are statistically significant when products are classified by market concentration and annual quantity, indicating that GenAI reduces search frictions in settings with greater product differentiation or limited sales history.

\begin{table}
\caption{Product HTE for Search Query Refinement} 
\label{table:product_hte_search_query_refinement}
\centering
\small
\scriptsize
\begin{threeparttable}
\begin{tabular}{m{2cm} >{\raggedright}m{1.5cm} >{\raggedright}m{1.5cm} >{\raggedright}m{0.1em} >{\raggedright}m{1.5cm} m{1.5cm} >{\raggedright}m{0.1em} >{\raggedright}m{1.5cm} m{1.5cm}}
\hline\hline
~ & (1) & (2) && (3) & (4) && (5) & (6)  \\\cline{2-3}\cline{5-6}\cline{8-9}
~ &  \multicolumn{2}{l}{Market Concentration} && \multicolumn{2}{l}{Annual Quantity}&& \multicolumn{2}{l}{Price}   \\ 
~ &  High & Low  && High & Low  && High & Low  \\ \hline
\multicolumn{2}{l}{\textit{Panel A: Sales}} & & & & & &  & \\ 
Treat        &   -0.00862    & 0.0734***   &&  -0.00381  & 0.0686*** && 0.0590**   & 0.00586   \\
             & (0.0174)   & (0.0253) && (0.0156)  & (0.0258) && (0.0296) & (0.00864) \\
\% Change    &     -0.80\%    & 6.49\%   &&   -0.47\%   & 4.92\%    && 3.79\%    & 0.89\%    \\
P-value & 0.0067 &&& 0.0332 &&& 0.1937 &  \\ 
~ & & & & & &&& \\ 
\multicolumn{2}{l}{\textit{Panel B: Conversion Rate}} & & & &  & & &   \\
Treat     & 0.000247  & 0.000823*** &&  0.0000005  & 0.00121*** && 0.000526* & 0.000548   \\
          & (0.000332)  & (0.000302) && (0.000305) & (0.000343) && (0.000301) & (0.000342)    \\
\% Change  & 0.45\%   & 1.85\%   &&  0.00\%   & 2.07\%  && 1.19\%    & 0.94\%     \\
P-value & 0.0888 &&& 0.0064 &&& 0.7493 &  \\ 
~ & & & & & &&& \\ 
\multicolumn{2}{l}{\textit{Panel C: Cart Value}} & ~ & ~ & ~ & ~ & & &  \\
Treat &  -0.292 & 1.106** && -0.098 & 0.587 && 0.717 & -0.032  \\ 
~ & (0.291) & (0.534) && (0.319) & (0.411) && (0.615) & (0.132)   \\
\%Change & -1.49\% & 4.35\%  && -0.55\% & 2.46\% && 2.04\% & -0.28\%  \\ 
P-value & 0.0208 &&& 0.2052 &&& 0.2583 &  \\ 
~ & & & & & &&& \\ 
Observations (Panel A\&B) & 1,849,382 & 1,849,382 && 1,849,382 & 1,849,382 && 1,849,382 & 1,849,382   \\ 
Observation (Panel C) & 101,975 & 82,912 && 84,799 & 109,265 && 82,272  & 108,485    \\ \hline\hline
\end{tabular}
\begin{tablenotes}   
\scriptsize  
\item[1] Refer to Table \ref{table:product_hte_pre_sale_service_chatbot} for detailed notes.
\end{tablenotes}
\end{threeparttable} 
\end{table}

\paragraph{Product Description} Table \ref{table:product_hte_product_description} shows that, for sales, statistically significant differences across product groups arise only when products are classified by price level. This pattern is consistent with GenAI being more effective at reducing information asymmetries in settings with higher decision stakes.

\begin{table}
\caption{Product HTE for Product Description} 
\label{table:product_hte_product_description}
\centering
\small
\scriptsize
\begin{threeparttable}
\begin{tabular}{m{2cm} >{\raggedright}m{1.5cm} >{\raggedright}m{1.5cm} >{\raggedright}m{0.1em} >{\raggedright}m{1.5cm} m{1.5cm} >{\raggedright}m{0.1em} >{\raggedright}m{1.5cm} m{1.5cm}}
\hline\hline
~ & (1) & (2) && (3) & (4) && (5) & (6)  \\\cline{2-3}\cline{5-6}\cline{8-9}
~ &  \multicolumn{2}{l}{Market Concentration} && \multicolumn{2}{l}{Annual Quantity}&& \multicolumn{2}{l}{Price}   \\ 
~ &  High & Low  && High & Low  && High & Low  \\ \hline
\multicolumn{2}{l}{\textit{Panel A: Sales}} & & & & & & & \\ 
Treat     &  0.00932*** & 0.00113  && 0.00497*  & 0.00547* && 0.00760***   & 0.00285    \\
          & (0.00299)  & (0.00284) && (0.00281)  & (0.00299) && (0.00291) & (0.00286)  \\
\% Change  &  3.10\%  & 0.55\%  &&  2.10\%  & 2.03\%  && 4.10\%    & 0.89\%    \\
P-value & 0.1309 &&& 0.9641 &&& 0.0703 &  \\ 
~ & & & & & &&& \\ 
\multicolumn{2}{l}{\textit{Panel B: Conversion Rate}} &  & & &  & & &   \\
Treat     & 0.000384**   & 0.000141 && 0.000456***  & 0.000166 && 0.000393*** & 0.000209  \\
          & (0.000151)  & (0.000120)  && (0.000134) & (0.000141) && (0.000100)  & (0.000164) \\
\% Change & 1.38\%   & 0.81\%   && 2.07\%    & 0.69\%  && 3.32\%  & 0.63\%   \\
P-value & 0.5014 &&& 0.0845 &&& 0.0047 &  \\ 
~ & & & & & &&& \\ 
\multicolumn{2}{l}{\textit{Panel C: Cart Value}} & ~ & ~ & ~ & ~ & & &  \\
Treat &  0.186** & -0.019 && 0.018 & 0.138 && 0.092 & 0.032  \\ 
~ & (0.089) & (0.141) && (0.108) & (0.104) && (0.204) & (0.071)   \\
\%Change & 1.72\% & -0.16\%  && 0.16\% & 1.24\% && 0.59\% & 0.33\%  \\ 
P-value & 0.1895 &&& 0.4258 &&& 0.8605 &  \\ 
~ & & & & & &&& \\ 
Observations (Panel A\&B) & 4,772,937 & 4,772,937 && 4,772,937 & 4,772,937 && 4,772,937 & 4,772,937   \\ 
Observation (Panel C) & 133,779 & 83,687 && 106,252 & 115,810 && 57,458  & 159,492    \\\hline\hline
\end{tabular}
\begin{tablenotes}   
\scriptsize  
\item[1] Refer to Table \ref{table:product_hte_pre_sale_service_chatbot} for detailed notes.
\end{tablenotes}
\end{threeparttable} 
\end{table}

\paragraph{Marketing Push Message}
Table \ref{table:product_hte_marketing_push_message} shows that, in most cases, we do not observe statistically significant differences between the two product groups.

\begin{table}
\caption{Product HTE for Marketing Push Message} 
\label{table:product_hte_marketing_push_message}
\centering
\small
\scriptsize
\begin{threeparttable}
\begin{tabular}{m{2cm} >{\raggedright}m{1.5cm} >{\raggedright}m{1.8cm} >{\raggedright}m{0.1em} >{\raggedright}m{1.5cm} >{\raggedright}m{1.5cm} >{\raggedright}m{0.1em} >{\raggedright}m{1.5cm} m{1.5cm} }
\hline\hline
~ & (1) & (2) && (3) & (4) && (5) & (6)  \\\cline{2-3}\cline{5-6}\cline{8-9}
~ & \multicolumn{2}{l}{Market Concentration} && \multicolumn{2}{l}{Annual Quantity}&& \multicolumn{2}{l}{Price}   \\ 
~ &   High & Low  && High & Low  && High & Low  \\ \hline
\multicolumn{2}{l}{\textit{Panel A: Sales}} & ~ & ~ & ~ & ~  & & &  \\ 
Treat    & 0.000146  & 0.000255  && -0.0000554  & 0.000457 && -0.000429  & 0.000831   \\
          & (0.000339) & (0.000742) && (0.000349)  & (0.000737) && (0.000511) & (0.000636)   \\
\% Change   & 3.03\%  &1.25\%   &&  -0.92\%  & 2.39\% && -5.21\%  & 4.90\%    \\
P-value &  0.8225   &&&   0.6347  &&&   0.1631  &  \\ 
~ & ~ & ~ & ~ & ~ & ~ & & &  \\ 
\multicolumn{2}{l}{\textit{Panel B: Conversion Rate}} & ~ & ~ & ~ & ~  & & &  \\
Treat   & 0.0000213** & 0.0000263  && 0.0000115  & 0.0000361* && 0.0000130 & 0.0000346*  \\
      & (0.00000921) & (0.0000198)  && (0.0000108)  & (0.0000190) && (0.00000976) & (0.0000195) \\
\% Change   & 7.59\% & 1.98\%  && 2.91\% & 2.97\% && 4.06\% & 2.68\%  \\
P-value &   0.1190  &&&  0.9813   &&&  0.6838   &  \\ 
\\
\multicolumn{2}{l}{\textit{Panel C: Cart Value}} & ~ & ~ & ~ & ~ & & &  \\
Treat         & -0.730  & -0.108 && -0.566 & -0.0896 && -2.292* & 0.285 \\
 & (1.038) & (0.504)  && (0.772) & (0.546) && (1.381) & (0.444) \\
\% Change    & -4.24\% & -0.71\%  && -3.71\% & -0.57\% && -8.92\% & 2.17\%  \\
P-value &   0.6095  &&&  0.6106   &&&  0.0825   &  \\ 
\\
&   &    &  &    &    &  &    &   \\
Observations (Panel A\&B) & 13,715,528 & 13,715,528 & ~ & 13,715,528 & 13,715,528& ~ & 13,715,528 & 13,715,528      \\ 
Observations (Panel C)   & 3,993  &  18,432   &&  5,498  &  16,927  &&  4,483    &  17,942    \\ 
~ & ~ & ~ & ~ & ~ & ~ & & &  \\ 
\hline\hline
\end{tabular}
\begin{tablenotes}   
\scriptsize  
\item[1] Refer to Table \ref{table:product_hte_pre_sale_service_chatbot} for detailed notes.
\end{tablenotes}
\end{threeparttable} 
\end{table}

\paragraph{Google Advertising Title}
Table \ref{table:product_hte_google_advertising_titles} shows no statistically significant differences between the two product groups.

\begin{table}
\caption{Product HTE for Google Advertising Title} 
\label{table:product_hte_google_advertising_titles}
\centering
\scriptsize
\begin{threeparttable}
\begin{tabular}{m{2cm} >{\raggedright}m{1.5cm} >{\raggedright}m{1.5cm} >{\raggedright}m{0.1em} >{\raggedright}m{1.5cm} m{1.5cm} >{\raggedright}m{0.1em} >{\raggedright}m{1.5cm} m{1.5cm}}
\hline\hline
~ & (1) & (2) && (3) & (4) && (5) & (6)  \\\cline{2-3}\cline{5-6}\cline{8-9}
~ &  \multicolumn{2}{l}{Market Concentration} && \multicolumn{2}{l}{Annual Quantity}&& \multicolumn{2}{l}{Price}   \\ 
~ &  High & Low  && High & Low  && High & Low  \\ \hline
\multicolumn{2}{l}{\textit{Panel A: Sales}} & & & & & &  & \\ 
Treat        & -0.00564  & -0.00651  && -0.00247  & -0.0109  && -0.00831  & -0.00282  \\
             & (0.00709) & (0.00811) && (0.00736) & (0.00766) && (0.00758) & (0.00727) \\
\% Change    & -4.16\%    & -5.10\%   &&   -1.71\%    & -9.39\%  && -5.98\%    & -2.30\%    \\
P-value &   0.9073 &&& 0.3350   &&& 0.6386 &  \\ 
~ & & & & & &&& \\ 
\multicolumn{2}{l}{\textit{Panel B: Conversion Rate}} & & & &  & & &   \\
Treat     & -0.0000708  & -0.000113 && -0.0000432  & -0.000154 && -0.0000788  & -0.000116   \\
          & (0.000151) & (0.000165) && (0.000156) & (0.000155) && (0.000135) & (0.000187) \\
\% Change  & -1.73\%  & -3.09\%  &&  -0.98\%   & -4.72\%  && -2.37\%   & -2.46\%  \\
P-value &   0.8177  &&&  0.5241  &&&  0.9891   &  \\ 
&   &    &  &    &    &  &    &   \\
\multicolumn{2}{l}{\textit{Panel C: Cart Value}} & ~ & ~ & ~ & ~ & & &  \\
Treat         & -0.819 & -0.723&& -0.242 & -1.733 && -1.546 & 0.0417 \\
 & (1.258) & (1.608) && (1.215) & (1.714) && (1.567) & (1.159) \\
\% Change    & -2.48\% & -2.07\% && -0.74\% & -4.87\% && -3.69\% & 0.16\%  \\
P-value &  0.9452   &&&  0.5023   &&&   0.4771  &  \\ 
\\
Observations (Panel A\&B) & 714,872    & 529,144    && 714,642    & 529,374    && 714,642    & 529,374    \\ 
Observations (Panel C)&  2,906   & 1,905     &&  3,126  &   1,685 &&  2,347    & 2,464     \\ 
\hline\hline
\end{tabular}
\begin{tablenotes}   
\scriptsize  
\item[1] Refer to Table \ref{table:product_hte_pre_sale_service_chatbot} for detailed notes.
\end{tablenotes}
\end{threeparttable} 
\end{table}

\end{document}